\documentclass[twocolumn,aps,pra,a4paper,superscriptaddress]{revtex4} %

\usepackage{graphicx}
\usepackage{amssymb}
\usepackage{amsmath}

\begin{document}

\title{Cascade atom in high-Q cavity:\\ The spectrum for non-Markovian decay}

\author{B.J.~Dalton} 
  \affiliation{ ARC Centre for Quantum-Atom Optics and
  Centre for Atom Optics and Ultrafast Spectroscopy,\\ Swinburne University of
  Technology, Melbourne, Victoria 3122, Australia}

\author{B.M.~Garraway}
 \affiliation{Department of Physics and Astronomy,
 University of Sussex, Falmer, Brighton, BN1 9QH, United Kingdom}

\date{\today}

\begin{abstract}
The spontaneous emission spectrum for a three level cascade configuration
atom in a single mode high-Q cavity coupled to a zero temperature reservoir
of continuum external modes is determined from the atom-cavity mode master
equation using the quantum regression theorem. Initially the atom is in its
upper state and the cavity mode empty of photons. Following Glauber, the
spectrum is defined via the response of a detector atom. Spectra are
calculated for the detector located inside the cavity (case~A), outside the
cavity end mirror (Case B---end emission), or placed for emission out the
side of the cavity (Case C). The spectra for case~A and case~B are found to
be essentially the same. In all the cases the predicted lineshapes are free
of instrumental effects and only due to cavity decay. Spectra are presented
for intermediate and strong coupling regime situations (where both atomic
transitions are resonant with the cavity frequency), for cases of non-zero
cavity detuning, and for cases where the two atomic transition frequencies
differ. The spectral features for Cases B(A) and C are qualitatively
similar, with six spectral peaks for resonance cases and eight for detuned
cases. These general features of the spectra can be understood via the
dressed atom model. However, Case B and C spectra differ in detail, with the
latter exhibiting a deep spectral hole at the cavity frequency due to
quantum interference effects.
\end{abstract}

\maketitle

\section{Introduction}

\label{Section 1}

Since the original paper of Purcell \cite{Purcell46a} the modification of
atomic radiative decay due to the presence of optical and microwave
cavities, photonic band gap (PBG)\ materials has become a familiar topic of
study in quantum optics. \ The presence of these systems structures the
modes of the electromagnetic (EM)\ field, resulting in situations where the
mode frequency dependences of the mode density or the atom-field coupling
constants (or both) are changed from the free space situation. The quantum
EM field acts as a structured reservoir whose behaviour depends on the
product of the mode density with the square of the coupling constants---the
reservoir structure functions. \ If the reservoir structure functions vary
slowly with mode frequency, the corresponding reservoir correlation time may
be short compared to the time scale over which the atomic state changes, and
the atomic density operator satisfies a Markovian master equation. This
results in the irreversible decay of excited atomic state populations,
though with a decay rate different to that for radiative decay in free
space. This regime is characterised by a weak coupling between the atom and
the EM\ field. On the other hand, if the reservoir structure functions vary
rapidly with mode frequency, the corresponding reservoir correlation time
may be long compared to the time scale over which the atomic state changes,
and the atomic density operator no longer satisfies a Markovian master
equation. A reversible oscillatory decay of excited atomic state populations
results, with a period related to the atom-field coupling constant. \ This
regime is characterised by a strong coupling between the atom and the EM\
field.

There are a large variety of effects associated with atoms in structured EM
fields, including modifications to the spontaneous emission, fluorescence
and probe absorption spectra, and comprehensive surveys may be found in
several reviews, for example \cite{Berman94a,Lambropoulos00a,Walther06a}.
Amongst the many effects is the modification of the spectrum of the
radiation emitted when an excited atom decays into initially empty field
modes. It has been known from the early days of quantum optics that for an
initially excited two-level atom (TLA) in free space, this so-called
spontaneous emission (SE) spectrum is a Lorentzian centred around the atomic
transition frequency and with a width given by the excited atom decay rate 
\cite{Weisskopf30a}. In the weak coupling regime the spectrum will remain
Lorentzian, but with a modified width reflecting the changed atomic decay
rate. The situation changes qualitatively for the strong coupling regime,
with the spectrum no longer remaining Lorentzian. The essential physics for
a TLA coupled to a single mode cavity in the strong coupling regime can be
described via the Jaynes-Cummings model \cite{JaynesCummings63a}, where the
energy eigenstates allowing for the atom-cavity mode coupling are the
so-called dressed atom states \cite{Cohen-Tannoudji77a}.

The first calculation for the SE spectrum for a TLA in an ideal lossless
cavity was carried out by Sanchez-Mondragon et al.\ \cite{Sanchez83a}. Using a
Heisenberg equation of motion approach and determining the spectrum from the
two-time correlation function for the atomic dipole operator, they found a
double peaked spectrum with a separation between peaks given by the
so-called one-photon Rabi frequency, that is the Rabi frequency associated
with one photon in the cavity mode. Line shapes were associated with the
spectrometer bandwidth \cite{Eberly77a}, since no cavity bandwidth was
included. Another early calculation for a TLA in a high-Q cavity was that of
Agarwal \cite{Agarwal86a}, who used the dressed atom model to predict the SE
and weak probe absorption spectra. This article determines the SE\ spectrum
for different positions of the spectrometer, the spectrum associated with
sideways emission via direct SE\ being given in terms of the two-time dipole
correlation function, whilst the spectrum associated with end emission via
the cavity output mirror is given in terms of the two-time correlation
function for cavity creation and annihilation operators. Although
qualitatively similar in exhibiting vacuum Rabi frequency splitting, the
spectra are quantitatively different. The SE spectra shown had lineshapes
determined for the situation where the spectrometer bandwidth was large
compared to the cavity decay rate. Non-Lorentzian SE spectra were also found
by Lewenstein et al.\ \cite{Lewenstein88a} for emission from a TLA into both
high Q cavities and PBG systems, here an essential states approach was used
and the spectrum defined in terms of the long time probability for finding
one photon in a cavity or background mode. Using a master equation and
quantum regression theorem \cite{Lax63a67a} approach, Carmichael et al.\ \cite%
{Carmichael89a} took into account both cavity damping and spontaneous
emission damping to determine the side emission SE spectrum for a TLA\ in a
high-Q cavity from the two-time dipole correlation function, based on a zero
bandwidth spectrometer. Further discussion and references on the theory of
the SE\ spectrum for a TLA in a high-Q cavity are given by Childs et al.\ \cite%
{Childs94a} and Carmichael et al.\ \cite{Carmichael94a} in the review \cite%
{Berman94a}.

No \emph{direct} experimental measurements of the SE\ spectrum for a single
TLA in a high-Q cavity appear to have been carried out. However there are
related experiments in the strong coupling regime, such as on the absorption
spectrum for a weak probe field that demonstrates single atom vacuum Rabi
splitting \cite{Thompson92a,Boca04a}, or single atom Rabi oscillations \cite%
{Brune96a}, or the collapse and revival experiments with a single TLA\ \cite%
{Rempe87a} that demonstrate cavity mode quantization \cite%
{Eberly80a,Barnett86a}. All these measurements can be interpreted in terms
of the Jaynes-Cummings model \cite{JaynesCummings63a} and the associated
dressed atom states \cite{Cohen-Tannoudji77a}. The direct measurement of SE\
spectra would probably only be possible in the optical regime, since SE in
the microwave regime would be too weak to detect. For the case of a single
TLA in a high-Q cavity or a PBG system which is also coupled to a strong
laser field, there is an extensive literature dealing with the related
spectra, in particular the spectra associated with resonance fluorescence or
probe absorption. The work of Mollow \cite{Mollow69a72a} for the free space
situation predicted a three-peaked (AC Stark effect) fluorescence spectrum,
and a dispersion-like probe absorption spectrum. For a TLA in an ideal
lossless cavity in a coherent state Agarwal et al.\ \cite{Agarwal91a}
predicted a large number of spectral lines in the fluorescence spectra,
reflecting the contributions from the numerous dressed atom states
involved.\ These spectra can also be interpreted via the dressed atom model,
see the recent reviews \cite{Berman94a,Lambropoulos00a,Walther06a}.

In contrast to the case of the SE\ spectrum for a TLA in a high-Q cavity or
a PBG system, the case of three level atoms (3LA) has received little
attention. The SE spectrum for a 3LA in a lambda configuration in a PBG
system has been treated by John et al.\ \cite{John94a} via the essential
states approach for various detunings from the band gap edge, the spectrum
(which shows doublet and window effects) being defined via the photon
emission probability. Ashraf \cite{Ashraf94a} determined the SE spectrum for
a 3LA in a lambda configuration in an ideal lossless cavity using the
dressed states approach, the spectrum (which shows a doublet structure
reflecting the one photon Rabi frequency) being given via the two-time
dipole correlation function. Line shapes were due to the spectrometer
bandwidth. For a 3LA in a cascade configuration in a PBG system, Bay et al.\ 
\cite{Bay98a} used the essential states approach to determine the SE\
spectrum for various detunings of one of the atomic transitions from the
band gap edge, the latter (which is strongly non-Lorentzian) being defined
via the photon emission probability. The treatment applies when the two
atomic frequencies are quite distinct, enabling only one at a time to
interact strongly with the PBG. Paspalakis et al.\ \cite{Paspalakis99a}
treated the case of SE from a 3LA in a lambda configuration in a PBG system,
using the same approach and spectrum definition as \cite{John94a}. Similar
spectral features were found, including the window not previously noted by
John et al. The case of a 3LA in a cascade configuration in an ideal
lossless cavity has been studied by Zhou et al.\ \cite{Zhou05a} using a
dressed atom approach and both end and side SE spectra were determined. Line
shapes were due to the spectrometer bandwidth, as no cavity damping was
included. The cavity mode was resonant with the average of the atomic
transition frequencies. For the situation where the two atomic transition
frequencies were equal, six peaks were found in both the side emission and
end emission SE spectra, though in the latter case two of these were
negligible. When the two atomic transition frequencies differed, there were
eight peaks in the side emission spectrum.

In a recent paper \cite{Garraway06a} we have considered the non-Markovian
decay of a three level cascade atom with both transitions coupled to a
single structured reservoir of quantized field modes for both a high-Q
cavity and in a PBG system. Based on the approach given in \cite{Dalton01a}
the dynamics of this system has been treated via the essential states
approach, using Laplace transform methods applied to the coupled amplitude
equations. Non-Markovian behaviour for the population dynamics of the atomic
system was found, such as oscillatory decay for the high-Q cavity case and
population trapping for the photonic band-gap case. A Markovian master
equation approach was also applied, in which the atomic system was augmented
by a small number of discrete quasimodes or pseudomodes, which in the
quasimode treatment themselves undergo Markovian relaxation into a flat
reservoir of continuum quasimodes. For the high-Q cavity case a single
discrete quasimode was involved, for the PBG case two coupled discrete
quasimodes were needed. The essential states and Markoff methods gave
identical results, showing that complicated non-Markovian behaviour can be
treated by enlarging the non-Markovian system, thereby turning a
non-Markovian problem into a Markovian one.

In the present paper we now consider the SE\ spectrum for the case of a
cascade atom in a high-Q cavity. The spectrum is much richer than for the
TLA case and quantum interference effects may now occur. 
Quantum coherence and interference phenomena are central in many
applications of fundamental quantum optics results, and a comprehensive
review of such effects may be found in a special issue of this journal on
quantum inteference \cite{Swain02a}. Unlike
previous work, we will consider the case of an ideal spectrometer with zero
bandwidth in order to display the actual SE\ spectral line shapes without
these being masked by instrumental effects. A master equation approach
incorporating cavity decay will be used to evaluate the spectra via the
quantum regression theorem. Both end (Case A(B)) and side (Case C) SE spectra
will be calculated, and we aim to exhibit the expected quantum interference
effects that can occur in a cascade system. We will show that the spectrum
in both cases are given in terms of the Laplace transforms of both the
atom-cavity mode density matrix elements and the evolution operator matrix
elements.

Before developing the mathematical formalism it may be useful to consider
the physical processes that can be involved in the decay of the cascade atom
in a high-Q cavity and the registration of a photon arrival in a suitable
detector---this will enable possible interference effects to be identified.
In a cascade atom initially in its upper state $\left\vert 2\right\rangle $
and coupled to a high Q cavity empty of photons, two photons can be emitted
before the atom makes a transition to its lowest state $\left\vert
0\right\rangle $ via the intermediate state $\left\vert 1\right\rangle $. If
there is a detector atom weakly coupled to the atom-cavity mode system and
prepared in its lower state $\left\vert A\right\rangle $, the detector atom
could absorb one of these photons and make a transition to its upper state $%
\left\vert B\right\rangle $. Following the approach of Glauber \cite%
{Glauber65a} the overall transition probability for the $\left\vert
A\right\rangle \rightarrow \left\vert B\right\rangle $ process considered as
a function of the detector atom transition frequency $\omega $ may be used
for an operational definition of the spontaneous emission spectrum for this
cascade atom in a high Q cavity---the cavity mode being coupled to empty
external modes via the cavity mirror. Since the detector atom is only weakly
coupled we need only consider processes with a single $\left\vert
A\right\rangle \rightarrow \left\vert B\right\rangle $ transition. Also,
steps where a photon is created in an external mode would not be reversible.
On the other hand if the atom-cavity mode coupling is strong, atomic
transitions accompanied by photon number changes in the cavity mode may be
reversible. If we consider the case (\emph{Case A}) where the detector atom
is placed inside the cavity, then the photon causing the detector atom
transition must have come from the cavity mode. However photons in the
cavity mode can also be transferred to an external mode via loss through the
cavity mirror, and if one photon is used to cause the $\left\vert
A\right\rangle \rightarrow \left\vert B\right\rangle $ transition in the
detector atom, the other will be transferred to an external mode. We
consider states of the combined detector atom, cavity mode, cascade atom,
external mode system of the product form $\left\vert D;n;\nu ;m\right\rangle 
$, where $D=A,B;$ $n=0,1,2,\hdots;$ $\nu =0,1,2;$ $m=0,1,2,\hdots$ specify the
detector atom state, the cavity photon number, the atomic state and the
photon number for a specific external mode respectively. Then the overall
process in which the cascade atom changes from upper state $\left\vert
2\right\rangle $ to lowest state $\left\vert 0\right\rangle $, the detector
atom changes from lower state $\left\vert A\right\rangle $ to upper state $%
\left\vert B\right\rangle $ and one photon appears in a specific external
mode, whilst the cavity mode is initially and finally empty of photons is
denoted $\left\vert A;0;2;0\right\rangle \rightarrow \left\vert
B;0;0;1\right\rangle $. However this overall process involving the overall
emission of two photons (one being absorbed by the detector atom, the other
appearing in an external mode) has a number of different quantum pathways: 
\begin{widetext}
\begin{eqnarray}
\left\vert A;0;2;0\right\rangle &\leftrightarrow &\left\vert
A;1;1;0\right\rangle \rightsquigarrow \left\vert A;0;1;1\right\rangle
\leftrightarrow \left\vert A;1;0;1\right\rangle \rightarrow \left\vert
B;0;0;1\right\rangle  \notag \\
\left\vert A;0;2;0\right\rangle &\leftrightarrow &\left\vert
A;1;1;0\right\rangle \rightarrow \left\vert B;0;1;0\right\rangle
\leftrightarrow \left\vert B;1;0;0\right\rangle \rightsquigarrow \left\vert
B;0;0;1\right\rangle  \notag \\
\left\vert A;0;2;0\right\rangle &\leftrightarrow &\left\vert
A;1;1;0\right\rangle \leftrightarrow \left\vert A;2;0;0\right\rangle
\rightsquigarrow \left\vert A;1;0;1\right\rangle \rightarrow \left\vert
B;0;0;1\right\rangle  \notag \\
\left\vert A;0;2;0\right\rangle &\leftrightarrow &\left\vert
A;1;1;0\right\rangle \leftrightarrow \left\vert A;2;0;0\right\rangle
\rightarrow \left\vert B;1;0;0\right\rangle \rightsquigarrow \left\vert
B;0;0;1\right\rangle
\end{eqnarray}%
Reversible atom-cavity mode transitions are designated $\leftrightarrow $,
irreversible cavity-external mode transitions $\rightsquigarrow $ and weak
detector atom transitions $\rightarrow $. In addition, each of the
reversible steps in the above pathways may involve further pathways if the
atom-cavity mode coupling is strong. For example, the first pathway 
\begin{equation}
\left\vert A;0;2;0\right\rangle \leftrightarrow \left\vert
A;1;1;0\right\rangle \rightsquigarrow \left\vert A;0;1;1\right\rangle
\leftrightarrow \left\vert A;1;0;1\right\rangle \rightarrow \left\vert
B;0;0;1\right\rangle
\end{equation}%
could branch into two different sub-pathways 
\begin{eqnarray}
\left\vert A;0;2;0\right\rangle &\leftrightarrow &\left\vert
A;1;1;0\right\rangle \leftrightarrow \left\vert A;2;0;0\right\rangle
\leftrightarrow \left\vert A;1;1;0\right\rangle \rightsquigarrow \left\vert
A;0;1;1\right\rangle \leftrightarrow  
\left\vert A;1;0;1\right\rangle \rightarrow \left\vert
B;0;0;1\right\rangle \\
\left\vert A;0;2;0\right\rangle &\leftrightarrow &\left\vert
A;1;1;0\right\rangle \leftrightarrow \left\vert A;0;2;0\right\rangle
\leftrightarrow \left\vert A;1;1;0\right\rangle \rightsquigarrow \left\vert
A;0;1;1\right\rangle \leftrightarrow  
\left\vert A;1;0;1\right\rangle \rightarrow \left\vert
B;0;0;1\right\rangle
\end{eqnarray}%
\end{widetext}
The possibilities for branching are endless. The transition amplitudes for
all these different pathways (and sub-pathways) are added to form the
overall transition amplitude, and hence the transition probability would be
expected to exhibit quantum interference effects. Thus the spontaneous
emission spectrum for a three level cascade atom should demonstrate
interesting interference phenomena.

Of course there will also be a detection processes in which the cascade atom
changes from upper state $\left\vert 2\right\rangle $ to intermediate state $%
\left\vert 1\right\rangle $, the detector atom changes from lower state $%
\left\vert A\right\rangle $ to upper state $\left\vert B\right\rangle $ but
no photon appears in a specific external mode, whilst the cavity mode is
initially and finally empty of photons. This is denoted $\left\vert
A;0;2;0\right\rangle \rightarrow \left\vert B;0;1;0\right\rangle $ and
involves only one photon emission, the emitted photon being absorbed by the
detector atom. Here 
\begin{equation}
\left\vert A;0;2;0\right\rangle \leftrightarrow \left\vert
A;1;1;0\right\rangle \rightarrow \left\vert B;0;1;0\right\rangle
\end{equation}%
is the only quantum pathway, though again there are numerous sub-pathways
such as
\begin{widetext}
\begin{eqnarray}
\left\vert A;0;2;0\right\rangle &\leftrightarrow &\left\vert
A;1;1;0\right\rangle \leftrightarrow \left\vert A;2;0;0\right\rangle
\leftrightarrow \left\vert A;1;1;0\right\rangle \rightarrow \left\vert
B;0;1;0\right\rangle  \notag \\
\left\vert A;0;2;0\right\rangle &\leftrightarrow &\left\vert
A;1;1;0\right\rangle \leftrightarrow \left\vert A;0;2;0\right\rangle
\leftrightarrow \left\vert A;1;1;0\right\rangle \rightarrow \left\vert
B;0;1;0\right\rangle
\end{eqnarray}
\end{widetext}
if the atom-cavity mode coupling is strong . For this overall process there
still may be quantum interference effects. Similar processes to these would
apply if the only atomic states considered were $\left\vert 2\right\rangle $%
\ and $\left\vert 1\right\rangle $, with the coupling to the lower state $%
\left\vert 0\right\rangle $\ being set to zero, and sub-pathways involving $%
\left\vert 0\right\rangle $ excluded. In all these cases the amplitudes for
different sub-pathways would combine to produce an overall transition
amplitude, and hence interference effects giving maxima and minima in the
detection probability (spectrum) could occur. Whether these interference
effects can be be demonstrated directly via altering the system coupling
constants, detunings etc.\ is a separate issue.

A similar discussion can be presented for the case where the detector atom
is placed just outside the cavity mirror (\emph{Case B}) to detect end
emission, or where the detector atom is placed outside the cavity to detect
the weak side emission (\emph{Case C}). In case B it is the external mode
photon absorption that is associated with the detector atom transition, in
case C it is the transition in the cascade atom itself, so the details of
the discussion will differ for these two cases.

In Section \ref{Section 2} the general features of the quasimode description
used to treat the cascade atom-cavity mode system are covered, along with a
brief outline of the approach for defining the spectrum and how the quantum
regression theorem is used in the calculations. Section \ref{Section 3}
covers the master equation for the cascade atom-cavity mode system, the
related evolution operator and their determination in terms of Laplace
transforms. Expressions for the SE spectra for three different positions of
the spectrometer atom---inside the cavity (Case A), outside the cavity to
detect end emission (Case B) and outside the cavity to detect side emission
(Case C) are given in Section \ref{Section 3}. Numerical results for the SE
spectrum for the cascade atom in a high-Q cavity are presented in Section %
\ref{Section 4} and conclusions given in Section \ref{Section 5}. Details
are presented in the Appendices.

\section{General Theory}

\label{Section 2}

\subsection{Atom-quasimode system}

The general system of interest is a radiating atom with energy states $%
\left\vert E_{\alpha }\right\rangle $ coupled to a discrete set of
quasimodes $i$ of the quantum EM\ field, which in turn are coupled to a
continuum set of quasimodes $\Delta $. The Hamiltonian $\widehat{H}$ for the
atom-quasimodes system is given as the sum of the atomic Hamiltonian $%
\widehat{H}_{A}$, the quasimodes Hamiltonian $\widehat{H}_{Q}$ and an
interaction between the atom and quasimodes $\widehat{H}_{AQ}$ \cite%
{Dalton01a} 
\begin{equation}
\widehat{H}=\widehat{H}_{A}+\widehat{H}_{Q}+\widehat{H}_{AQ} ,
\end{equation}%
where%
\begin{eqnarray}
\widehat{H}_{A} &=&\sum\limits_{\alpha }\hbar \omega _{\alpha }\left\vert
E_{\alpha }\right\rangle \left\langle E_{\alpha }\right\vert \\
\widehat{H}_{Q} &=&\sum\limits_{i}\hbar \nu _{i}\,\widehat{a}_{i}^{\dag }%
\widehat{a}_{i}
+\sum\limits_{i\neq j}\hbar V_{ij}\,\widehat{a}_{i}^{\dag }%
\widehat{a}_{j}  \notag \\
&&+\int d\Delta \,\rho _{C}(\Delta )\,\hbar \Delta \,\widehat{b}(\Delta
)^{\dag }\widehat{b}(\Delta )
\notag \\ && %
+\sum\limits_{i}\int d\Delta \,\rho _{C}(\Delta
)\,(\hbar W_{i}(\Delta )\widehat{a}_{i}^{\dag }\widehat{b}(\Delta )+H.C.) 
\notag \\ && \\ %
\widehat{H}_{AQ} &=&\sum\limits_{E_{\alpha }>E_{\beta
}}\sum\limits_{i}(\hbar \lambda _{i;\alpha \beta }^{\ast }\,\widehat{a}%
_{i}\,\left\vert E_{\alpha }\right\rangle \left\langle E_{\beta }\right\vert
+H.C.) .
\end{eqnarray}%
In these equations $E_{\alpha }=\hbar \omega _{\alpha }$ is the atomic
energy, $\nu _{i}$ and $\Delta $ are the frequencies of the discrete and
continuum quasimodes, $\widehat{a}_{i}^{\dag },\widehat{a}_{i}$ and $%
\widehat{b}(\Delta )^{\dag },\widehat{b}(\Delta )$ are the standard
creation, annihilation operators for these quasimodes with commutation rules 
$[\widehat{a}_{i},\widehat{a}_{j}^{\dag }]=\delta _{ij},[\widehat{b}(\Delta
),\widehat{b}(\Delta ^{\prime })^{\dag }]=\delta (\Delta -\Delta ^{\prime
})/\rho _{C}(\Delta )$, $\hbar V_{ij}$ is the coupling energy between
discrete quasimodes, $\hbar W_{i}(\Delta )$ describes the coupling between
discrete and continuum quasimodes and $\hbar \lambda _{i;\alpha \beta }$
specifies the coupling between the discrete quasimode and the atomic
transition between states $\left\vert E_{\alpha }\right\rangle ,\left\vert
E_{\beta }\right\rangle $. All interactions are given via the rotating wave
approximation. The continuum mode density is $\rho _{C}(\Delta )$.

In the case where $W_{i}(\Delta )$ and $\rho _{C}(\Delta )$ are slowly
varying functions of the continuum quasimode frequency, the system
consisting of the atom plus discrete quasimodes is described by a reduced
density operator $\widehat{\rho }$ satisfying a Markovian master equation 
\cite{Dalton01a}%
\begin{equation}
\frac{\partial }{\partial t}\widehat{\rho }=-\frac{i}{\hbar }[\widehat{H}%
_{S},\widehat{\rho }]+\sum\limits_{ij}\pi \rho _{C}W_{i}W_{j}^{\ast }([%
\widehat{a}_{j},\widehat{\rho }\,\widehat{a}_{i}^{\dag }]+[\widehat{a}_{j}\,%
\widehat{\rho },\,\widehat{a}_{i}^{\dag }])
,
\end{equation}%
where the first term involves the Hamiltonian for the atom-discrete
quasimode system $\widehat{H}_{S}$ and the second term describes relaxation
due to the interaction between the discrete quasimodes and a reservoir
consisting of the continuum quasimodes. The Hamiltonian $\widehat{H}_{S}$ is
given by 
\begin{eqnarray}
\widehat{H}_{S} &=&\widehat{H}_{A}+\widehat{H}_{QD} \\
\widehat{H}_{QD} &=&\sum\limits_{i}\hbar \nu _{i}\,\widehat{a}_{i}^{\dag }%
\widehat{a}_{i}+\sum\limits_{i\neq j}\hbar V_{ij}\,\widehat{a}_{i}^{\dag }%
\widehat{a}_{j}
.
\end{eqnarray}

\subsection{Spectrum}

\begin{figure}
\centerline{\includegraphics[width=8cm]{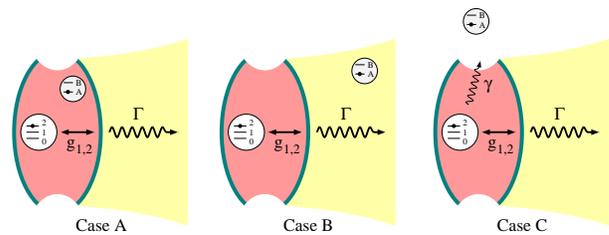}}
\caption[fig1]{
The cascade atom in a high-Q cavity with the detector
atom in various locations. In Case A the detector atom is inside the cavity,
in Case B it is outside the cavity output mirror and positioned to detect
end emission and in Case C it is outside the cavity positioned to detect
side emission.} \label{Figure1} 
\end{figure}

The spectrum $S(\omega )$ is defined operationally \cite%
{Glauber65a,Cresser83a} in terms of the long time transition probability to
the upper state $\left\vert B\right\rangle $ for a two-level atom detector
with transition frequency $\omega $ that is initially in its lower state $%
\left\vert A\right\rangle $ and which is weakly coupled to the EM\ field,
and where the final state of the atom-quasimode system is unobserved. The
Hamiltonian $\widehat{\mathcal{H}}$ for the combined system of detector atom
and the radiating atom-quasimode system is then%
\begin{equation}
\widehat{\mathcal{H}}=\widehat{H}+\widehat{H}_{D}+\widehat{V}_{SD}
\end{equation}%
where the Hamiltonians for the detector atom and its coupling to the
radiating atom-quasimode system are 
\begin{eqnarray}
\widehat{H}_{D} &=&\hbar \omega \widehat{S}_{Z} \\
\widehat{V}_{SD} &=&\hbar (\widehat{V}_{+}\widehat{S}_{-}+\widehat{V}_{-}%
\widehat{S}_{+})
.
\end{eqnarray}%
Here $\widehat{S}_{+}=\left\vert B\right\rangle \left\langle A\right\vert $, 
$\widehat{S}_{-}=\left\vert A\right\rangle \left\langle B\right\vert $ and $%
\widehat{S}_{Z}=\frac{1}{2}(\left\vert B\right\rangle \left\langle
B\right\vert -\left\vert A\right\rangle \left\langle A\right\vert )$ are the
usual atomic spin operators. The operators $\widehat{V}_{+}$, $\widehat{V}%
_{-}$ depend on where the detector atom is situated and hence what region of
the quantum EM\ field it samples. In all cases $\widehat{V}_{+}$, $\widehat{V%
}_{-}$ are proportional to the negative, positive frequency components $%
\widehat{\mathcal{E}}_{-}$, $\widehat{\mathcal{E}}_{+}$ of the electric
field operator at the detector atom. For the case where the detector atom is
situated inside a high Q cavity or is situated outside the output mirror,
the EM\ field can be conveniently described in terms of quasimodes \cite%
{Dalton99a,Brown01a,Dalton01a}. In the case $(Case\;A)$ when it is inside a
high Q cavity---see figure \ref{Figure1}---the detector atom samples the
discrete quasimodes, in another case $(Case\;B)$ when it is situated outside
a high Q cavity to detect end emission---see figure \ref{Figure1}---it
samples the continuum quasimodes. For cases when the detector atom is
situated outside a high Q cavity to detect side emission---see figure \ref%
{Figure1} $(Case\;C)$---the EM\ field is conveniently described in terms of
true modes \cite{Dalton96a,Dalton97a}. In the three cases we have
\begin{eqnarray}
\widehat{V}_{-} &=&\sum\limits_{i}\mu _{i}^{\ast }\,\widehat{a}_{i}=(%
\widehat{V}_{+})^{\dag }
\qquad
\qquad Case\;A \\
\widehat{V}_{-} &=&\int d\Delta \,\rho _{C}(\Delta )\,\mu ^{\ast }(\Delta )%
\widehat{b}(\Delta )=(\widehat{V}_{+})^{\dag }
\notag \\ && %
\qquad\qquad\qquad\qquad\qquad
\qquad Case\;B \\
\widehat{V}_{-} &=&\sum\limits_{E_{\alpha }<E_{\beta }}R_{\alpha \beta
}^{\ast }\,\left\vert E_{\alpha }\right\rangle \left\langle E_{\beta
}\right\vert =(\widehat{V}_{+})^{\dag }
\notag \\ && %
\qquad\qquad\qquad\qquad\qquad
\qquad Case\;C
\, ,
\end{eqnarray}%
where $\mu _{i}$, $\mu (\Delta )$ and $R_{\alpha \beta }$ are weak coupling
constants. For simplicity we will ignore retardation effects---these could
be included by incorporating the position of the detector atom in $\mu
_{i}^{\ast }$, $\mu ^{\ast }(\Delta )$ and $R_{\alpha \beta }$ and that of
the radiating atom in the $\lambda _{i;\alpha \beta }^{\ast }$, and by
arranging the detector atom to begin responding when the emitted EM$\ $\
field first reaches it. In Case C the expression for $\widehat{V}_{-}$ is
based on the long distance radiating atom contribution to the EM$\ $field
positive frequency component $\widehat{\mathcal{E}}_{+}$ being related to
the electric dipole operator of the radiating atom (ignoring the free
evolution term, which gives a zero contribution to the spectrum), and $%
\widehat{V}_{-}$ can be expressed in terms of the downward atomic transition
operators $\left\vert E_{\alpha }\right\rangle \left\langle E_{\beta
}\right\vert $ ($E_{\alpha }<E_{\beta }$).

If the initial state of the atom-quasimodes system is a pure state $%
\left\vert \Phi _{i}\right\rangle $ and the detector is in state $\left\vert
A\right\rangle $ then the probability amplitude for the transition to an
atom-quasimode system state $\left\vert \Phi _{f}\right\rangle $ and the
detector in state $\left\vert B\right\rangle $ is given by%
\begin{equation}
A_{FI}=\left\langle F\right\vert \widehat{U}(t)\left\vert I\right\rangle
,
\end{equation}%
with $\left\vert I\right\rangle =\left\vert \Phi _{i}\right\rangle $ $%
\left\vert A\right\rangle $ and $\left\vert F\right\rangle =\left\vert \Phi
_{f}\right\rangle $ $\left\vert B\right\rangle $ and $\widehat{U}(t)=\exp (-i%
\widehat{\mathcal{H}}t/\hbar )$ the evolution operator for the combined
atom-quasimode-detector system. If the atom-quasimode system is initially in
a mixed state with density operator $\widehat{\rho }_{I}=
\sum\limits_{i}p_{i}\left\vert \Phi _{i}\right\rangle \left\langle \Phi
_{i}\right\vert $ then the transition probability to the upper detector atom
state and to any final state of the atom-quasimode system is given by 
\begin{equation}
P_{BA}=\sum\limits_{f}\sum\limits_{i}p_{i}|\left\langle F\right\vert 
\widehat{U}(t)\left\vert I\right\rangle |^{2}
,
\end{equation}%
and for long times $t$ this will be taken as the un-normalised spectrum $%
S(\omega )$.

A straightforward quantum treatment \cite{Louisell73a} which involves
expressing the evolution operator as a sum of terms with increasing orders
of $\widehat{V}_{SD}$ then gives 
\begin{widetext}
\begin{eqnarray}
&&\left\langle F\right\vert \widehat{U}(t)\left\vert I\right\rangle  
=
\frac{1}{i\hbar }\exp (-i\omega t)\int_{0}^{t}dt_{1}\exp (i\omega
t_{1})\left\langle \Phi _{f}\right\vert \exp (-i\widehat{\mathcal{H}}%
(t-t_{1})/\hbar )\widehat{V}_{-}\exp (-i\widehat{\mathcal{H}}(t_{1})/\hbar
)\left\vert \Phi _{i}\right\rangle  
,
\end{eqnarray}%
\end{widetext}
and finally as $t\rightarrow \infty $%
\begin{eqnarray}
S(\omega )&=&\frac{1}{\hbar ^{2}} \iint_{0}^{\infty } dt_{1}dt_{2}\exp i\omega
(t_{1}-t_{2})
\nonumber\\&&\times
Tr\,\left( \widehat{\rho }_{I}\,\widehat{V}_{+}(t_{2})\,\,%
\widehat{V}_{-}(t_{1})\right)  \label{Eq.OperationalSpectrum}
,
\end{eqnarray}%
where
\begin{eqnarray}
\widehat{V}_{-}(t) &=&\exp (+i\widehat{H}\,t/\hbar )\,\widehat{V}_{-}\,\exp
(-i\widehat{H}\,t)/\hbar )  \notag \\
\widehat{V}_{+}(t) &=&\exp (+i\widehat{H}\,t/\hbar )\,\widehat{V}_{+}\,\exp
(-i\widehat{H}\,t)/\hbar )
\nonumber\\ %
&=&(\widehat{V}_{-}(t))^{\dag }
\end{eqnarray}%
are the system Heisenberg picture operators associated with $\widehat{V}%
_{-}\,$and $\widehat{V}_{+}$. The trace is over the radiating
atom-quasimodes system. The expression for the spectrum is real, as
expected. This approach defines the so-called \emph{operational spectrum}.\ 

A rather different approach to defining the spectrum is based on the
essential states approximation. Here the state vector for the full atom and
EM\ field modes system---the modes are still described via quasi-modes---is
expanded in a basis set $\left\vert E_{\alpha };n_{i};m(\Delta
)\right\rangle $ where $n_{i}$ and $m(\Delta )$ are the numbers of photons
in the $i$th discrete and $\Delta $ continuum quasimodes. Thus%
\begin{eqnarray}
\left\vert \Psi (t)\right\rangle &=&\sum\limits_{\alpha
}\sum\limits_{n_{i}}\int d\Delta \,\rho _{C}(\Delta )\,  
\sum\limits_{m(\Delta )}b(E_{\alpha };n_{i};m(\Delta );t)
\notag \\ &&\times 
\exp(-i\{\omega _{\alpha }+n_{i}\nu _{i}\,+m(\Delta )\Delta \}t)
\notag \\ &&\times 
\left\vert E_{\alpha };n_{i};m(\Delta )\right\rangle  
,
\end{eqnarray}%
with interaction picture amplitudes $b(E_{\alpha };n_{i};m(\Delta );t).$ The
long time one photon emission spectrum would be defined in this approach via
the long time probability of finding one photon in a continuum mode of
frequency $\Delta $%
\begin{equation}
S_{I}(\Delta )=\sum\limits_{\alpha }\sum\limits_{n_{i}}|\,b(E_{\alpha
};n_{i};1(\Delta );\infty )\,|^{2}  \label{Eq.IdealSpectrum}
,
\end{equation}%
in which the continuum quasimode frequency acts as the spectral variable.
This approach may be referred to as defining the \emph{ideal spectrum},
corresponding to quantities arising naturally from fundamental
considerations of what can in principle be measured in quantum mechanics,
rather than to the analysis of the behaviour of a model spectrometer. An
alternative definition of an ideal spectrum \cite{Linington06a} could be
based on true mode states---which also form a continuum---rather than
continuum quasimodes states. The interrelationship between the different
approaches to defining spectra has been considered by Cresser \cite%
{Cresser83a}. Calculations of the ideal spectrum would involve solving the
coupled amplitude equations.

\subsection{Quantum regression theorem}

Determination of the operational spectrum requires the evaluation of
two-time correlation functions, and here the quantum regression theorem is
used. The usual statement of the quantum regression theorem \cite%
{Lax63a67a,Walls94a,Dalton79a} is that for a system operator $\widehat{Y}%
_{i}(t)$\ in the Heisenberg picture whose average satisfies the linear
equation $(t\geqslant 0)$%
\begin{equation}
\frac{d}{dt}\left\langle \widehat{Y}_{i}(t)\right\rangle
=\sum\limits_{j}G_{ij}(t)\left\langle \widehat{Y}_{j}(t)\right\rangle
,
\end{equation}%
with matrix elements $G_{ij}(t)$, then we can assert that the two time
correlation functions $(\tau \geqslant 0)$\ satisfy a similar linear
equation involving the same matrix elements 
\begin{eqnarray}
\frac{d}{d\tau }\left\langle \widehat{Y}_{i}(t+\tau )\widehat{Y}%
_{l}(t)\right\rangle &=&\sum\limits_{j}G_{ij}(\tau )\left\langle \widehat{Y}%
_{j}(t+\tau )\widehat{Y}_{l}(t)\right\rangle  
\nonumber\\ && \\
\frac{d}{d\tau }\left\langle \widehat{Y}_{l}(t)\widehat{Y}_{i}(t+\tau
)\right\rangle &=&\sum\limits_{j}G_{ij}(\tau )\left\langle \widehat{Y}_{l}(t)%
\widehat{Y}_{j}(t+\tau )\right\rangle
.
\nonumber \\ && 
\end{eqnarray}

A useful result giving the two time correlation function $\left\langle 
\widehat{S}_{\beta \alpha }(t+\tau )\widehat{S}_{\delta \gamma
}(t)\right\rangle $\ for system transition operators in terms of matrix
elements of the evolution operator and density matrix elements \cite%
{Dalton79a} can be derived from this form of the quantum regression theorem
by considering the case where $\widehat{Y}_{i}$\ is a system transition
operator $\widehat{Y}_{i}\equiv \widehat{S}_{\beta \alpha }=\left\vert \beta
\right\rangle \left\langle \alpha \right\vert $. In this case $\left\langle 
\widehat{Y}_{i}(t)\right\rangle \equiv \left\langle \widehat{S}_{\beta
\alpha }(t)\right\rangle =\rho _{\alpha \beta }(t)$, the density matrix
element which satisfies the master equation. In general we can write the
solution to the master equation in terms of matrix elements $U_{\alpha \beta
;\,\gamma \delta }\,(t)$ of the evolution operator
\begin{equation}
\rho _{\alpha \beta }(t)=\sum\limits_{\gamma \delta }U_{\alpha \beta
;\,\gamma \delta }\,(t)\,\rho _{\gamma \delta }(0)
.
\end{equation}
We then can show that for $(t\geqslant 0,\tau \geqslant 0)$\ the two
important results%
\begin{eqnarray}
\left\langle \widehat{S}_{\beta \alpha }(t+\tau )\widehat{S}_{\delta \gamma
}(t)\right\rangle &=&\sum\limits_{\mu }U_{\alpha \beta ;\,\delta \mu
}\,(\tau )\,\rho _{\gamma \mu }(t)  \label{Eq.QRThm1} \\
\left\langle \widehat{S}_{\beta \alpha }(t)\widehat{S}_{\delta \gamma
}(t+\tau )\right\rangle &=&\sum\limits_{\mu }U_{\gamma \delta ;\,\mu \alpha
}\,(\tau )\,\rho _{\mu \beta }(t)  
.
\label{Eq.QRThm2}
\end{eqnarray}%
These results give the two time correlation function $\left\langle \widehat{S}%
_{\beta \alpha }(t+\tau )\widehat{S}_{\delta \gamma }(t)\right\rangle $\ for
system transition operators in terms of matrix elements of the evolution
operator and density matrix elements.

\section{Three level cascade atom in single mode cavity}

\label{Section 3}

\subsection{Master equation}

\begin{figure}
\centerline{\includegraphics[width=6cm]{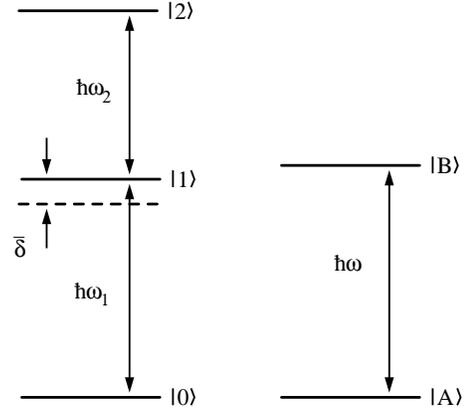}}
\caption[fig2]{
Energy levels and energy differences in the three level
cascade atom (left). The offset $\overline{\delta }$ represents the
difference between the energy of level 1 and the midpoint between levels 0
and 2. The energy levels for the two state detector atom are also shown
(right).} 
\label{Figure2} 
\end{figure}

The master equation for a three level cascade atom in a single mode high Q
cavity coupled to the reservoir of continuum quasimodes at zero temperature
is given by%
\begin{equation}
\frac{\partial }{\partial t}\widehat{\rho }=-\frac{i}{\hbar }[\widehat{H}%
_{S},\widehat{\rho }]+\frac{1}{2}\Gamma ([\widehat{a},\widehat{\rho }\,%
\widehat{a}^{\dag }]+[\widehat{a}\,\widehat{\rho },\,\widehat{a}^{\dag }])
,
\label{Eq.MasterEqnCascadeSingleModeCavity}
\end{equation}%
where the Hamiltonian for the three level atom plus cavity mode system in
the rotating wave approximation is%
\begin{eqnarray}
\widehat{H}_{S}&=&\hbar \omega _{c}\widehat{a}^{\dag }\widehat{a}+\hbar
(\omega _{0}-\overline{\delta })\widehat{\sigma }_{22}-\hbar (\omega _{0}+%
\overline{\delta })\widehat{\sigma }_{00}
\nonumber\\ &&
+\hbar \,[\widehat{a}^{\dag
}\,(g_{2}\widehat{\sigma }_{2}^{-}+g_{1}\widehat{\sigma }_{1}^{-})+(g_{2}%
\widehat{\sigma }_{2}^{+}+g_{1}\widehat{\sigma }_{1}^{+})\,\widehat{a}].
\nonumber\\ &&
\label{Eq.HamiltonianCascadeSingleModeCavity}
\end{eqnarray}%
Here $\widehat{a}\,,\widehat{a}^{\dag }$ are the mode annihilation, creation
operators for the single cavity quasimode with frequency $\omega _{c}$. The
operators $\widehat{\sigma }_{2}^{+}=\left\vert 2\right\rangle \left\langle
1\right\vert $, $\widehat{\sigma }_{2}^{-}=\left\vert 1\right\rangle
\left\langle 2\right\vert $, $\widehat{\sigma }_{1+}^{+}=\left\vert
1\right\rangle \left\langle 0\right\vert $, $\widehat{\sigma }%
_{1-}^{-}=\left\vert 0\right\rangle \left\langle 1\right\vert $ and $%
\widehat{\sigma }_{22}$, $\widehat{\sigma }_{11}$, $\widehat{\sigma }_{00}$
are atomic transition and population operators respectively, involving the
upper state $\left\vert 2\right\rangle $, the intermediate state $\left\vert
1\right\rangle $ and the lower state $\left\vert 0\right\rangle $. The lower 
$(1\leftrightarrow 0)$ and upper $(2\leftrightarrow 1)$ atomic transition
frequencies are $\omega _{1}$, $\omega _{2}$ respectively. The average
atomic transition frequency is $\omega _{0}$ and $\overline{\delta }$ is
half the difference between lower and upper transition frequencies. Thus $%
\omega _{0}=(\omega _{1}+\omega _{2})/2$, $\overline{\delta }=(\omega
_{1}-\omega _{2})/2$. The atomic term in the Hamiltonian is based on adding
a constant $\lambda (\widehat{\sigma }_{22}+\widehat{\sigma }_{11}+\widehat{%
\sigma }_{00})$ to the original Hamiltonian so as to make the energy of the
intermediate state $\left\vert 1\right\rangle $ equal to zero. The atomic
states are illustrated in figure \ref{Figure2}, along with the states of the
detector atom. The one photon Rabi frequencies are $g_{2}$ and $g_{1}$ for
the two transitions and $\Gamma $ is the cavity decay rate, where we have
used the notation $\lambda _{21}\rightarrow g_{2}$, $%
\lambda _{10}\rightarrow g_{1}$. The coupling constants $g_{1}$\ and $g_{2}$%
\ are proportional to the scalar product of the vector dipole matrix
elements $\left\langle 2\right\vert \,\underrightarrow{\widehat{d}}%
\,\left\vert 1\right\rangle $, $\left\langle 1\right\vert \,\underrightarrow{%
\widehat{d}}\,\left\vert 0\right\rangle $ between the upper and intermediate
states or the intermediate and lower states with the polarization unit
vector for the cavity quasimode, and by a suitable choice of phase can be
taken as real and positive. The reduced density operator is $\widehat{\rho }$%
. In terms of the flat continuum mode density $\rho _{C}$ and
discrete-continuum quasimodes coupling constant $W$, we have $\Gamma =2\pi
\,\rho _{C}\,|W|^{2}$. At zero temperature the continuum quasimodes are
empty of photons. In the present case the atomic spontaneous emission rate $%
\gamma $ into sideways EM\ field modes will be ignored in comparison to the
cavity loss rate $\Gamma $. Thus we have $\Gamma \gg \gamma $. The case of
non-negligible $\gamma $ is discussed by \cite{Carmichael89a} for the case
of a two level atom.

For density matrix elements in the basis $\left\vert n;\nu \right\rangle $
(where $n=0,1,2, \hdots$ is the cavity photon number and $\nu =2,1,0$ specifies the
upper, intermediate, and lowest atomic states) the master equation can be used to derive a set
of coupled linear equations for the density matrix elements $\rho _{n\nu
;\,m\mu }$ in which the coefficients depend on the cavity frequency $\omega
_{c}$, the detuning $\delta $ between the cavity frequency and the
average atomic transition frequency $(\delta =\omega _{c}-\omega _{0})$, the
half difference between lower and upper transition frequencies $\overline{%
\delta }$, the coupling constants $g_{1}$, $g_{2}$ and the cavity decay rate 
$\Gamma $. The coupled density matrix equations are set out in a suitable
form in Appendix \ref{Appendix 1}. The equations make up almost independent
coupled sets of density matrix elements, if coupling only via the reversible
processes associated with $g_{1}$, $g_{2}$\ is considered. In general we
designate the $(n,m)$\ set for all $n,m$\ to include the density matrix
element $\rho _{n0;\,m0}$\ (associated with the atom being in the lowest
state $\left\vert 0\right\rangle $) and any density matrix elements coupled
to it via reversible processes associated with the atom-cavity mode
interaction terms. As we will see, relaxation due to cavity decay $\Gamma $
couples a set $(n,m)$ to other sets. \ 

The sets with $(n,m\geqslant 2)$\ consist of $9$\ elements---$\rho _{%
\overline{n-2}2;\,\overline{m-2}2}$, $\rho _{\overline{n-2}2;\,\overline{m-1}%
1}$, $\rho _{\overline{n-2}2;\,m0}$, $\rho _{\overline{n-1}1;\,\overline{m-2}%
2}$, $\rho _{\overline{n-1}1;\,\overline{m-1}1}$, $\rho _{\overline{n-1}%
1;\,m0}$, $\rho _{n0;\,\overline{m-2}2}$, $\rho _{n0;\,\overline{m-1}1}$\
and $\rho _{n0;\,m0}$\ and may be represented as a $1\times 9$\ column
matrix 
\begin{equation}
\mathbf{\rho }(n,m)=%
\begin{bmatrix}
\rho _{\overline{n-2}2;\,\overline{m-2}2} \\ 
\rho _{\overline{n-2}2;\,\overline{m-1}1} \\ 
\rho _{\overline{n-2}2;\,m0} \\ 
\rho _{\overline{n-1}1;\,\overline{m-2}2} \\ 
\rho _{\overline{n-1}1;\,\overline{m-1}1} \\ 
\rho _{\overline{n-1}1;\,m0} \\ 
\rho _{n0;\,\overline{m-2}2} \\ 
\rho _{n0;\,\overline{m-1}1} \\ 
\rho _{n0;\,m0}%
\end{bmatrix}
.
\label{Eq.(NM)set}
\end{equation}%
This set is based on the three states of the form $\left\vert \overline{%
n-2};2\right\rangle $, $\left\vert \overline{n-1};1\right\rangle $ and $%
\left\vert n;0\right\rangle $ $(n\geqslant 2)$ which are coupled via the
reversible processes associated with the atom-cavity mode interaction.
However the terms arising from the irreversible processes associated with
the relaxation term $\Gamma \widehat{a}\,\widehat{\rho }\,\widehat{a}^{\dag
} $ couple in $\rho _{\overline{n-1}2;\,\overline{m-1}2}$, $\rho _{\overline{%
n-1}2;\,m1}$, $\rho _{\overline{n-1}2;\,\overline{m+1}0}$, $\rho _{n1;\,%
\overline{m-1}2}$, $\rho _{n1;\,m1}$, $\rho _{n1;\,\overline{m+1}0}$, $\rho
_{\overline{n+1}0;\,\overline{m-1}2}$, $\rho _{\overline{n+1}0;\,m1}$ and $%
\rho _{\overline{n+1}0;\,\overline{m+1}0}$, which are the density matrix
elements in the $(n+1,m+1)$ set. Such irreversible processes are of the form 
$\left\vert n+1;\nu \right\rangle \rightarrow \left\vert n;\nu \right\rangle 
$ and these link in the three states $\left\vert \overline{n-1}%
;2\right\rangle $ $\left\vert n;1\right\rangle $ and $\left\vert \overline{%
n+1};0\right\rangle $ on which are based the $(n+1,m+1)$ set of density
matrix elements.

Special consideration is required for the density matrix elements associated
with the state $\left\vert 0;0\right\rangle $ (where there are no cavity
photons and the atom is in the lowest state). This state is not coupled to
other states via the atom-cavity mode reversible interaction processes, and
its only link with other states is via the irreversible $\left\vert
1;0\right\rangle \rightarrow \left\vert 0;0\right\rangle $ relaxation
process. The density matrix equations associated with the $\left\vert
0;0\right\rangle $ state take on special forms. Similarly, special
consideration is required for the density matrix elements associated with
the states $\left\vert 0;1\right\rangle $ and $\left\vert 1;0\right\rangle $
(where there are no cavity photons and the atom is in the intermediate state
or there is one cavity photon and the atom is in the lowest state). These
states are coupled together but not coupled to other states via the
atom-cavity mode reversible interaction processes. Their only link with
other states is via the irreversible $\left\vert 1;1\right\rangle
\rightarrow \left\vert 0;1\right\rangle $ and $\left\vert 2;0\right\rangle
\rightarrow \left\vert 1;0\right\rangle $ relaxation processes, and $%
\left\vert 1;0\right\rangle $ is linked to $\left\vert 0;0\right\rangle $
via the $\left\vert 1;0\right\rangle \rightarrow \left\vert 0;0\right\rangle 
$ relaxation process. The density matrix equations associated with the $%
\left\vert 0;1\right\rangle $ and $\left\vert 1;0\right\rangle $ states also
take on special forms. We still find that the density matrix elements break
up into separate coupled sets associated with the reversible atom-cavity
mode interaction processes. Following the general procedure for designating
the sets via the photon quantum numbers $n,m$ associated with the density
matrix element $\rho _{n0;\,m0}$ included in the set, the $(0,0)$ set only
contains the single density matrix element $\rho _{00;\,00}$, the $(0,1)$
set contains two elements $\rho _{00;\,01},\rho _{00;\,10\text{,}}$ with the 
$(1,0)$ set also containing two elements $\rho _{01;\,00},\rho _{10;\,00}$.
The $(0,2)$ set contains three elements $\rho _{00;\,02},\rho _{00;\,11}$
and $\rho _{00;\,20}$, the $(2,0)$ set also contains three elements $\rho
_{02;\,00},\rho _{11;\,00}$ and $\rho _{20;\,00}$. Next there is the $(1,1)$
set with four elements $\rho _{01;\,01},\rho _{01;\,10},\rho _{10;\,01}$ and 
$\rho _{10;\,10}$. Finally, there are two sets each with six elements, the $%
(1,2)$ set consisting of $\rho _{01;\,02},\rho _{01;\,11},\rho
_{01;\,20},\rho _{10;\,02},\rho _{10;\,11}$ and $\rho _{10;\,20}$, and the $%
(2,1)$ set consisting of $\rho _{02;\,01},\rho _{02;\,10},\rho
_{11;\,01},\rho _{11;\,10},\rho _{20;\,01}$and $\rho _{20;\,10}$. Instead of
there being $9$ coupled equations as for the $(n,m)$ set $(n,m\geqslant 2)$,
there are $1,2,2,3,3,4,6$ and $6$ coupled equations respectively for the $%
(0,0),(0,1),(1,0),(0,2),(2,0),(1,1),(1,2)$ and $(2,1)$ coupled sets. The
complete sets of equations can be obtained from the master equation.

In all cases the density matrix equations can be conveniently be written in
matrix form as%
\begin{equation}
\frac{\partial }{\partial t}\mathbf{\rho }(n,m)=-iA(n,m)\mathbf{\rho }%
(n,m)+iB(n,m)\mathbf{\rho }(n+1,m+1)
.
\end{equation}%
Expressions for the column vectors $\rho (n,m)$ and the matrices $A(n,m)$\
and $B(n,m)$ which specify the coupling within the $(n,m)$\ set and the
coupling to the $(n+1,m+1)$ set respectively, are set out in Appendix \ref%
{Appendix 1} for $(n,m\geqslant 2)$ and for the special $(0,0)$, $(1,0)$, $%
(1,1)$, $(2,1)$ and $(2,2)$ sets.

Taking the Laplace transform of the equations for $\mathbf{\rho }(n,m)$ gives%
\begin{equation}
(s+iA(n,m))\,\widetilde{\mathbf{\rho }}(n,m)-iB(n,m)\,\widetilde{\mathbf{%
\rho }}(n+1,m+1)=\mathbf{\rho }(n,m,0)
,
\end{equation}%
where $s$ is the Laplace variable. The Laplace transform $\widetilde{A}(s)$
of a function $A(t)$ which is bounded for $t\geqslant 0$ is defined as for
complex $s$\ in the right half plane via%
\begin{equation}
\widetilde{A}(s)=\int_{0}^{\infty }dt\,\exp (-st)\,A(t)\qquad \text{Re}%
\,s\geqslant 0
\, .
\end{equation}%
The form of these equations indicates that if relaxation is \emph{also}
taken into account, \emph{certain }collections of the $\mathbf{\rho }(n,m)$
form independent coupled sets, the set containing a specific $\mathbf{\rho }%
(n,m)$ also includes $\mathbf{\rho }(n\pm 1,m\pm 1), \mathbf{\rho }(n\pm
2,m\pm 2), \hdots ,\mathbf{\rho }(n\pm k,m\pm k), \hdots $, where $k$ is a positive
integer such that $n\pm k\geqslant 0,m\pm k\geqslant 0$. Thus $\mathbf{\rho }%
(0,0),$ $\mathbf{\rho }(1,1),$ $\mathbf{\rho }(2,2), \hdots $are all coupled, as
are $\mathbf{\rho }(1,0),$ $\mathbf{\rho }(2,1),$ $\mathbf{\rho }(3,2), \hdots $%
etc. We see from the Laplace transform equations that if $\mathbf{\rho }%
(n>N,m>M,0)$ are all zero, then for $n>N$ and $m>M$ $(s+iA(n,m))\,\widetilde{%
\mathbf{\rho }}(n,m)-iB(n,m)\,\widetilde{\mathbf{\rho }}(n+1,m+1)=0$, and
hence a solution to the equations given by $\,\widetilde{\mathbf{\rho }}%
(n,m)=0$ for $n>N$ and $m>M$ exists. Since linear first order equations have
unique solutions, then $\,\rho (n,m,t)=0$ for $n>N$ and $m>M$. This feature
is the consequence of the relaxation processes occuring in one direction,
with only the $\left\vert n+1;\nu \right\rangle \rightarrow \left\vert n;\nu
\right\rangle $ transition (and not its reverse) taking place.

The solution to the density matrix equations at time $t$ can be related to 
\emph{any} initial density matrix in terms of the matrices $U_{n,m::l,k}(t)$
representing the evolution super-operator $\widehat{\mathbf{U}}(t)$ for the
master equation as%
\begin{equation}
\mathbf{\rho }(n,m,t)=\sum\limits_{l,k}U_{n,m::l,k}(t)\,\mathbf{\rho }(l,k,0)
,
\end{equation}%
where the $U_{n,m::l,k}(t)$ satisfy the same matrix equations as $\mathbf{%
\rho }(n,m,t)$, but have an initial condition involving a unit matrix $E$ 
\begin{equation}
U_{n,m::l,k}(0)=\delta _{(n,m),(l,k)}E_{(n,m)}.
\end{equation}%
The dimensionality of the matrices $U_{n,m::l,k}(t)$ and $E_{(n,m)}$ depends
on the sets $(n,m)$ and $(l,k)$. For example, if both $(n,m)$ and $(l,k)$
are such that $(n,m\geqslant 2)$ and $(l,k\geqslant 2)$ then the matrix is $%
9\times 9$. On the other hand for $(n,m\geqslant 2)$ and $(l,k)=(0,1)$ or $%
(1,0)$ the matrix is $9\times 2$, whilst if $(l,k)=(0,0)$ the matrix is $%
9\times 1$. Other cases follow similar lines. In full for $(n,m\geqslant 2)$
and $(l,k\geqslant 2)$ the $81$ matrix elements of $U_{n,m::l,k}(t)$ are $%
U_{(n\nu ),(m\lambda )::(l\beta ),(k\alpha )}(t)$, where $(n\nu )\equiv (%
\overline{n-2}2,\overline{n-1}1,n0),(m\lambda )\equiv (\overline{m-2}2,%
\overline{m-1}1,m0),(l\beta )\equiv (\overline{l-2}2,\overline{l-1}1,l0)$
and $(k\alpha )\equiv (\overline{k-2}2,\overline{k-1}1,k0)$. Other cases
follow similar lines.

For the evolution matrices we can also obtain equations for their Laplace
transforms, and we have%
\begin{eqnarray}
(s+iA(n,m))\,\widetilde{U}(n,m::l,k)
&&\nonumber\\   %
-iB(n,m)\,\widetilde{U}%
(n+1,m+1::l,k)&=&\delta _{(n,m),(l,k)}E_{(n,m)}
.
\nonumber\\ &&  %
\end{eqnarray}%
As we will see, Laplace transforms $\widetilde{U}(n,m::l,k)$ occur in the
final expression for the spectrum. Note that as the $\mathbf{\rho }(n,m)$
form independent coupled sets with relaxation only linking neighbouring sets 
$(n,m)$ and $(n\pm 1,m\pm 1)$, the only non-zero $U_{n,m::l,k}(t)$ are such
that $n-m=l-k$.

For the initial condition of interest the atom is in the upper state $%
\left\vert 2\right\rangle $ and the single mode is in the vacuum state and
the reservoir of continuum quasimodes in the vacuum state we have 
\begin{eqnarray}
\widehat{\rho }_{I} &=&\widehat{\rho }_{SI}\,\widehat{\rho }_{RI} \\
\widehat{\rho }_{SI} &=&\left\vert 0;2\right\rangle \left\langle
0;2\right\vert \\
\widehat{\rho }_{RI} &=&\prod_{\Delta }(\left\vert 0\right\rangle
\left\langle 0\right\vert )_{\Delta }
\, ,
\end{eqnarray}%
and hence the only non-zero $\mathbf{\rho }(n,m,0)$ is for $n=m=2$ 
\begin{equation}
\mathbf{\rho }(2,2,0)=%
\begin{bmatrix}
1 \\ 
0 \\ 
0 \\ 
0 \\ 
0 \\ 
0 \\ 
0 \\ 
0 \\ 
0%
\end{bmatrix}%
.
\end{equation}

For the discussion above regarding the irreversible linking of the coupled
sets of density matrix elements, the unique solution for this initial
condition is such that all $\mathbf{\rho }(n,m,t)$ are zero except $\mathbf{%
\rho }(2,2,t)$, $\mathbf{\rho }(1,1,t)$ and $\mathbf{\rho }(0,0,t)$, which
are obtained from their Laplace transforms 
\begin{eqnarray}
\widetilde{\mathbf{\rho }}(2,2,s) &=&(s+iA(2,2))^{-1}\,\mathbf{\rho }(2,2,0)
\notag \\
\widetilde{\mathbf{\rho }}(1,1,s) &=&(s+iA(1,1))^{-1}\,iB(1,1)\,\widetilde{%
\mathbf{\rho }}(2,2,s)  \notag \\
\widetilde{\mathbf{\rho }}(0,0,s) &=&(s+iA(0,0))^{-1}\,iB(0,0)\,\widetilde{%
\mathbf{\rho }}(1,1,s)  
.
\label{Eq.DensityMatrixCascade}
\end{eqnarray}%
This solution involves inverting a $9\times 9$ and a $4\times 4$ matrix.%
\emph{\ }For this case of finite initial energy, the spectrum will be be
that associated with a finite photon pulse in the cavity, but as we will see
the spectrum will be different to that for spontaneous emission from an
excited three-level cascade atom in free space.

For the evolution matrices that are required for determining the spectra we
have the solutions
\begin{eqnarray}
\widetilde{U}(1,0::1,0;s)&=&(s+iA(1,0))^{-1}  \notag \\
\,\widetilde{U}(n+1,n::1,0;s)&=&0\qquad (n>1)
,
\label{Eq.EvolutionMatrixCascadeTerm1}
\end{eqnarray}
and
\begin{eqnarray}
\widetilde{U}(2,1::2,1;s)&=&(s+iA(2,1))^{-1}  \notag \\
\widetilde{U}(1,0::2,1;s)&=&(s+iA(1,0))^{-1}
\nonumber\\ &&\times  %
iB(1,0)\,(s+iA(2,1))^{-1} 
\notag \\
\,\widetilde{U}(n+1,n::2,1;s)&=&0\qquad (n>1).
\label{Eq.EvolutionMatrixCascadeTerm2}
\end{eqnarray}
This solution corresponds to the previous result that an initial density
matrix with only non-zero elements in the $(N,M)$\ set or below cannot
evolve into a density matrix with non-zero elements in sets such as $%
(N+1,M+1)$, $(N+2,M+2), \hdots $\ , due to the irreversible nature of the
relaxation processes. Further details are in Appendix \ref{Appendix 2}.

\subsection{Detector atom inside cavity---Case A}

To evaluate the spectrum for the case where the detector atom is inside the
high Q cavity ($\emph{CaseA}$) and with initial conditions of the atom
excited and the cavity mode empty of photons we take%
\begin{eqnarray}
\widehat{V}_{-} &=&\mu ^{\ast }\,\widehat{a}  \notag \\
&=&\mu ^{\ast }\sum\limits_{n\nu }\sqrt{n}\,\left\vert n-1;\nu \right\rangle
\left\langle n;\nu \right\vert \\
\widehat{V}_{+} &=&\mu \,\widehat{a}^{\dag }  \notag \\
&=&\mu \sum\limits_{m\lambda }\sqrt{m}\,\left\vert m;\lambda \right\rangle
\left\langle m-1;\lambda \right\vert .
\end{eqnarray}
The two time correlation function is then
\begin{eqnarray}
&&
Tr_{SR}\,\left( \widehat{\rho }_{I}\,\widehat{V}_{+}(t_{2})\,\,\widehat{V}%
_{-}(t_{1})\right) 
\nonumber\\&&
=|\mu |^{2}\sum\limits_{m\lambda }\sum\limits_{n\nu }\sqrt{m}\sqrt{n}%
\,\left\langle \widehat{S}_{m\lambda ;\,\overline{m-1}\lambda }(t_{2})%
\widehat{S}_{\overline{n-1}\nu ;\,n\nu }(t_{1})\right\rangle
,
\nonumber\\&&
\end{eqnarray}%
where the transition operators $\widehat{S}_{m\lambda ;\,n\nu }(t)\equiv
\left\vert m;\lambda \right\rangle \left\langle n;\nu \right\vert $ are
Heisenberg operators at time $t$, and the trace is over both system and
reservoir states.\ The two time correlation functions can be evaluated using
the quantum regression theorem \cite{Lax63a67a,Walls94a,Dalton79a}, the
result for which is given in terms of the evolution operator matrix elements 
$U_{m\lambda ;\,n\nu ::l\beta ;\,k\alpha }(\tau )$ associated with the
density matrix equations and considered as a function of the time difference 
$\tau =|t_{1}-t_{2}|\,\geqslant \,0$, and density matrix elements $\rho
_{m\lambda ;\,n\nu }(t_{1,2})$ considered as a function of the smaller of
the two times $t_{1}$ and $t_{2}$. The derivation of the expression for the
spectrum for Case A is given in Appendix \ref{Appendix 2}.

The spectrum is given by 
\begin{equation}
  S(\omega ) =S_{2}(\omega )+S_{6}(\omega )
,
\label{Eq.SpectrumResultCascadeCaseA} 
\end{equation}
with
\begin{widetext}
\begin{eqnarray}
S_{2}(\omega ) &=&2\,\text{Re}\,\frac{|\mu |^{2}}{\hbar ^{2}}\sqrt{1}\sqrt{1}%
\, 
\times \{\widetilde{\rho }_{10;\,10}(\epsilon ^{\prime })\widetilde{U}%
_{10;\,00::01;\,00}(-i\omega +\epsilon )\,+ \widetilde{\rho }%
_{01;\,10}(\epsilon ^{\prime }) \widetilde{U}_{10;\,00::10;\,00}(-i\omega
+\epsilon )\,\}   \\
S_{6}(\omega ) &=&2\,\text{Re}\,\frac{|\mu |^{2}}{\hbar ^{2}}\,  
\lbrack \sqrt{1}\,\widetilde{\rho }_{02;\,11}(\epsilon ^{\prime })\{%
\sqrt{1}\widetilde{U}_{10,00::02;\,01}(-i\omega +\epsilon )+\sqrt{1}%
\widetilde{U}_{11,01::02;\,01}(-i\omega +\epsilon ) 
+\sqrt{2}\widetilde{U}_{20,10::02;\,01}(-i\omega +\epsilon )\}  \notag \\
&&+\sqrt{2}\,\widetilde{\rho }_{02;\,20}(\epsilon ^{\prime })\{\sqrt{1}%
\widetilde{U}_{10,00::02;\,10}(-i\omega +\epsilon )+\sqrt{1}\widetilde{U}%
_{11,01::02;\,10}(-i\omega +\epsilon ) 
+\sqrt{2}\widetilde{U}_{20,10::02;\,10}(-i\omega +\epsilon )\}  \notag \\
&&+\sqrt{1}\,\widetilde{\rho }_{11;\,11}(\epsilon ^{\prime })\{\sqrt{1}%
\widetilde{U}_{10,00::11;\,01}(-i\omega +\epsilon )+\sqrt{1}\widetilde{U}%
_{11,01::11;\,01}(-i\omega +\epsilon ) 
+\sqrt{2}\widetilde{U}_{20,10::11,01}(-i\omega +\epsilon )\}  \notag \\
&&+\sqrt{2}\,\widetilde{\rho }_{11;\,20}(\epsilon ^{\prime })\{\sqrt{1}%
\widetilde{U}_{10,00::11,10}(-i\omega +\epsilon )+\sqrt{1}\widetilde{U}%
_{11,01::11,10}(-i\omega +\epsilon ) 
+\sqrt{2}\widetilde{U}_{20,10::11;\,10}(-i\omega +\epsilon )\}  \notag \\
&&+\sqrt{1}\,\widetilde{\rho }_{20;\,11}(\epsilon ^{\prime })\{\sqrt{1}%
\widetilde{U}_{10,00::20;\,01}(-i\omega +\epsilon )+\sqrt{1}\widetilde{U}%
_{11,01::20;\,01}(-i\omega +\epsilon ) 
+\sqrt{2}\widetilde{U}_{20,10::20;\,01}(-i\omega +\epsilon )\}  
\notag \\&&
+\sqrt{2}\,\widetilde{\rho }_{20;\,20}(\epsilon ^{\prime })\{\sqrt{1}%
\widetilde{U}_{10,00::20,10}(-i\omega +\epsilon )+\sqrt{1}\widetilde{U}%
_{11,01::20,10}(-i\omega +\epsilon )  
+\sqrt{2}\widetilde{U}_{20,10::20,10}(-i\omega +\epsilon )\}]  
.
\notag \\ &&%
\end{eqnarray}
\end{widetext}
We see that in Case A, the spectrum involves Laplace transforms of the
evolution operator from the $U_{1,0::1,0}(t)$, $U_{1,0::2,1}(t)$ and $%
U_{2,1::2,1}(t)$ matrices and from the $\rho (2,2,t)$ and $\rho (1,1,t)$
density matrix elements. The spectrum is calculated from the equation (\ref%
{Eq.SpectrumResultCascadeCaseA}) using the results from equations (\ref%
{Eq.DensityMatrixCascade}), (\ref{Eq.EvolutionMatrixCascadeTerm1}) and (\ref%
{Eq.EvolutionMatrixCascadeTerm2}). The spectral variable $\omega $ occurs in
the $\widetilde{U}$\ matrices associated with $(1,0)$ and $(2,1)$ sets. From
the equations determining these, it is easy to see that the spectral
variable appears via the factor $\Delta \omega =(\omega -\omega _{c})$, so
the spectra can be shown as functions of the detuning of the spectral
variable from the cavity frequency. The calculation requires inverting $%
6\times 6$ and $2\times 2$ matrices as a function of the spectral variable $%
\omega $ to give the $\widetilde{U}$ terms and inverting $9\times 9$ and $%
4\times 4$ matrices to give the $\,\widetilde{\rho }$ terms.

\subsection{Detector atom outside cavity---Case B (end emission)}

To evaluate the spectrum in the case where the detector atom is outside the
cavity to detect end emission and thus samples the continuum cavity
quasimodes (\emph{Case B}) and with the same initial conditions as before we
now take 
\begin{equation}
\widehat{V}_{-}=\int d\Delta \,\rho _{C}(\Delta )\,\mu ^{\ast }(\Delta )%
\widehat{b}(\Delta )
,
\end{equation}%
and will assume that $\mu ^{\ast }(\Delta )$ is also a slowly varying
function of $\Delta $.

Starting from the formal solutions of Heisenberg equations of motion for the
Heisenberg operators $\widehat{b}(\Delta ,t)$ and $\widehat{a}(t)$ and
making a Markoff approximation based on slowly varying factors $\rho
_{C}(\Delta ),\,\mu ^{\ast }(\Delta )$ and$\,W^{\ast }(\Delta )$, we can
show that $\widehat{V}_{-}$ is the sum of a free field term $\widehat{V}%
_{-}^{F}$ and a cavity term $\widehat{V}_{-}^{C}$, where%
\begin{eqnarray}
\widehat{V}_{-} &=&\widehat{V}_{-}^{F}+\widehat{V}_{-}^{C} \\
\widehat{V}_{-}^{F} &=&\int d\Delta \,\rho _{C}(\Delta )\,\mu ^{\ast
}(\Delta )\,\exp (-i\Delta t)\,\widehat{b}(\Delta ) \\
\widehat{V}_{-}^{C} &=&M^{\ast }\widehat{a}
,
\end{eqnarray}%
and where
\begin{equation}
M^{\ast }=\int d\Delta \,\frac{\rho _{C}(\Delta )\,\mu ^{\ast }(\Delta
)\,W^{\ast }(\Delta )}{\omega _{c}-\Delta +i\epsilon }
\end{equation}%
is an effective dipole coupling constant. Thus the cavity contribution $%
\widehat{V}_{-}^{C}$ to the spectral quantity $\widehat{V}_{-}$ is the same
as for Case A, apart from a constant of proportionality. The details are
given in Appendix \ref{Appendix 3}.

The expression for the two-time correlation function will involve free
evolution continuum quasimode contributions of the form $Tr_{SR}\,\left( 
\widehat{\rho }_{I}\,\widehat{V}_{+}^{F}(t_{2})\,\,\widehat{V}%
_{-}^{F}(t_{1})\right) $, cross terms involving the continuum and cavity
quasimode contributions of the form $Tr_{SR}\,\left( \widehat{\rho }_{I}\,%
\widehat{V}_{+}^{F}(t_{2})\,\,\widehat{V}_{-}^{C}(t_{1})\right)
,Tr_{SR}\,\left( \widehat{\rho }_{I}\,\widehat{V}_{+}^{C}(t_{2})\,\,\widehat{%
V}_{-}^{F}(t_{1})\right) $ and purely cavity quasimode contributions of the
form $Tr_{SR}\,\left( \widehat{\rho }_{I}\,\widehat{V}_{+}^{C}(t_{2})\,\,%
\widehat{V}_{-}^{C}(t_{1})\right) $. Since the continuum quasimodes are in
the vacuum state it is not difficult to see using results such as $\widehat{b%
}(\Delta )(\left\vert 0\right\rangle \left\langle 0\right\vert )_{\Delta }=0$
and $(\left\vert 0\right\rangle \left\langle 0\right\vert )_{\Delta }%
\widehat{b}^{\dag }(\Delta )=0$, that the continuum quasimode term and the
two cross terms all give zero, leaving behind only the cavity quasimode
contribution%
\begin{equation}
Tr_{SR}\,\left( \widehat{\rho }_{I}\,\widehat{V}_{+}(t_{2})\,\,\widehat{V}%
_{-}(t_{1})\right) =Tr_{SR}\,\left( \widehat{\rho }_{I}\,\widehat{V}%
_{+}^{C}(t_{2})\,\,\widehat{V}_{-}^{C}(t_{1})\right)
,
\end{equation}%
where%
\begin{eqnarray}
\widehat{V}_{-}^{C} &=&M^{\ast }\sum\limits_{n\nu }\sqrt{n}\,\left\vert
n-1;\nu \right\rangle \left\langle n;\nu \right\vert \\
\widehat{V}_{+}^{C} &=&M\sum\limits_{m\lambda }\sqrt{m}\,\left\vert
m;\lambda \right\rangle \left\langle m-1;\lambda \right\vert
.
\end{eqnarray}%
The spectrum for Case B will therefore have the same form as that for Case A.

\subsection{Detector atom outside cavity---Case C (side emission)}

In the case where the detector atom is outside the cavity to detect side
emission, and thus responds to the atomic transition operators (\emph{Case C}%
), we have%
\begin{eqnarray}
\widehat{V}_{-} &=&R_{2}^{\ast }\,\widehat{\sigma }_{2}^{-}+R_{1}^{\ast }\,%
\widehat{\sigma }_{1}^{-} \\
&=&R_{2}^{\ast }\sum_{n}\,\left\vert n;1\right\rangle \left\langle
n;2\right\vert +R_{1}^{\ast }\sum_{n}\,\left\vert n;0\right\rangle
\left\langle n;1\right\vert \\
\widehat{V}_{+} &=&R_{2}\,\widehat{\sigma }_{2}^{+}+R_{1}\,\widehat{\sigma }%
_{1}^{+} \\
&=&R_{2}\sum_{m}\,\left\vert m;2\right\rangle \left\langle m;1\right\vert
+R_{1}\sum_{m}\,\left\vert m;1\right\rangle \left\langle m;0\right\vert
\, ,
\end{eqnarray}%
where we have used the notation $R_{12}\rightarrow R_{2}$, $%
R_{01}\rightarrow R_{1}$. Since both $R_{2}$ and $g_{2}$\ are both
proportional to the scalar product of the vector dipole matrix element $%
\left\langle 2\right\vert \,\underrightarrow{\widehat{d}}\,\left\vert
1\right\rangle $ between the upper and intermediate state with either the
polarization unit vector for the spontaneous emission modes or the cavity
quasimode, and similarly for $R_{1}$ and $g_{1}$ in regard to the vector
dipole matrix element $\left\langle 1\right\vert \,\underrightarrow{\widehat{%
d}}\,\left\vert 0\right\rangle $ between the intermediate and lower state,
it follows that 
\begin{equation}
\frac{R_{2}}{R_{1}}=\frac{g_{2}}{g_{1}}
.
\end{equation}%
Since we have chosen $g_{1}$ and $g_{2}$\ to be real we will make the same
choice for $R_{1}$ and $R_{2}$\ from now on.

Hence the two-time correlation function is 
\begin{eqnarray}
&&
Tr_{SR}\,\left( \widehat{\rho }_{I}\,\widehat{V}_{+}(t_{2})\,\,\widehat{V}%
_{-}(t_{1})\right)  
\notag \\
&=&R_{2}^{2}\sum\limits_{m}\sum\limits_{n}\,\left\langle \widehat{S}%
_{m2;\,m1}(t_{2})\widehat{S}_{n1;\,n2}(t_{1})\right\rangle
\nonumber\\&& %
+R_{1}^{2}\sum\limits_{m}\sum\limits_{n}\,\left\langle \widehat{S}%
_{m1;\,m0}(t_{2})\widehat{S}_{n0;\,n1}(t_{1})\right\rangle  \notag \\
&&+R_{2}R_{1}\sum\limits_{m}\sum\limits_{n}\,\left\langle \widehat{S}%
_{m2;\,m1}(t_{2})\widehat{S}_{n0;\,n1}(t_{1})\right\rangle
\nonumber\\&& %
+R_{1}R_{2}\sum\limits_{m}\sum\limits_{n}\,\left\langle \widehat{S}%
_{m1;\,m0}(t_{2})\widehat{S}_{n1;\,n2}(t_{1})\right\rangle  
.
\end{eqnarray}%
Note the four contributions to the spectrum, two involve just one atomic
transition, the other two involve both transitions. The possibility of
interference terms is clear. The derivation of the expression for the
spectrum is given in Appendix \ref{Appendix 4}

The spectrum is given by
\begin{widetext}
\begin{eqnarray}
S(\omega ) &=2\,\text{Re}\,\frac{R_{2}^{2}}{\hbar ^{2}}  
\lbrack & 
\widetilde{U}_{02;\,01::20;\,01}(-i\omega +\epsilon )\,%
\widetilde{\rho }_{20;\,02}(\epsilon ^{\prime })+\widetilde{U}%
_{02;\,01::11;\,01}(-i\omega +\epsilon )\,\widetilde{\rho }%
_{11;\,02}(\epsilon ^{\prime })  \notag \\
&&+\widetilde{U}_{02;\,01::02;\,01}(-i\omega +\epsilon )\,\widetilde{\rho }%
_{02;\,02}(\epsilon ^{\prime })]  \notag \\
&+2\,\text{Re}\,\frac{R_{1}^{2}}{\hbar ^{2}}  
\lbrack &
\widetilde{U}_{01;\,00::01;\,00}(-i\omega +\epsilon )\,%
\widetilde{\rho }_{01;\,01}(\epsilon ^{\prime })+\widetilde{U}%
_{01;\,00::10;\,00}(-i\omega +\epsilon )\,\widetilde{\rho }%
_{10;\,01}(\epsilon ^{\prime })  \notag \\
&&+\{\widetilde{U}_{01;\,00::02;\,10}(-i\omega +\epsilon )\,+\widetilde{U}%
_{11;\,10::02;\,10}(-i\omega +\epsilon )\}\,\widetilde{\rho }%
_{02;\,11}(\epsilon ^{\prime })  \notag \\
&&+\{\widetilde{U}_{01;\,00::11;\,10}(-i\omega +\epsilon )+\widetilde{U}%
_{11;\,10::11;\,10}(-i\omega +\epsilon )\}\,\widetilde{\rho }%
_{11;\,11}(\epsilon ^{\prime })  \notag \\
&&+\{\widetilde{U}_{01;\,00::20;\,10}(-i\omega +\epsilon )+\widetilde{U}%
_{11;\,10::20;\,10}(-i\omega +\epsilon )\}\,\widetilde{\rho }%
_{20;\,11}(\epsilon ^{\prime })]  \notag \\
&+2\,\text{Re}\,\frac{R_{2}R_{1}}{\hbar ^{2}}  
\lbrack &
\{\widetilde{U}_{01;\,00::02;\,01}(-i\omega +\epsilon )+%
\widetilde{U}_{11;\,10::02;\,01}(-i\omega +\epsilon )\}\,\widetilde{\rho }%
_{02;\,02}(\epsilon ^{\prime })  \notag \\
&&+\{\widetilde{U}_{01;\,00::11;\,01}(-i\omega +\epsilon )+\widetilde{U}%
_{11;\,10::11;\,01}(-i\omega +\epsilon )\}\,\widetilde{\rho }%
_{11;\,02}(\epsilon ^{\prime })  
 \notag \\ &&
+\{\widetilde{U}_{01;\,00::20;\,01}(-i\omega +\epsilon )+\widetilde{U}%
_{11;\,10::20;\,01}(-i\omega +\epsilon )\}\,\widetilde{\rho }%
_{20;\,02}(\epsilon ^{\prime })]  \notag \\
&+2\,\text{Re}\,\frac{R_{1}R_{2}}{\hbar ^{2}}  
\lbrack &
\widetilde{U}_{02;\,01::02;\,10}(-i\omega +\epsilon )\,%
\widetilde{\rho }_{02;\,11}(\epsilon ^{\prime })+\widetilde{U}%
_{02;\,01::11;\,10}(-i\omega +\epsilon )\,\widetilde{\rho }%
_{11;\,11}(\epsilon ^{\prime })  
\notag \\ &&
+\widetilde{U}_{02;\,01::20;\,10}(-i\omega +\epsilon )\,\widetilde{\rho }%
_{20;\,11}(\epsilon ^{\prime })]  
.
\label{Eq.SpectrumResultCascadeCaseC}
\end{eqnarray}%
\end{widetext}
We see that as in Case A (and Case B), the spectrum involves Laplace
transforms of the evolution operator from the $U_{1,0::2,1}(t)$, $%
U_{2,1::2,1}(t)$\ and $U_{1,0::1,0}(t)$, and from the density matrix
elements $\rho (2,2,t)$\ and $\rho (1,1,t)$. However, in Case C different
elements are involved, so although the spectra will be qualitatively similar
in the two cases there will be quantitative differences. The spectrum is
calculated from the equation (\ref{Eq.SpectrumResultCascadeCaseC}) using the
results from equations (\ref{Eq.DensityMatrixCascade}), (\ref%
{Eq.EvolutionMatrixCascadeTerm1}) and (\ref{Eq.EvolutionMatrixCascadeTerm2}%
). The spectral variable $\omega $ occurs in the $\widetilde{U}$\ matrices
associated with $(1,0)$ and $(2,1)$ sets. From the equations determining
these, it is easy to see that the spectral variable appears via the factor $%
\Delta \omega =(\omega -\omega _{c})$,\ so again the spectra can be shown as
functions of the detuning of the spectral variable from the cavity frequency.

\section{Cascade atom spectrum: numerical results}

\label{Section 4}

\begin{figure}[t]
\centerline{\includegraphics[width=8cm]{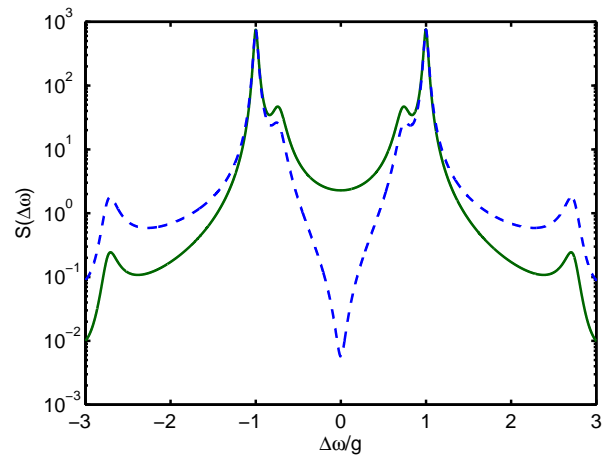}}
\caption[fig3]{
Cascade atom in high-Q cavity. SE spectra 
$S(\omega ) $ versus spectral detuning from cavity frequency $\Delta \omega
=(\omega -\omega _{c})$. The case of resonance is shown, where each
transition is resonant with the cavity frequency, $\delta =\overline{\delta }%
=0$. The coupling constants are $g_{1}=g_{2}=1$ and the detector coupling
constants are $\mu =R_{1}=R_{2}=1$. The cavity decay is $\Gamma =0.1$. The
solid line is for Case A(B)---end emission---and the dashed line is for Case C%
---side emission.} 
\label{Figure3} 
\end{figure}

Results for the spontaneous emission spectra from the three level cascade
atom in a damped high-Q cavity are presented in figures \ref{Figure3}--\ref%
{Figure11} and figures \ref{Figure13}--\ref{Figure14}.
In all cases the spectrum $S(\omega )$ is shown as a function of
the detuning $\Delta \omega =\omega -\omega _{c}$ of the spectral variable $%
\omega $ from the cavity frequency $\omega _{c}$. A logarithmic scale is
used for $S(\omega )$ in order to highlight the numerous spectral peaks, and 
$\Delta \omega $ is in units of $g=g_{2}$. The spectrum for Case A(B)%
---detector in cavity or end emission---is calculated from equation (\ref%
{Eq.SpectrumResultCascadeCaseA}) and the spectrum for Case C---side emission
is calculated from equation (\ref{Eq.SpectrumResultCascadeCaseC}). In all
cases, except figures \ref{Figure13} and \ref{Figure14},
the values of the coupling constants are $g_{1}=g_{2}=1.0$ for
numerical computation. For Case A(B) the detector atom-cavity mode coupling
constant is $\mu =1.0$ ($M=1.0$), whilst for Case C the detector
atom-cascade atom coupling constants are $R_{1}=R_{2}=1.0$
(except in figures \ref{Figure13} and \ref{Figure14}).
The spectra are
shown for various choices of the cavity detuning $\delta $, the half
difference between the two atomic transition frequencies $\overline{\delta }$,
and the cavity decay rate $\Gamma $. Strong coupling cases $g\gg \Gamma $
and intermediate coupling cases $g\sim \Gamma $ are displayed in the figures.

Spectral calculations can also be based on a derivation of the spectra using
density matrix elements defined in terms of dressed atom states, rather than
the uncoupled states that we have used here. This can result in simpler
expressions in the strong coupling regime, where line width factors are
small compared to dressed atom level splittings. Analytical formulae for
positions, heights and widths of spectral peaks can be often obtained in the
strong coupling regime. However, for calculations covering the weak,
intermediate and strong coupling regimes using the dressed atom basis does
not gain much advantage and for the present the uncoupled basis will be used.

Figure \ref{Figure3} is for the case of resonance, with $\overline{\delta }%
=\delta =0.0$, where the two atomic transition frequencies are equal and the
same as the cavity frequency. A strong coupling situation with $\Gamma =0.1$
is presented for both Case A(B) and Case C. A symmetrical six peak spectrum
is shown. For Case C the spectral intensity becomes very small when the
spectral frequency equals the cavity frequency, i.e.\ when $\Delta \omega =0 
$. This effect is not seen for Case A(B) and indicates the presence of
interference effects.

\begin{figure}[t!]
\centerline{\includegraphics[width=8cm]{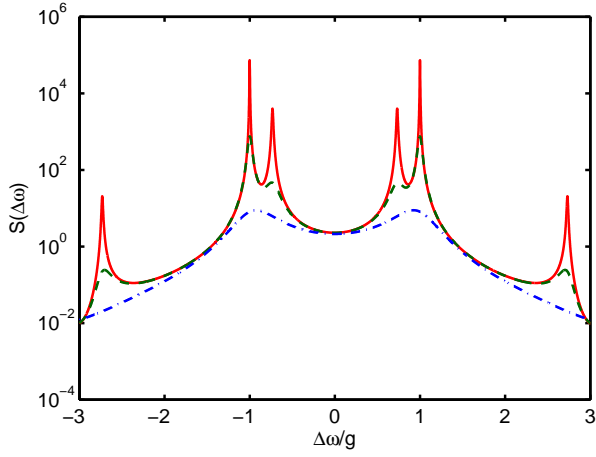}}
\caption[fig4]{
Cascade atom in high-Q cavity. SE
spectra $S(\omega ) $ versus spectral detuning from cavity frequency $\Delta
\omega =(\omega -\omega _{c})$. The case of resonance is shown, where each
transition is resonant with the cavity frequency, $\delta =\overline{\delta }%
=0$. The coupling constants are $g_{1}=g_{2}=1$ and the detector coupling
constant is $\mu =1$. The cavity decay $\Gamma =0.01$ is the solid line, $%
\Gamma =0.1$ is the dashed line and $\Gamma =1.0$ is the chained line. The
spectrum for Case A(B)---end emission---is shown.} \label{Figure4} 
\end{figure}

\begin{figure}[t!]
\centerline{\includegraphics[width=8cm]{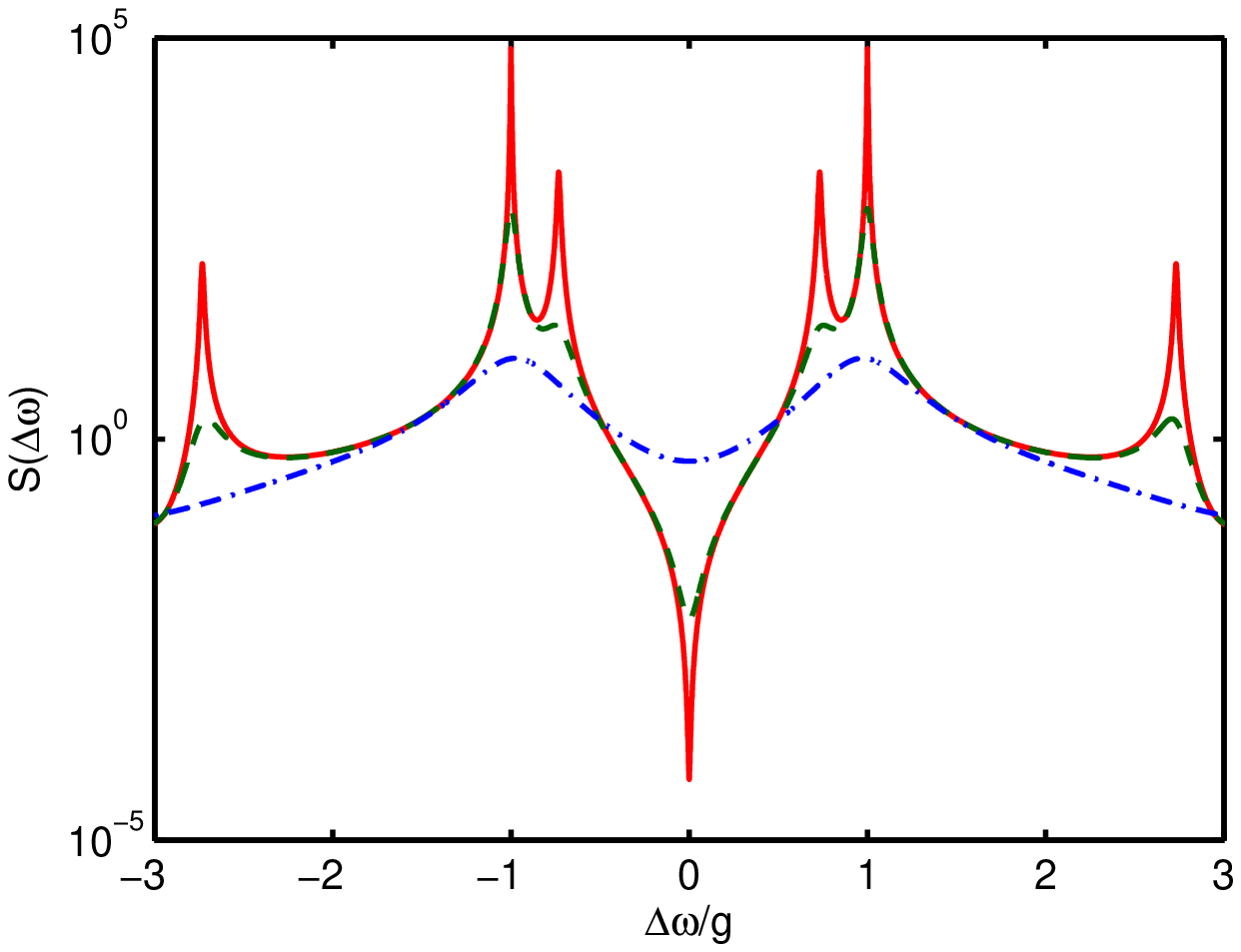}}
\caption[fig5]{
Cascade atom in high-Q cavity. SE
spectra $S(\omega ) $ versus spectral detuning from cavity frequency $\Delta
\omega =(\omega -\omega _{c})$. The case of resonance is shown, where each
transition is resonant with the cavity frequency, $\delta =\overline{\delta }%
=0$. The coupling constants are $g_{1}=g_{2}=1$ and the detector coupling
constants are $R_{1}=R_{2}=1$. The cavity decay $\Gamma =0.01$ is the solid
line, $\Gamma =0.1$ is the dashed line and $\Gamma =1.0$ is the chained
line. The spectrum for Case C---side emission---is shown.} 
\label{Figure5} 
\end{figure}

Figures \ref{Figure4} and \ref{Figure5} are also for the same resonance
situation as in figure \ref{Figure3}, $\overline{\delta }=\delta =0.0$ for
Case A(B) and Case C respectively. In these figures the cavity decay rate $%
\Gamma $ takes on the values $\Gamma =1.0$, $0.1$ and $0.01$, traversing the
regime from intermediate coupling to very strong coupling. In the strong and
very strong coupling cases $\Gamma =0.1$ and $0.01$ the six peaks are very
clearly seen, whilst for intermediate coupling only two peaks remain due to
the line broadening effect of the larger cavity decay. Again, for Case C
(figure~\ref{Figure5}) the spectral intensity tends to zero for $\Delta
\omega =0$ and $\Gamma =0.1$ and $0.01$. However, for $\Gamma =1.0$, the
spectral hole is suppressed and in fact the spectrum in Case C is very
similar to that of Cases A(B) for $\Gamma =1.0$ as seen in figure \ref%
{Figure4}. We know that for a two-level system in the low-Q limit the cavity
field operators can be adiabatically eliminated in favour of the atomic
operators, i.e.\ the cavity field `follows' the atomic state. For this
reason we might expect that spectra based on field operators (case A) and
spectra based on the atomic operators (case C) would become similar in this
limit.

\begin{figure}[t!]
\centerline{\includegraphics[width=8cm]{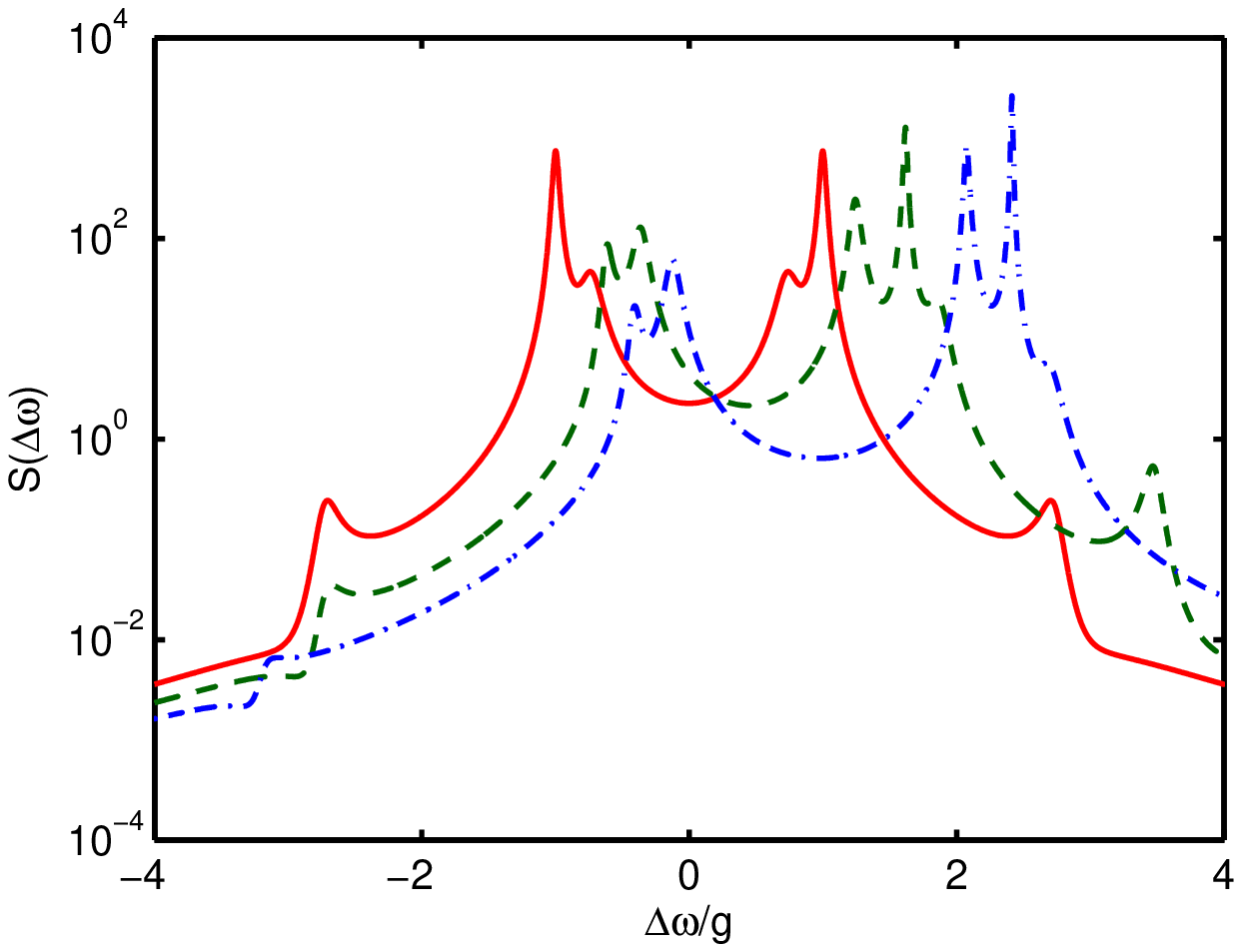}}
\caption[fig6]{
Cascade atom in high-Q cavity. SE
spectra $S(\omega ) $ versus spectral detuning from cavity frequency $\Delta
\omega =(\omega -\omega _{c})$. The case of equal atomic transition
frequencies is shown, but with the cavity frequency is detuned from the
average of the two transition frequencies. $\overline{\delta }=0$ but $%
\delta $ may be non-zero. The coupling constants are $g_{1}=g_{2}=1$ and the
detector coupling constant is $\mu =1$. The cavity decay is $\Gamma =0.1$.
The cavity detuning $\delta =0$ is the solid line, $\delta =-1$ is the
dashed line, $\delta =-2$ is the chained line. The spectrum for Case A(B)%
---end emission---is shown.} 
\label{Figure6}
\end{figure}

\begin{figure}[t!]
\centerline{\includegraphics[width=8cm]{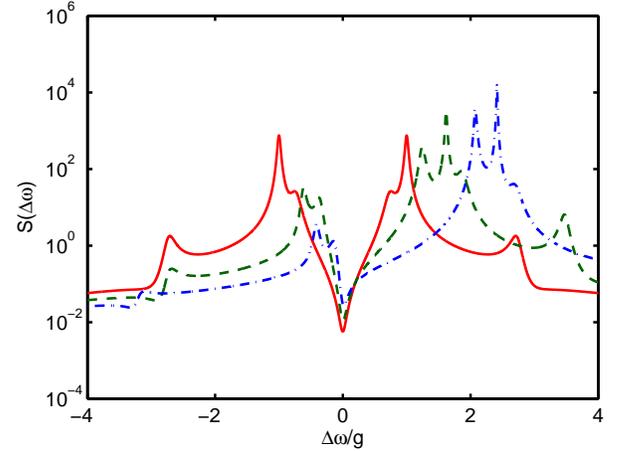}}
\caption[fig7]{
Cascade atom in high-Q cavity. SE
spectra $S(\omega ) $ versus spectral detuning from cavity frequency $\Delta
\omega =(\omega -\omega _{c})$. The case of equal atomic transition
frequencies is shown, but with the cavity frequency is detuned from the
average of the two transition frequencies. $\overline{\delta }=0$, but $%
\delta $ may be non-zero. The coupling constants are $g_{1}=g_{2}=1$ and the
detector coupling constants are $R_{1}=R_{2}=1$. The cavity decay is $\Gamma
=0.1$. The cavity detuning $\delta =0$ is the solid line, $\delta =-1$ is
the dashed line, $\delta =-2$ is the chained line. The spectrum for Case C%
---side emission---is shown.} \label{Figure7} 
\end{figure}

Figures \ref{Figure6} and \ref{Figure7} for Cases A(B) and Case C
respectively, apply to the situation where the two atomic transition
frequencies are the same $\overline{\delta }=0.0$, but where the cavity
detuning $\delta $ may be non-zero. Cases where $\delta =0.0$ (resonance), $%
-1.0$ and $-2.0$ are shown. A strong coupling situation with $\Gamma =0.1$
applies in both figures. The effect of cavity detuning is that the six peak
symmetrical spectra, applying for resonance, are replaced by asymmetrical
spectra still with six peaks, but with the hint that other peaks may be
hidden in the relatively broad features. Again the hole in the intensity for
Case C near $\Delta \omega =0$ remains when the detuning is non-zero, though
it is not as deep as in figure \ref{Figure5} for $\Gamma =0.1$ (dashed
curve).

\begin{figure}[t!]
\centerline{\includegraphics[width=8cm]{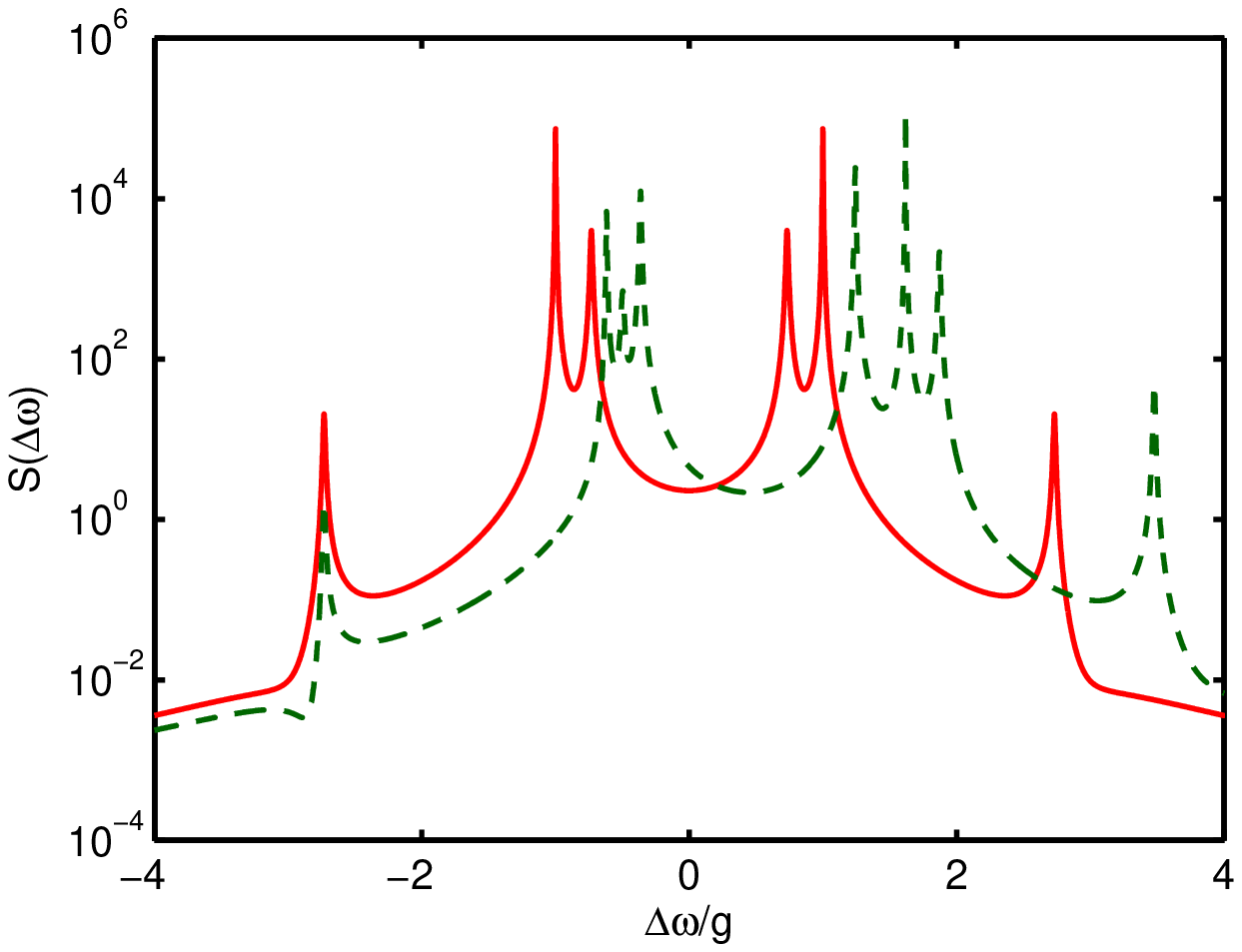}}
\caption[fig8]{
Cascade atom in high-Q cavity. SE
spectra $S(\omega ) $ versus spectral detuning from cavity frequency $\Delta
\omega =(\omega -\omega _{c})$. The case of equal atomic transition
frequencies is shown, but with the cavity frequency is detuned from the
average of the two transition frequencies. $\overline{\delta }=0$, but $%
\delta $ may be non-zero. The coupling constants are $g_{1}=g_{2}=1$ and the
detector coupling constant is $\mu =1$. The cavity decay is $\Gamma =0.01$.
The cavity detuning $\delta =0$ is the solid line, $\delta =-1$ is the
dashed line. The spectrum for Case A(B)---end emission---is shown.} 
\label{Figure8} 
\end{figure}

\begin{figure}[t!]
\centerline{\includegraphics[width=8cm]{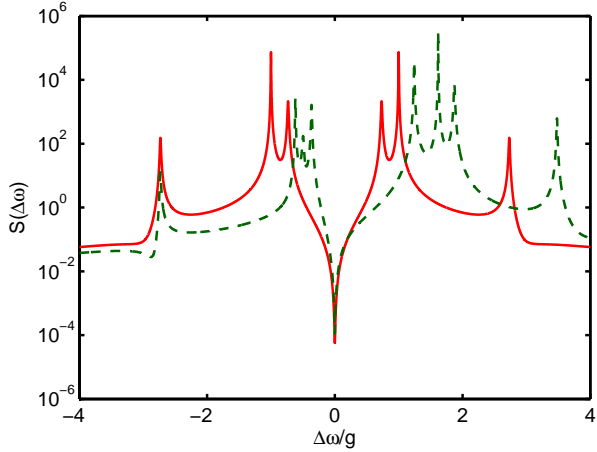}}
\caption[fig9]{
Cascade atom in high-Q cavity. SE
spectra $S(\omega ) $ versus spectral detuning from cavity frequency $\Delta
\omega =(\omega -\omega _{c})$. The case of equal atomic transition
frequencies is shown, but with the cavity frequency is detuned from the
average of the two transition frequencies. $\overline{\delta }=0$, but $%
\delta $ may be non-zero. The coupling constants are $g_{1}=g_{2}=1$ and the
detector coupling constants are $R_{1}=R_{2}=1$. The cavity decay is $\Gamma
=0.01$. The cavity detuning $\delta =0$ is the solid line, $\delta =-1$ is
the dashed line. The spectrum for Case C---side emission---is shown.} \label%
{Figure9} 
\end{figure}

Figures \ref{Figure8} and \ref{Figure9} for Cases A(B) and Case C
respectively, also apply to the situation where the two atomic transition
frequencies are the same, $\overline{\delta }=0.0$, but where the cavity
detuning $\delta $ may be non-zero. Now a very strong coupling situation
with $\Gamma =0.01$ applies in both figures, so that the spectral peaks will
become narrower. Cases where $\delta =0.0$ (resonance) and $-1.0$ are shown.
It is now clearly shown that the effect of detuning is to replace the six
peak symmetrical spectrum for the resonance case, with eight peaks. We also
note that in the detuned case one of the peaks displays a distinctly
dispersion-like line shape. The spectral hole at $\Delta \omega =0$ for Case
C now reaches an even lower intensity ($10^{-4}$ units) than for figure \ref%
{Figure7} ($10^{-2}$ units), and is equally sharp for the detuned case as
for resonance.

\begin{figure}[t!]
\centerline{\includegraphics[width=8cm]{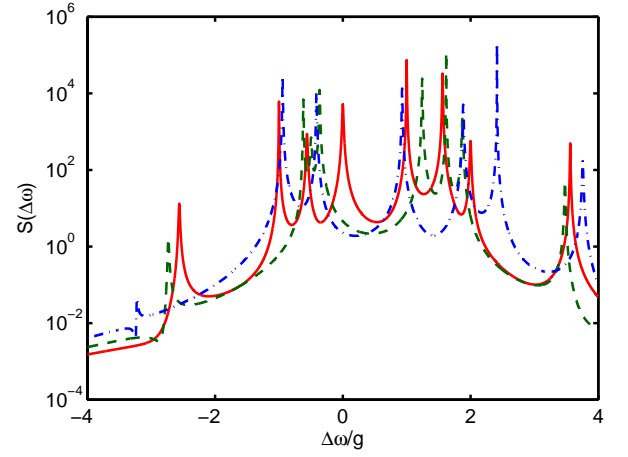}}
\caption[fig10]{
Cascade atom in high-Q cavity. SE
spectra $S(\omega )$ versus spectral detuning from cavity frequency $\Delta
\omega =(\omega -\omega _{c})$. The case of cavity detuning from the average
of the two transition frequencies $\delta =-1$. The atomic transition
frequencies may be different. The coupling constants are $g_{1}=g_{2}=1$ and
the detector coupling constant is $\mu =1$. The cavity decay is $\Gamma =0.01
$. The half atomic transition frequency difference $\overline{\delta }%
=-1=+\delta $ is the solid line, $\overline{\delta }=0$ is the dashed line
and $\overline{\delta }=1=-\delta $ is the chained line. For $\overline{%
\delta }=+\delta ,-\delta $ the cavity frequency coincides with the lower,
upper atomic transition frequency respectively. The spectrum for Case A(B)%
---end emission---is shown.} \label{Figure10} 
\end{figure}

\begin{figure}[t!]
\centerline{\includegraphics[width=8cm]{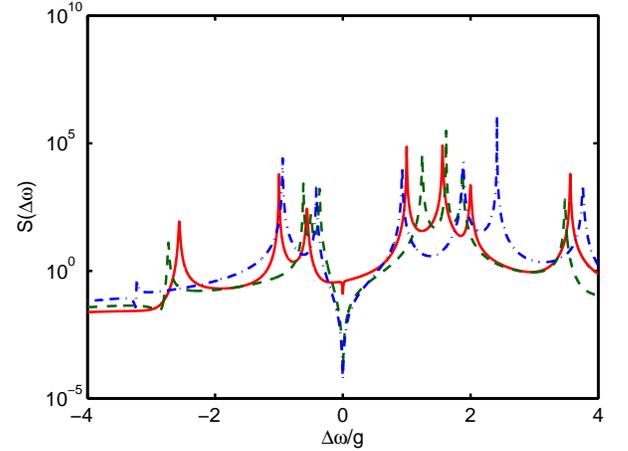}}
\caption[fig11]{
Cascade atom in high-Q cavity. SE
spectra $S(\omega )$ versus spectral detuning from cavity frequency $\Delta
\omega =(\omega -\omega _{c})$. The case of cavity detuning from the average
of the two transition frequencies $\delta =-1$. The atomic transition
frequencies may be different. The coupling constants are $g_{1}=g_{2}=1$ and
the detector coupling constants are $R_{1}=R_{2}=1$. The cavity decay is $%
\Gamma =0.01$. The half atomic transition frequency difference $\overline{%
\delta }=-1=+\delta $ is the solid line, $\overline{\delta }=0$ is the
dashed line and $\overline{\delta }=1=-\delta $ is the chained line. For $%
\overline{\delta }=+\delta ,-\delta $ the cavity frequency coincides with
the lower, upper atomic transition frequency respectively. The spectrum for
Case C---side emission---is shown.} \label{Figure11} 
\end{figure}

Figures \ref{Figure10} and \ref{Figure11} for Cases A(B) and Case C
respectively, apply to the situation where the cavity frequency is detuned
from the average of the atomic transition frequencies, and where these
atomic transition frequencies may themselves be different. The cavity
detuning is $\delta =-1.0$ in both figures. A very strong coupling situation
with $\Gamma =0.01$ applies in both figures so that the spectral peaks will
become narrower. Cases where $\overline{\delta }=-1.0=+\delta $, $\overline{%
\delta }=0.0$ (both atomic transition frequencies equal) and $\overline{%
\delta }=+1.0=-\delta $ are shown. Note that for the case $\overline{\delta }%
=-\delta $, the cavity frequency coincides with the upper atomic
transition frequency ($\omega _{c}=\omega _{1}$), whilst the case $\overline{%
\delta }=+\delta $, the cavity frequency coincides with the lower atomic
transition frequency ($\omega _{c}=\omega _{2}$). In both Case A(B) and Case
C up to eight spectral peaks occur. For Case C (see figure \ref{Figure11})
an interesting feature is that the spectral hole at $\Delta \omega =0$ is
not present when the cavity frequency is resonant with the lower atomic
transition frequency ($\overline{\delta }=+\delta $), but is still present
when the cavity frequency is resonant with the upper atomic transition
frequency.

The same general features for the present Case A(B) and Case C situation in
terms of the numbers of spectral features were found by Zhou et al.\ \cite%
{Zhou05a}, but the presence of the spectral hole at $\Delta \omega =0$ for
Case C (which they referred to as the emission spectrum) and its
relationship to quantum interference was not reported in their paper. Their
lineshapes were associated only with the finite spectrometer bandwidth
incorporated in their definition of the spectra and did not reflect cavity
decay---which was assumed to be zero.\

\begin{figure}
\centerline{\includegraphics[width=5cm]{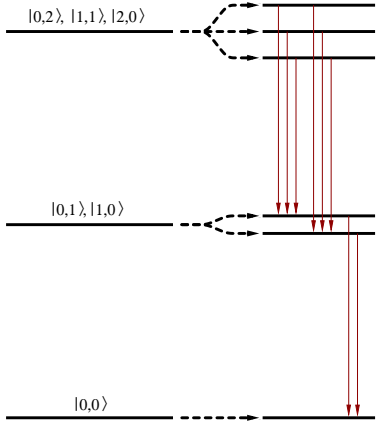}}
\caption[fig12]{
The uncoupled atom-cavity mode energy levels and the
dressed atom energy levels allowing for atom-cavity mode coupling. The
situation shown is that of resonance and with equidistant cascade atom
levels. $\delta =\overline{\delta }=0$. The uncoupled states are denoted $%
\left\vert n;\nu \right\rangle $, where $n=0,1,2, \hdots $ gives the number of
photons in the cavity mode and $\nu =0,1,2$ lists the cascade atom states.
The three lowest dressed atom multiplets (singlet, doublet, triplet) are
shown. Allowed transitions between the dressed atom states are indicated.} 
\label{Figure12} 
\end{figure}

The spectral features can be interpreted in terms of the dressed atom model 
\cite{Cohen-Tannoudji77a}. Figure \ref{Figure12} shows the uncoupled
atom-cavity mode states and their splitting into dressed atom states for the
three lowest dressed atom multiplets in the case of resonance $\delta =%
\overline{\delta }=0$. The lowest multiplet is a singlet and is based on the $%
\left\vert 0;0\right\rangle $\ uncoupled states. The next lowest multiplet
is a doublet and is based on the $\left\vert 0;1\right\rangle $\ and $%
\left\vert 1;0\right\rangle $\ uncoupled states. The next lowest multiplet
is a triplet and is based on the $\left\vert 0;2\right\rangle $, $\left\vert
1;1\right\rangle $\ and $\left\vert 2;0\right\rangle $\ uncoupled states.
These are of course the only states that could be occupied in the cascade
atom decay process. The uncoupled levels are separated by an energy $\hbar
\omega _{c}=\hbar \omega _{0}$. For the resonance case the averages of
dressed atom multiplet levels does not change from the uncoupled energies,
and the triplet levels are equispaced. For the
resonance case, the energies of the three triplet states relative to the
uncoupled energy are $+\hbar \sqrt{2g_{1}^{2}+g_{2}^{2}}$, $0$, $-\hbar 
\sqrt{2g_{1}^{2}+g_{2}^{2}}$, whilst those for the doublet state are $+\hbar
g_{1}$, $-\hbar g_{1}$ and $0$ for the singlet state. The position for the
spectral features is in accord with these dressed atom energies. 
For non-zero $\delta $, $\overline{\delta}$ the
dressed atom multiplets do not have these symmetry features. The selection
rules for electric dipole processes only allow transitions between
neighbouring dressed atom multiplets. It is therefore easy to see from figure %
\ref{Figure12} that there will be up to eight different possible transition
frequencies, all clustered around $\omega =\omega _{c}$, and in non-zero $%
\delta $, $\overline{\delta }$ situations all eight frequencies could be
present. In the resonance case $\delta =\overline{\delta }=0$\ the symmetry
feature results in the middle to bottom dressed atom transition frequencies
coinciding with two of the upper to middle dressed atom transition
frequencies, thereby reducing the number of distinct lines to six in this
case. The fact that photons of the same frequency could have come from
either an upper to middle or a middle to lower dressed atom transition, and
that these processes cannot be separately measured, is likely to result in
quantum interference effects.\

\begin{figure}
\centerline{\includegraphics[width=8cm]{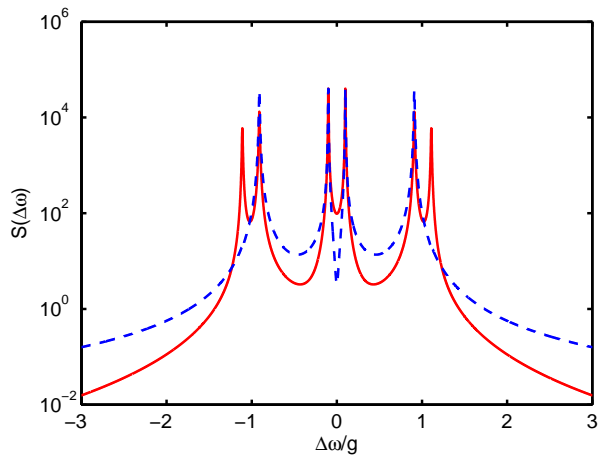}}
\caption[fig13]{
Cascade atom in high-Q cavity. SE
spectra $S(\omega )$ versus spectral detuning from cavity frequency $\Delta
\omega =(\omega -\omega _{c})$. The case of resonance is shown, where each
transition is resonant with the cavity frequency, $\delta =\overline{\delta }%
=0$. The coupling constants are $g_{1}=0.1$, $g_{2}=1$ and the detector
coupling constants are $R_{1}=0.1$, $R_{2}=1$. The cavity decay is $\Gamma
=0.01$. The spectrum for Case A(B)---end emission---is shown as the solid
line, that for Case C---side emission---is shown as the dashed line.}
\label{Figure13} 
\end{figure}

\begin{figure}
\centerline{\includegraphics[width=8cm]{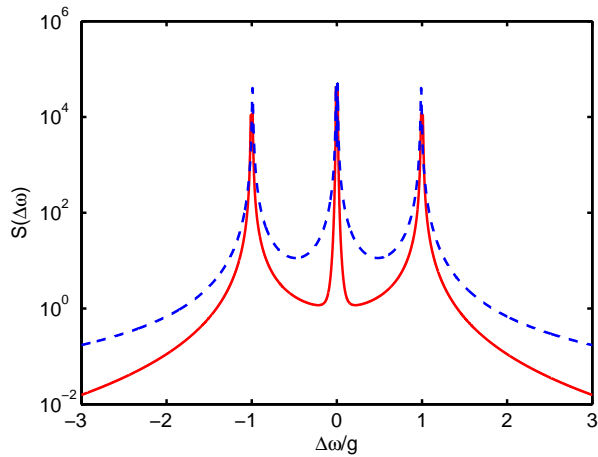}}
\caption[fig14]{
Cascade atom in high-Q cavity. SE
spectra $S(\omega )$ versus spectral detuning from cavity frequency $\Delta
\omega =(\omega -\omega _{c})$. The case of resonance is shown, where each
transition is resonant with the cavity frequency, $\delta =\overline{\delta }%
=0$. The coupling constants are $g_{1}=0.01$, $g_{2}=1$ and the detector
coupling constants are $R_{1}=0.01$, $R_{2}=1$. The cavity decay is $\Gamma
=0.01$. The spectrum for Case A(B)---end emission---is shown as the solid
line, that for Case C---side emission---is shown as the dashed line. }
\label{Figure14} 
\end{figure}

To demonstrate the role of both upper
to middle and middle to lower dressed atom transitions in the interference
process, figures \ref{Figure13}, and \ref{Figure14} show the effect for the
resonance case of switching off the middle to lower dressed atom
transitions. The figures show the spectrum for Case A(B)---end emission---as
the solid lines and for Case C---side emission---as dashed lines. In these two
figures the two atomic transition frequencies are equal and the same as the
cavity frequency $\overline{\delta }=\delta =0.0$, the cavity decay rate is $%
\Gamma =0.01$, the value of the upper transition coupling constant is $%
g_{2}=1.0 = g$. For Case A(B) the detector atom-cavity mode coupling constant is 
$\mu =1.0$ ($M=1.0$), whilst for Case C the detector atom-cascade atom upper
transition coupling constant is $R_{2}=1.0$. In figure \ref{Figure13} the
lower transition and detector atom coupling constants are $g_{1}=R_{1}=0.1$,
for figure \ref{Figure14} the values are $g_{1}=R_{1}=0.01$. The solid lines
in previous figures \ref{Figure4} (Case A(B)) and \ref{Figure5} (Case C)
show the spectra for the resonance case with $g_{1}=R_{1}=1.0$, where the
lower transition and detector atom coupling constants are the same as those
for the upper transition. The interference hole at $\Delta \omega =0$ for
Case C is present in figure \ref{Figure5} is much less prominent in figure %
\ref{Figure13} and has disappeared in figure \ref{Figure14}, as the lower
dressed atom transitions have been essentially switched off. In fact there
is now a spectral peak at $\Delta \omega =0$ for both Cases A(B) and C for
the situation in figure \ref{Figure14}. This shows that both upper to middle
and middle to lower dressed atom transitions are needed to produce the
destructive interference effect associated with the spectral hole.

The situation when the cavity frequency is resonant with the lower atomic
transition frequency and significantly detuned from the upper atomic
transition frequency is associated with a dressed atom level scheme in which
the upper triplet is replaced by a doublet associated with the splitting of
the degenerate uncoupled states $\left\vert 1;1\right\rangle $\ and $%
\left\vert 2;0\right\rangle $ together with a separate dressed atom state
which is essentially $\left\vert 0;2\right\rangle $. The doublet based on
the $\left\vert 0;1\right\rangle $\ and $\left\vert 1;0\right\rangle $\
uncoupled states is still present, as is the singlet based on $\left\vert
0;0\right\rangle $. In this situation for Case C (side emission), the
spectral hole at $\Delta \omega =0$ is absent. For the situation when the
cavity frequency is resonant with the upper atomic transition frequency and
significantly detuned from the lower atomic transition frequency is
associated with a dressed atom level scheme in which the upper triplet is
replaced by a doublet associated with the splitting of the degenerate
uncoupled states $\left\vert 1;1\right\rangle $\ and $\left\vert
0;2\right\rangle $ together with a separate dressed atom state which is
essentially $\left\vert 2;0\right\rangle $. The doublet based on the $%
\left\vert 0;1\right\rangle $\ and $\left\vert 1;0\right\rangle $\ uncoupled
states is now replaced by two singlets, each associated with these uncoupled
states. The singlet based on $\left\vert 0;0\right\rangle $ is unchanged. In
this situation for Case C (side emission), the spectral hole at $\Delta
\omega =0$ is prominent. Why this is the case and what accounts for the
general absence of the spectral hole at $\Delta \omega =0$ for the Case A(B)
(end emission) and its usual occurrence in Case C (side emission) spectra is
not yet clear and a more extensive analysis is needed to find the answer,
possibly based on dressed atom states. For the strong
coupling regime, the use of dressed atom states may enable transition
amplitudes for different processes---beginning with the same initial state
with the atom in the upper state $\left\vert 2\right\rangle $, no photons
present in any mode and the detector in state $\left\vert A\right\rangle $
and ending in specific final states with the detector in state $\left\vert
B\right\rangle $, to be evaluated analytically using a perturbative approach
for the weak detector atom and cavity decay transitions, but treating the
strong coupling processes involving the coupling constants $g_{1},g_{2}$
without such an approximation. Having analytic expressions for the various
quantum pathways would facilitate an understanding of the interference
effects that emerge in the spectra calculated numerically via the master
equation-quantum regression theorem approach. 

\section{Conclusion}

\label{Section 5}

In this paper we have examined the spontaneous emission spectrum from a
cascade atom located inside a cavity. That spectrum was examined for three
different physical situations regarding the location of an idealised
detector atom. In case A the detector atom was located inside the cavity and
directly influenced by the cavity field. In case B the detector atom was
exposed to the field emerging from a partially transmitting cavity mirror.
In case C the detector atom responded to sideways emission from the cascade
atom inside the cavity. The cascade atom was initially in its uppermost
level and the cavity mode empty of photons, and thus a system with two basic
excitations was studied. The spectrum was found from the weakly coupled
detector atom response and involved in each case Laplace transforms of the
density matrix and matrix elements of the evolution operator in a
super-operator form. The spectral line-shapes only reflect cavity decay,
since the weakly coupled detector atom spectrometer had zero bandwidth.
Spectra have been presented for intermediate and strong coupling regime
situations where both atomic transitions are resonant with the cavity
frequency, for cases of non-zero cavity detuning and for cases where the two
atomic transition frequencies differ.

The spectra for Case A and Case B were found to be essentially the same. The
spectral features for Cases B(A) and C were qualitatively similar, with six
spectral peaks for resonance cases and eight for detuned cases. These
general features of the spectra could be explained via the dressed atom
model. However, Case B(A) and Case C spectra differed in detail. In
particular, the spectra for Case C exhibited a deep spectral hole when the
spectral frequency was equal to the cavity mode frequency, a feature that
persisted, for strong coupling, over a wide range of detuning and coupling
conditions. However, this spectral hole was absent for Case A, a feature
that cannot yet be explained. The spectral hole in Case C was particularly
prominent for resonance conditions, when both atomic transition frequencies
were the same and resonant with the cavity frequency. In this resonance
situation, photons associated with the middle to bottom dressed atom
transitions coincide in frequency with photons associated with two of the
upper to middle dressed atom transitions, and therefore the origin of a
photon detected in the spectrometer cannot be identified. This suggests that
an interference effect is occuring, and this is
confirmed by showing that the hole disappears when only the upper to middle
dressed atom transitions are present. However, why the interference hole is
absent in Case B(A) spectra is still unclear. 
One clue is that the spectral hole in Case C spectra disappeared in
figure \ref{Figure11} when the atomic transition frequencies were quite
different and the cavity frequency was resonant with the lower atomic
transition frequency. This disappearance was sensitive to both the detunings
having opposite sign and a magnitude close to $g=g_{1}=g_{2}$. The
explanation is expected to be found from an examination of the dressed
states in the strong coupling limit. This analysis is currently being
undertaken.

\begin{acknowledgments}
This paper is in honour of the 60th birthday of Professor Sir Peter Knight
FRS, whose major contributions in the field of multi-photon physics, quantum
optics and quantum information it is a pleasure to acknowledge.
\end{acknowledgments}

\appendix

\vspace*{1em}

\section{Coupled density matrix sets}

\label{Appendix 1}

For the $(n,m)$ set $(n,m\geqslant 2)$ we define the $1\times 9$ column
matrix $\mathbf{\rho }(n,m)$ of density matrix elements in the $(n,m)$ set as
\begin{equation}
\mathbf{\rho }(n,m)=
\begin{bmatrix}
\rho _{\overline{n-2}2;\,\overline{m-2}2} \\ 
\rho _{\overline{n-2}2;\,\overline{m-1}1} \\ 
\rho _{\overline{n-2}2;\,m0} \\ 
\rho _{\overline{n-1}1;\,\overline{m-2}2} \\ 
\rho _{\overline{n-1}1;\,\overline{m-1}1} \\ 
\rho _{\overline{n-1}1;\,m0} \\ 
\rho _{n0;\,\overline{m-2}2} \\ 
\rho _{n0;\,\overline{m-1}1} \\ 
\rho _{n0;\,m0}
\end{bmatrix}
,
\end{equation}%
\begin{widetext}
and introduce suitable $9\times 9$ matrices $A(n,m),B(n,m)$ to incorporate
the couplings, detunings and decay terms%
\begin{eqnarray}
&&A(n,m)=\omega _{c}(n-m)\left[ 
\begin{smallmatrix}
1 & 0 & 0 & 0 & 0 & 0 & 0 & 0 & 0 \\ 
0 & 1 & 0 & 0 & 0 & 0 & 0 & 0 & 0 \\ 
0 & 0 & 1 & 0 & 0 & 0 & 0 & 0 & 0 \\ 
0 & 0 & 0 & 1 & 0 & 0 & 0 & 0 & 0 \\ 
0 & 0 & 0 & 0 & 1 & 0 & 0 & 0 & 0 \\ 
0 & 0 & 0 & 0 & 0 & 1 & 0 & 0 & 0 \\ 
0 & 0 & 0 & 0 & 0 & 0 & 1 & 0 & 0 \\ 
0 & 0 & 0 & 0 & 0 & 0 & 0 & 1 & 0 \\ 
0 & 0 & 0 & 0 & 0 & 0 & 0 & 0 & 1%
\end{smallmatrix}%
\right]   \notag \\
&&\hspace*{-2cm}+\left[ 
\begin{smallmatrix}
-\frac{1}{2}i\Gamma (p-4) & -g_{2}\,\sqrt{m-1} & 0 & g_{2}\,\sqrt{n-1} & 0 & 
0 & 0 & 0 & 0 \\ 
-g_{2}\,\sqrt{m-1} & -(\delta +\overline{\delta })-\frac{1}{2}i\Gamma (p-3)
& -g_{1}\,\sqrt{m} & 0 & g_{2}\,\sqrt{n-1} & 0 & 0 & 0 & 0 \\ 
0 & -g_{1}\,\sqrt{m} & -2\delta -\frac{1}{2}i\Gamma (p-2) & 0 & 0 & g_{2}\,%
\sqrt{n-1} & 0 & 0 & 0 \\ 
g_{2}\,\sqrt{n-1} & 0 & 0 & (\delta +\overline{\delta })-\frac{1}{2}i\Gamma
(p-3) & -g_{2}\,\sqrt{m-1} & 0 & g_{1}\,\sqrt{n} & 0 & 0 \\ 
0 & g_{2}\,\sqrt{n-1} & 0 & -g_{2}\,\sqrt{m-1} & -\frac{1}{2}i\Gamma (p-2) & 
-g_{1}\,\sqrt{m} & 0 & g_{1}\,\sqrt{n} & 0 \\ 
0 & 0 & g_{2}\,\sqrt{n-1} & 0 & -g_{1}\,\sqrt{m} & (-\delta +\overline{%
\delta })-\frac{1}{2}i\Gamma (p-1) & 0 & 0 & g_{1}\,\sqrt{n} \\ 
0 & 0 & 0 & g_{1}\,\sqrt{n} & 0 & 0 & +2\delta -\frac{1}{2}i\Gamma (p-2) & 
-g_{2}\,\sqrt{m-1} & 0 \\ 
0 & 0 & 0 & 0 & g_{1}\,\sqrt{n} & 0 & -g_{2}\,\sqrt{m-1} & -(-\delta +%
\overline{\delta })-\frac{1}{2}i\Gamma (p-1) & -g_{1}\,\sqrt{m} \\ 
0 & 0 & 0 & 0 & 0 & g_{1}\,\sqrt{n} & 0 & -g_{1}\,\sqrt{m} & -\frac{1}{2}%
i\Gamma p%
\end{smallmatrix}%
\right]   \notag \\
&&
\end{eqnarray}%
with $p=n+m$, and 
\begin{equation}
\hspace*{-2.5cm}B(n,m)=-i\Gamma \left[ 
\begin{smallmatrix}
\sqrt{n-1}\sqrt{m-1} & 0 & 0 & 0 & 0 & 0 & 0 & 0 & 0 \\ 
0 & \sqrt{n-1}\sqrt{m} & 0 & 0 & 0 & 0 & 0 & 0 & 0 \\ 
0 & 0 & \sqrt{n-1}\sqrt{m+1} & 0 & 0 & 0 & 0 & 0 & 0 \\ 
0 & 0 & 0 & \sqrt{n}\sqrt{m-1} & 0 & 0 & 0 & 0 & 0 \\ 
0 & 0 & 0 & 0 & \sqrt{n}\sqrt{m} & 0 & 0 & 0 & 0 \\ 
0 & 0 & 0 & 0 & 0 & \sqrt{n}\sqrt{m+1} & 0 & 0 & 0 \\ 
0 & 0 & 0 & 0 & 0 & 0 & \sqrt{n+1}\sqrt{m-1} & 0 & 0 \\ 
0 & 0 & 0 & 0 & 0 & 0 & 0 & \sqrt{n+1}\sqrt{m} & 0 \\ 
0 & 0 & 0 & 0 & 0 & 0 & 0 & 0 & \sqrt{n+1}\sqrt{m+1}%
\end{smallmatrix}%
\right] 
.
\end{equation}
\end{widetext}

For the $(0,0)$\ set we have%
\begin{equation}
\mathbf{\rho }(0,0)=%
\begin{bmatrix}
\rho _{00;\,00}%
\end{bmatrix}%
\end{equation}%
\begin{equation}
A(0,0)=\omega _{c}(0)%
\begin{bmatrix}
1%
\end{bmatrix}%
\end{equation}%
\begin{equation}
B(0,0)=%
\begin{bmatrix}
0 & 0 & 0 & -i\Gamma%
\end{bmatrix}%
.
\end{equation}

For the $(1,0)$\ set we have%
\begin{equation}
\mathbf{\rho }(1,0)=%
\begin{bmatrix}
\rho _{01;\,00} \\ 
\rho _{10;\,00}%
\end{bmatrix}%
\end{equation}%
\begin{eqnarray}
A(1,0) &=&\omega _{c}(1)%
\begin{bmatrix}
1 & 0 \\ 
0 & 1%
\end{bmatrix}
+
\begin{bmatrix}
-\delta +\overline{\delta } & g_{1} \\ 
g_{1} & -\frac{1}{2}i\Gamma 
\end{bmatrix}%
\end{eqnarray}
\begin{equation}
B(1,0)=\left[ 
\begin{tabular}{llllll}
$0$ & $0$ & $0$ & $-i\Gamma $ & $0$ & $0$ \\ 
$0$ & $0$ & $0$ & $0$ & $0$ & $-i\sqrt{2}\Gamma $%
\end{tabular}%
\right]
.
\end{equation}

For the $(1,1)$ set we have%
\begin{equation}
\mathbf{\rho }(1,1)=\left[ 
\begin{tabular}{l}
$\rho _{01;\,01}$ \\ 
$\rho _{01;\,10}$ \\ 
$\rho _{10;\,01}$ \\ 
$\rho _{10;\,10}$%
\end{tabular}%
\right] 
\end{equation}%
\begin{eqnarray}
A(1,1) &=&\omega _{c}(0)\left[ 
\begin{tabular}{llll}
$1$ & $0$ & $0$ & $0$ \\ 
$0$ & $1$ & $0$ & $0$ \\ 
$0$ & $0$ & $1$ & $0$ \\ 
$0$ & $0$ & $0$ & $1$%
\end{tabular}%
\right]   
\notag \\ && %
+\left[ 
\begin{tabular}{llll}
$0$ & $-g_{1}$ & $g_{1}$ & $0$ \\ 
$-g_{1}$ & $-\delta +\overline{\delta }-\frac{1}{2}i\Gamma $ & $0$ & $g_{1}$
\\ 
$g_{1}$ & $0$ & $+\delta -\overline{\delta }-\frac{1}{2}i\Gamma $ & $-g_{1}$
\\ 
$0$ & $g_{1}$ & $-g_{1}$ & $-i\Gamma $%
\end{tabular}%
\right] 
\notag \\ && %
\end{eqnarray}%
\begin{equation}
B(1,1)=\left[ 
\begin{tabular}{lllllllll}
$0$ & $0$ & $0$ & $0$ & $-i\Gamma $ & $0$ & $0$ & $0$ & $0$ \\ 
$0$ & $0$ & $0$ & $0$ & $0$ & $-i\sqrt{2}\Gamma $ & $0$ & $0$ & $0$ \\ 
$0$ & $0$ & $0$ & $0$ & $0$ & $0$ & $0$ & $-i\sqrt{2}\Gamma $ & $0$ \\ 
$0$ & $0$ & $0$ & $0$ & $0$ & $0$ & $0$ & $0$ & $-i2\,\Gamma $%
\end{tabular}%
\right] 
.
\end{equation}

For the $(2,1)$ set we have%
\begin{equation}
\mathbf{\rho }(2,1)=\left[ 
\begin{tabular}{l}
$\rho _{02;\,01}$ \\ 
$\rho _{02;\,10}$ \\ 
$\rho _{11;\,01}$ \\ 
$\rho _{11;\,10}$ \\ 
$\rho _{20;\,01}$ \\ 
$\rho _{20;\,10}$%
\end{tabular}%
\right] 
\end{equation}%
\begin{eqnarray}
A(2,1) &=&\omega _{c}(1)\left[ 
\begin{smallmatrix}
1 & 0 & 0 & 0 & 0 & 0 \\ 
0 & 1 & 0 & 0 & 0 & 0 \\ 
0 & 0 & 1 & 0 & 0 & 0 \\ 
0 & 0 & 0 & 1 & 0 & 0 \\ 
0 & 0 & 0 & 0 & 1 & 0 \\ 
0 & 0 & 0 & 0 & 0 & 1%
\end{smallmatrix}%
\right]  
 \notag \\ && %
+\left[ 
\begin{smallmatrix}
-\delta +\overline{\delta } & -g_{1} & g_{2} & 0 & 0 & 0 \\ 
-g_{1} & -2\delta -\frac{1}{2}i\Gamma  & 0 & g_{2} & 0 & 0 \\ 
g_{2} & 0 & -\frac{1}{2}i\Gamma  & -g_{1} & \sqrt{2}g_{1} & 0 \\ 
0 & g_{2} & -g_{1} & -\delta +\overline{\delta }-i\Gamma  & 0 & \sqrt{2}g_{1}
\\ 
0 & 0 & \sqrt{2}g_{1} & 0 & +\delta -\overline{\delta }-i\Gamma  & -g_{1} \\ 
0 & 0 & 0 & \sqrt{2}g_{1} & -g_{1} & -\frac{3}{2}i\Gamma 
\end{smallmatrix}%
\right]   \notag \\
&&
\end{eqnarray}%
\begin{eqnarray}
B(2,1) &=&\left[ 
\begin{smallmatrix}
0 & -i\Gamma  & 0 & 0 & 0 & 0 & 0 & 0 & 0 \\ 
0 & 0 & -i\sqrt{2}\Gamma  & 0 & 0 & 0 & 0 & 0 & 0 \\ 
0 & 0 & 0 & 0 & -i\sqrt{2}\Gamma  & 0 & 0 & 0 & 0 \\ 
0 & 0 & 0 & 0 & 0 & -i2\,\Gamma  & 0 & 0 & 0 \\ 
0 & 0 & 0 & 0 & 0 & 0 & 0 & -i\sqrt{3}\Gamma  & 0 \\ 
0 & 0 & 0 & 0 & 0 & 0 & 0 & 0 & -i\sqrt{6}\Gamma 
\end{smallmatrix}%
\right]   
.
\notag \\ &&
\end{eqnarray}

For the $(2,2)$ set we have%
\begin{equation}
\mathbf{\rho }(2,2)=%
\begin{bmatrix}
\rho _{02;\,02} \\ 
\rho _{02;\,11} \\ 
\rho _{02;\,20} \\ 
\rho _{11;\,02} \\ 
\rho _{11;\,11} \\ 
\rho _{11;\,20} \\ 
\rho _{20;\,02} \\ 
\rho _{20;\,11} \\ 
\rho _{20;\,20}%
\end{bmatrix}%
\end{equation}%
\begin{widetext}
\begin{eqnarray}
A(2,2) &=&\omega _{c}(0)\left[ 
\begin{smallmatrix}
1 & 0 & 0 & 0 & 0 & 0 & 0 & 0 & 0 \\ 
0 & 1 & 0 & 0 & 0 & 0 & 0 & 0 & 0 \\ 
0 & 0 & 1 & 0 & 0 & 0 & 0 & 0 & 0 \\ 
0 & 0 & 0 & 1 & 0 & 0 & 0 & 0 & 0 \\ 
0 & 0 & 0 & 0 & 1 & 0 & 0 & 0 & 0 \\ 
0 & 0 & 0 & 0 & 0 & 1 & 0 & 0 & 0 \\ 
0 & 0 & 0 & 0 & 0 & 0 & 1 & 0 & 0 \\ 
0 & 0 & 0 & 0 & 0 & 0 & 0 & 1 & 0 \\ 
0 & 0 & 0 & 0 & 0 & 0 & 0 & 0 & 1%
\end{smallmatrix}%
\right]   \notag \\
&&+\left[ 
\begin{smallmatrix}
0 & -g_{2}\, & 0 & g_{2}\, & 0 & 0 & 0 & 0 & 0 \\ 
-g_{2}\, & -(\delta +\overline{\delta })-\frac{1}{2}i\Gamma  & -g_{1}\,\sqrt{%
2} & 0 & g_{2}\, & 0 & 0 & 0 & 0 \\ 
0 & -g_{1}\,\sqrt{2} & -2\delta -i\Gamma  & 0 & 0 & g_{2}\, & 0 & 0 & 0 \\ 
g_{2}\, & 0 & 0 & +(\delta +\overline{\delta })-\frac{1}{2}i\Gamma  & 
-g_{2}\, & 0 & g_{1}\,\sqrt{2} & 0 & 0 \\ 
0 & g_{2}\, & 0 & -g_{2}\, & -i\Gamma  & -g_{1}\,\sqrt{2} & 0 & g_{1}\,\sqrt{%
2} & 0 \\ 
0 & 0 & g_{2}\, & 0 & -g_{1}\,\sqrt{2} & (-\delta +\overline{\delta })-\frac{%
3}{2}i\Gamma  & 0 & 0 & g_{1}\,\sqrt{2} \\ 
0 & 0 & 0 & g_{1}\,\sqrt{2} & 0 & 0 & +2\delta -i\Gamma  & -g_{2}\, & 0 \\ 
0 & 0 & 0 & 0 & g_{1}\,\sqrt{2} & 0 & -g_{2}\, & -(-\delta +\overline{\delta 
})-\frac{3}{2}i\Gamma  & -g_{1}\,\sqrt{2} \\ 
0 & 0 & 0 & 0 & 0 & g_{1}\,\sqrt{2} & 0 & -g_{1}\,\sqrt{2} & -i2\,\Gamma 
\end{smallmatrix}%
\right]   \notag \\
&&
\end{eqnarray}%
\begin{eqnarray}
B(2,2) &=&\left[ 
\begin{smallmatrix}
-i\Gamma  & 0 & 0 & 0 & 0 & 0 & 0 & 0 & 0 \\ 
0 & -i\Gamma \sqrt{2} & 0 & 0 & 0 & 0 & 0 & 0 & 0 \\ 
0 & 0 & -i\Gamma \sqrt{3} & 0 & 0 & 0 & 0 & 0 & 0 \\ 
0 & 0 & 0 & -i\Gamma \sqrt{2} & 0 & 0 & 0 & 0 & 0 \\ 
0 & 0 & 0 & 0 & -i\Gamma \,2 & 0 & 0 & 0 & 0 \\ 
0 & 0 & 0 & 0 & 0 & -i\Gamma \sqrt{6} & 0 & 0 & 0 \\ 
0 & 0 & 0 & 0 & 0 & 0 & -i\Gamma \sqrt{3} & 0 & 0 \\ 
0 & 0 & 0 & 0 & 0 & 0 & 0 & -i\Gamma \sqrt{6} & 0 \\ 
0 & 0 & 0 & 0 & 0 & 0 & 0 & 0 & -i\Gamma \,3%
\end{smallmatrix}%
\right]   
.
\notag \\ &&
\end{eqnarray}
\end{widetext}

\section{Spectrum for Case A}

\label{Appendix 2}

For the cascade case A situation we take%
\begin{eqnarray}
\widehat{V}_{-} &=&\mu ^{\ast }\,\widehat{a}  \notag \\
&=&\mu ^{\ast }\sum\limits_{n\nu }\sqrt{n}\,\left\vert n-1;\nu \right\rangle
\left\langle n;\nu \right\vert \\
\widehat{V}_{+} &=&\mu \,\widehat{a}^{\dag }  \notag \\
&=&\mu \sum\limits_{m\lambda }\sqrt{m}\,\left\vert m;\lambda \right\rangle
\left\langle m-1;\lambda \right\vert .
\end{eqnarray}%
The two time correlation function is then%
\begin{eqnarray}
&&Tr_{SR}\,\left( \widehat{\rho }_{I}\,\widehat{V}_{+}(t_{2})\,\,\widehat{V}%
_{-}(t_{1})\right) 
\nonumber\\ &&
=|\mu |^{2}\sum\limits_{m\lambda }\sum\limits_{n\nu }\sqrt{m}\sqrt{n}%
\,\left\langle \widehat{S}_{m\lambda ;\,\overline{m-1}\lambda }(t_{2})%
\widehat{S}_{\overline{n-1}\nu ;\,n\nu }(t_{1})\right\rangle
,
\nonumber\\ &&
\end{eqnarray}
where the transition operators $\widehat{S}_{m\lambda ;\,n\nu }(t)\equiv
\left\vert m;\lambda \right\rangle \left\langle n;\nu \right\vert $ are
Heisenberg operators at time $t$, and the trace is over both system and
reservoir states.\ The two time correlation functions can be evaluated using
the quantum regression theorem \cite{Lax63a67a,Walls94a,Dalton79a}, the
results for which is given in terms of the evolution operator matrix
elements $U_{m\lambda ;\,n\nu ::l\beta ;\,k\alpha }(\tau )$ associated with
the density matrix equations and considered as a function of the time
difference $\tau =|t_{1}-t_{2}|\,\geqslant \,0$, and density matrix elements 
$\rho _{m\lambda ;\,n\nu }(t_{1,2})$ considered as a function of the smaller
of the two times $t_{1}$ and $t_{2}$.

\begin{widetext}
We find using the result (\ref{Eq.QRThm1}) for $t_{2}=t_{1}+\tau \geqslant
t_{1}$%
\begin{equation}
Tr_{SR}\,\left( \widehat{\rho }_{I}\,\widehat{V}_{+}(t_{2})\,\,\widehat{V}%
_{-}(t_{1})\right) 
=|\mu |^{2}\sum\limits_{m\lambda }\sum\limits_{n\nu }\sqrt{m}\sqrt{n}\, 
\sum\limits_{k\alpha }U_{\overline{m-1}\lambda ;\,m\lambda ::\overline{n-1}%
\nu ;\,k\alpha }(\tau )\,\rho _{n\nu ;\,k\alpha }(t_{1})
,
\end{equation}%
and using (\ref{Eq.QRThm2}) for $t_{1}=t_{2}+\tau \geqslant t_{2}$ 
\begin{eqnarray}
Tr_{SR}\,\left( \widehat{\rho }_{i}\,\widehat{V}_{+}(t_{2})\,\,\widehat{V}%
_{-}(t_{1})\right) 
&=&|\mu |^{2}\sum\limits_{m\lambda }\sum\limits_{n\nu }\sqrt{m}\sqrt{n}\, 
\sum\limits_{k\alpha }U_{n\nu ;\,\overline{n-1}\nu ::k\alpha ;\,\overline{%
m-1}\lambda }(\tau )\,\rho _{k\alpha ;\,m\lambda }(t_{2})  \notag \\
&=&|\mu |^{2}\sum\limits_{m\lambda }\sum\limits_{n\nu }\sqrt{m}\sqrt{n}\, 
\sum\limits_{k\alpha }U_{m\lambda ;\,\overline{m-1}\lambda ::k\alpha ;\,%
\overline{n-1}\nu }(\tau )\,\rho _{k\alpha ;\,n\nu }(t_{2})
,
\end{eqnarray}%
where to obtain the last result we have made the interchange $m\lambda
\leftrightarrow n\nu $.

Substituting into the result for the spectrum we have%
\begin{eqnarray}
S(\omega ) &=&\frac{1}{\hbar ^{2}}\left[ \iint_{0}^{\infty }dt_{1}dt_{2}\exp
i\omega (t_{1}-t_{2})Tr_{S}\,\left( \widehat{\rho }_{i}\,\widehat{V}%
_{+}(t_{2})\,\,\widehat{V}_{-}(t_{1})\right) \right] _{t_{1}\geqslant
\,t_{2}}  \notag \\
&&+\frac{1}{\hbar ^{2}}\left[ \iint_{0}^{\infty }dt_{1}dt_{2}\exp i\omega
(t_{1}-t_{2})Tr_{S}\,\left( \widehat{\rho }_{i}\,\widehat{V}_{+}(t_{2})\,\,%
\widehat{V}_{-}(t_{1})\right) \right] _{_{t_{2}\geqslant \,t_{1}}}  \notag \\
S(\omega ) &=&\frac{|\mu |^{2}}{\hbar ^{2}}\sum\limits_{m\lambda
}\sum\limits_{n\nu }\sqrt{m}\sqrt{n}\,\sum\limits_{k\alpha }  
\int_{0}^{\infty }dt_{2}\int_{t_{2}}^{\infty }dt_{1}\exp i\omega
(t_{1}-t_{2})\,U_{m\lambda ;\,\overline{m-1}\lambda ::k\alpha ;\,\overline{%
n-1}\nu }(\tau )\,\rho _{k\alpha ;\,n\nu }(t_{2})  \notag \\
&&+\frac{|\mu |^{2}}{\hbar ^{2}}\sum\limits_{m\lambda }\sum\limits_{n\nu }%
\sqrt{m}\sqrt{n}\,\sum\limits_{k\alpha }  
\int_{0}^{\infty }dt_{1}\int_{t_{1}}^{\infty }dt_{2}\exp i\omega
(t_{1}-t_{2})\,U_{\overline{m-1}\lambda ;\,m\lambda ::\overline{n-1}\nu
;\,k\alpha }(\tau )\,\rho _{n\nu ;\,k\alpha }(t_{1})  \notag \\
S(\omega ) &=&\frac{|\mu |^{2}}{\hbar ^{2}}\sum\limits_{m\lambda
}\sum\limits_{n\nu }\sqrt{m}\sqrt{n}\,\sum\limits_{k\alpha }  
\int_{0}^{\infty }dt_{2}\int_{0}^{\infty }d\tau \,\exp (i\omega
\tau )\,U_{m\lambda ;\,\overline{m-1}\lambda ::k\alpha ;\,\overline{n-1}\nu
}(\tau )\,\rho _{k\alpha ;\,n\nu }(t_{2})  \notag \\
&&+\frac{|\mu |^{2}}{\hbar ^{2}}\sum\limits_{m\lambda }\sum\limits_{n\nu }%
\sqrt{m}\sqrt{n}\,\sum\limits_{k\alpha }  
\int_{0}^{\infty }dt_{1}\int_{0}^{\infty }d\tau \,\exp (-i\omega
\tau )\,U_{\overline{m-1}\lambda ;\,m\lambda ::\overline{n-1}\nu ;\,k\alpha
}(\tau )\,\rho _{n\nu ;\,k\alpha }(t_{1})  
.
\notag \\ &&
\end{eqnarray}%
\end{widetext}
To obtain the last expression, the integration variables have been changed
in the first region $(t_{1}\geqslant t_{2})$ to $\tau ,t_{2}$ via the
transformation $t_{1}=\tau +t_{2},t_{2}=t_{2}$ (so the Jacobian equals $+1$)
and in the second region $(t_{2}\geqslant t_{1})$ to $t_{1},\tau $ via the
transformation $t_{1}=t_{1},t_{2}=\tau +t_{1}$ (so the Jacobian equals $+1$).

Since by taking the complex conjugate of the density matrix equations we
have 
\begin{eqnarray}
\rho _{k\alpha ;\,n\nu }(t) &=&\rho _{n\nu ;\,k\alpha }(t)^{\ast } \\
U_{m\lambda ;\,\overline{m-1}\lambda ::k\alpha ;\,\overline{n-1}\nu }(\tau )
&=&U_{\overline{m-1}\lambda ;\,m\lambda ::\overline{n-1}\nu ;\,k\alpha
}(\tau )^{\ast }
\, ,
\nonumber\\&& %
\end{eqnarray}%
it is not difficult to see that the second term in the last equation is just
the complex conjugate of the first, thus proving that our result for the
spectrum is real.

For each contribution the double integrals factorise, each giving a Laplace
transform---albeit for $s$ on the imaginary axis (which may need to be
written as a limiting process). We find that%
\begin{widetext}
\begin{eqnarray}
S(\omega ) &=&\frac{|\mu |^{2}}{\hbar ^{2}}\sum\limits_{m\lambda
}\sum\limits_{n\nu }\sqrt{m}\sqrt{n}\,\sum\limits_{k\alpha }\widetilde{U}%
_{m\lambda ;\,\overline{m-1}\lambda ::k\alpha ;\,\overline{n-1}\nu
}(-i\omega +\epsilon )\,\widetilde{\rho }_{k\alpha ;\,n\nu }(\epsilon
^{\prime })  \notag \\
&&+\frac{|\mu |^{2}}{\hbar ^{2}}\sum\limits_{m\lambda }\sum\limits_{n\nu }%
\sqrt{m}\sqrt{n}\,\sum\limits_{k\alpha }\widetilde{U}_{\overline{m-1}\lambda
;\,m\lambda ::\overline{n-1}\nu ;\,k\alpha }(+i\omega +\epsilon )\,%
\widetilde{\rho }_{n\nu ;\,k\alpha }(\epsilon ^{\prime })  \notag \\
&=&2\,\text{Re}\,\frac{|\mu |^{2}}{\hbar ^{2}}\sum\limits_{m\lambda
}\sum\limits_{n\nu }\sqrt{m}\sqrt{n}\,\sum\limits_{k\alpha }\widetilde{U}%
_{m\lambda ;\,\overline{m-1}\lambda ::k\alpha ;\,\overline{n-1}\nu
}(-i\omega +\epsilon )\,\widetilde{\rho }_{k\alpha ;\,n\nu }(\epsilon
^{\prime }) 
,
\end{eqnarray}%
\end{widetext}
where the limits $\epsilon ,\epsilon ^{\prime }\rightarrow 0$ are
understood. The last result follows from the second term being the complex
conjugate of the first.

The initial conditions will lead to restrictions on the terms to be summed
over. Of course the quantities $\alpha ,\lambda ,\nu $ already only sum over
the three atomic states. As shown previously for the present initial
conditions, the only non zero $\rho _{k\alpha ;\,n\nu }(t)$ are those in the 
$(2,2)$, $(1,1)$ and $(0,0)$ coupled sets---for $(2,2)$ these are $\rho
_{02;\,02}$, $\rho _{02;\,11}$, $\rho _{02;\,20}$, $\rho _{11;\,02}$, $\rho
_{11;\,11}$, $\rho _{11;\,20}$, $\rho _{20;\,02}$, $\rho _{20;\,11}$ and $%
\rho _{20;\,20}$; for the $(1,1)$ set we have $\rho _{01;\,01},\rho
_{01;\,10},\rho _{10;\,01}$ and $\rho _{10;\,10}$ and for the $(0,0)$ set we
have $\rho _{00;\,00}(t)$---corresponding to the system states $\left\vert
0;2\right\rangle ,\left\vert 1;1\right\rangle $ and $\left\vert
0;1\right\rangle $ being the only ones populated during the decay process.
This will restrict the sums over $k,n$, and the $\widetilde{\rho }_{k\alpha
;\,n\nu }(\epsilon ^{\prime })$ that are involved are all Laplace transforms
of members of the $(2,2)$, $(1,1)$ and $(0,0)$ sets for the $\mathbf{\rho }%
(n,m,t)$. In fact given that we also require $n\geqslant 1$ due to the $%
\sqrt{n}$, factor, the only $\widetilde{\rho }_{k\alpha ;\,n\nu }(\epsilon
^{\prime })$ that are possible are $\widetilde{\rho }_{02;\,11}$, $%
\widetilde{\rho }_{02;\,20}$, $\widetilde{\rho }_{11;\,11}$, $\widetilde{%
\rho }_{11;\,20}$, $\widetilde{\rho }_{20;\,11}$, $\widetilde{\rho }%
_{20;\,20}$, which are all in the $(2,2)$ coupled set and $\widetilde{\rho }%
_{01;\,10}$, $\widetilde{\rho }_{10;\,10}$ which are in the $(1,1)$ coupled
set. This then restricts the $\widetilde{U}_{m\lambda ;\,\overline{m-1}%
\lambda ::k\alpha ;\,\overline{n-1}\nu }(-i\omega +\epsilon )$ to be of the
form $\widetilde{U}_{m\lambda ;\,\overline{m-1}\lambda ::02;\,01}(-i\omega
+\epsilon ),$ $\widetilde{U}_{m\lambda ;\,\overline{m-1}\lambda
::02;\,10}(-i\omega +\epsilon ),$ $\widetilde{U}_{m\lambda ;\,\overline{m-1}%
\lambda ::11;\,01}(-i\omega +\epsilon ),$ $\widetilde{U}_{m\lambda ;\,%
\overline{m-1}\lambda ::11;\,10}(-i\omega +\epsilon ),$ $\widetilde{U}%
_{m\lambda ;\,\overline{m-1}\lambda ::20;\,01}(-i\omega +\epsilon ),$ $%
\widetilde{U}_{m\lambda ;\,\overline{m-1}\lambda ::20;\,10}(-i\omega
+\epsilon )$ and $\widetilde{U}_{m\lambda ;\,\overline{m-1}\lambda
::01;\,00}(-i\omega +\epsilon ),$ $\widetilde{U}_{m\lambda ;\,\overline{m-1}%
\lambda ::10;\,00}(-i\omega +\epsilon )$ because of the $k,n$ restriction
(note the $\overline{n-1}\nu $ in the subscript). Since $\rho _{02;\,01},$ $%
\rho _{02;\,10},$ $\rho _{11;\,01},$ $\rho _{11;\,10},$ $\rho _{20;\,01},$ $%
\rho _{20;\,10}$ are all in the $(2,1)$ set these are Laplace transforms of
the $U_{n,m::2,1}(t)$. The other terms $\rho _{01;\,00},$ $\rho _{10;\,00}$
are both in the $(1,0)$ set these are Laplace transforms of the $%
U_{n,m::1,0}(t)$. The spectrum can be broken up into two contributions, $%
S_{6}(\omega )$\ which is associated with $\widetilde{\rho }_{k\alpha
;\,n\nu }(\epsilon ^{\prime })$\ for the $(2,2)$\ coupled set and $%
S_{2}(\omega )$\ which is associated with $\widetilde{\rho }_{k\alpha
;\,n\nu }(\epsilon ^{\prime })$\ for the $(1,1)$\ coupled set. This gives 
\begin{widetext}
\begin{eqnarray}
S(\omega ) &=&2\,\text{Re}\,\frac{|\mu |^{2}}{\hbar ^{2}}\sum\limits_{m%
\geqslant 1,\lambda }\sum\limits_{n\geqslant 1,\nu }\sqrt{m}\sqrt{n}%
\,\sum\limits_{k\alpha }\widetilde{U}_{m\lambda ;\,\overline{m-1}\lambda
::k\alpha ;\,\overline{n-1}\nu }(-i\omega +\epsilon )\,\widetilde{\rho }%
_{k\alpha ;\,n\nu }(\epsilon ^{\prime })  \notag \\
&=&S_{6}(\omega )+S_{2}(\omega ) \\
S_{6}(\omega ) &=&2\,\text{Re}\,\frac{|\mu |^{2}}{\hbar ^{2}}%
\sum\limits_{m\geqslant 1,\lambda }\sqrt{m}\sqrt{1}\widetilde{U}_{m\lambda
;\,\overline{m-1}\lambda ::02;\,01}(-i\omega +\epsilon )\,\widetilde{\rho }%
_{02;\,11}(\epsilon ^{\prime })  \notag \\
&&+2\,\text{Re}\,\frac{|\mu |^{2}}{\hbar ^{2}}\sum\limits_{m\geqslant
1,\lambda }\sqrt{m}\sqrt{2}\widetilde{U}_{m\lambda ;\,\overline{m-1}\lambda
::02;\,10}(-i\omega +\epsilon )\,\widetilde{\rho }_{02;\,20}(\epsilon
^{\prime })  \notag \\
&&+2\,\text{Re}\,\frac{|\mu |^{2}}{\hbar ^{2}}\sum\limits_{m\geqslant
1,\lambda }\sqrt{m}\sqrt{1}\widetilde{U}_{m\lambda ;\,\overline{m-1}\lambda
::11;\,01}(-i\omega +\epsilon )\,\widetilde{\rho }_{11;\,11}(\epsilon
^{\prime })  \notag \\
&&+2\,\text{Re}\,\frac{|\mu |^{2}}{\hbar ^{2}}\sum\limits_{m\geqslant
1,\lambda }\sqrt{m}\sqrt{2}\widetilde{U}_{m\lambda ;\,\overline{m-1}\lambda
::11;\,10}(-i\omega +\epsilon )\,\widetilde{\rho }_{11;\,20}(\epsilon
^{\prime })  \notag \\
&&+2\,\text{Re}\,\frac{|\mu |^{2}}{\hbar ^{2}}\sum\limits_{m\geqslant
1,\lambda }\sqrt{m}\sqrt{1}\widetilde{U}_{m\lambda ;\,\overline{m-1}\lambda
::20;\,01}(-i\omega +\epsilon )\,\widetilde{\rho }_{20;\,11}(\epsilon
^{\prime })  \notag \\
&&+2\,\text{Re}\,\frac{|\mu |^{2}}{\hbar ^{2}}\sum\limits_{m\geqslant
1,\lambda }\sqrt{m}\sqrt{2}\widetilde{U}_{m\lambda ;\,\overline{m-1}\lambda
::20;\,10}(-i\omega +\epsilon )\,\widetilde{\rho }_{20;\,20}(\epsilon
^{\prime })  \notag \\
S_{2}(\omega ) &=&2\,\text{Re}\,\frac{|\mu |^{2}}{\hbar ^{2}}%
\sum\limits_{m\geqslant 1,\lambda }\sqrt{m}\sqrt{1}\,\widetilde{U}_{m\lambda
;\,\overline{m-1}\lambda ::01;\,00}(-i\omega +\epsilon )\,\widetilde{\rho }%
_{01;\,10}(\epsilon ^{\prime })  \notag \\
&&+2\,\text{Re}\,\frac{|\mu |^{2}}{\hbar ^{2}}\sum\limits_{m\geqslant
1,\lambda }\sqrt{m}\sqrt{1}\,\widetilde{U}_{m\lambda ;\,\overline{m-1}%
\lambda ::10;\,00}(-i\omega +\epsilon )\,\widetilde{\rho }%
_{10;\,10}(\epsilon ^{\prime })
.
\end{eqnarray}%
\end{widetext}
The six $S_{6}(\omega )$ terms involve $\widetilde{U}_{n,m::2,1}(-i\omega
+\epsilon )$ and the two $S_{2}(\omega )$ terms involve $\widetilde{U}%
_{n,m::1,0}(-i\omega +\epsilon )$. Then because the possible $%
U_{n,m::2,1}(t) $ and $U_{n,m::1,0}(t)$ must be such that $n=m+1$, we see
that the non-zero $\widetilde{U}_{m\lambda ;\overline{m-1}\lambda
::k\alpha ;\overline{n-1}\nu }(-i\omega +\epsilon )$ for the first six $%
S_{6}(\omega )$ terms $\widetilde{U}_{n,m::2,1}(-i\omega +\epsilon )$ must
all be Laplace transforms from the sets of matrices 
$U_{1,0::2,1}(t), U_{2,1::2,1}(t), U_{3,2::2,1}(t), 
U_{4,3::2,1}(t), \hdots$ and
the the non-zero $\widetilde{U}_{m\lambda ;\overline{m-1}\lambda ::k\alpha
;\overline{n-1}\nu }(-i\omega +\epsilon )$ for the last two $S_{2}(\omega
) $ terms $\widetilde{U}_{n,m::1,0}(-i\omega +\epsilon )$ must all be
Laplace transforms from the sets of matrices $%
U_{1,0::1,0}(t), U_{2,1::1,0}(t), U_{3,2::1,0}(t), U_{4,3::1,0}(t), \hdots $

However, for the first six $S_{6}(\omega )$ terms we see from the Laplace
transforms 
\begin{eqnarray}
(s+iA(1,0))\,\widetilde{U}(1,0::2,1)-iB(1,0)\,\widetilde{U}(2,1::2,1)&=&0 
\notag \\
(s+iA(2,1))\,\widetilde{U}(2,1::2,1)-iB(2,1)\,\widetilde{U}%
(3,2::2,1)&=&E_{6}  \notag \\
(s+iA(3,2))\,\widetilde{U}(3,2::2,1)-iB(3,2)\,\widetilde{U}(4,3::2,1)&=&0 
\notag \\
(s+iA(4,3))\,\widetilde{U}(4,3::2,1)-iB(4,3)\,\widetilde{U}(5,4::2,1)&=&0 
\notag \\
&& \hdots
\end{eqnarray}%
and it is easy to see that the solution to these equations is 
\begin{eqnarray}
\widetilde{U}(2,1::2,1;s)&=&(s+iA(2,1))^{-1}  \notag \\
\widetilde{U}(1,0::2,1;s)&=&(s+iA(1,0))^{-1}
\nonumber\\ && \times
iB(1,0)\,(s+iA(2,1))^{-1} 
\notag \\
\,\widetilde{U}(n+1,n::2,1;s)&=&0\qquad (n>1).
\end{eqnarray}%
This solution corresponds to the previous result that an initial density
matrix with only non-zero elements in the $(N,M)$ set or below cannot evolve
into a density matrix with non-zero elements in sets such as $(N+1,M+1)$, $%
(N+2,M+2), \hdots $ , due to the irreversible nature of the relaxation processes.
This means that for the first six $S_{6}(\omega )$ terms the $\widetilde{U}%
_{m\lambda ;\,\overline{m-1}\lambda ::k\alpha ;\,\overline{n-1}\nu
}(-i\omega +\epsilon )$ must be from the $U_{2,1::2,1}(t)$ and $%
U_{1,0::2,1}(t)$ matrix elements only, and this places a restriction on $m$
and $\lambda $. Since the only density matrix element in the $(1,0)$ set of
the form $\rho _{m\lambda ;\,\overline{m-1}\lambda }$ is $\rho _{10;\,00}$
then in fact the only terms are for $m=1$ and $\lambda =0$ (namely $%
\widetilde{U}_{10,00::02;\,01}(-i\omega +\epsilon )$, $\widetilde{U}%
_{10,00::02;\,10}(-i\omega +\epsilon )$, $\widetilde{U}_{10,00::11;\,01}(-i%
\omega +\epsilon )$, $\widetilde{U}_{10,00::11;\,10}(-i\omega +\epsilon )$, $%
\widetilde{U}_{10,00::20;\,01}(-i\omega +\epsilon )$ and $\widetilde{U}%
_{10,00::20,10}(-i\omega +\epsilon )$). Since the only density matrix
elements in the $(2,1)$ set of the form $\rho _{m\lambda ;\,\overline{m-1}%
\lambda }$ are $\rho _{11;\,01}$ and $\rho _{20;\,10}$ then the only terms
are for $m=1$ and $\lambda =1$ or $m=2$ and $\lambda =0$, (namely $%
\widetilde{U}_{11;\,01::02;\,01}(-i\omega +\epsilon ),$ $\widetilde{U}%
_{11;\,01::02;\,10}(-i\omega +\epsilon )$, $\widetilde{U}_{11;\,01::11;%
\,01}(-i\omega +\epsilon )$, $\widetilde{U}_{11;\,01::11;\,10}(-i\omega
+\epsilon )$, $\widetilde{U}_{11;\,01::20;\,01}(-i\omega +\epsilon )$ and $%
\widetilde{U}_{11;\,01::20,10}(-i\omega +\epsilon )$ for $m=1$ and $\lambda
=1;$ then $\widetilde{U}_{20,10::02;\,01}(-i\omega +\epsilon )$, $\widetilde{%
U}_{20,10::02;\,10}(-i\omega +\epsilon )$, $\widetilde{U}_{20,10::11;%
\,01}(-i\omega +\epsilon )$, $\widetilde{U}_{20,10::11;\,10}(-i\omega
+\epsilon )$, $\widetilde{U}_{20,10::20;\,01}(-i\omega +\epsilon )$ and $%
\widetilde{U}_{20,10::20,10}(-i\omega +\epsilon )$ for $m=2$ and $\lambda =0$
). Thus each of the first six $S_{6}(\omega )$ terms produces three
contributions.

Also, for the last two $S_{2}(\omega )$ terms we see from the Laplace
transform equations%
\begin{eqnarray}
(s+iA(1,0))\,\widetilde{U}(1,0 &:&:1,0)-iB(1,0)\,\widetilde{U}%
(2,1::1,0)=E_{2}  \notag \\
(s+iA(2,1))\,\widetilde{U}(2,1 &:&:1,0)-iB(2,1)\,\widetilde{U}(3,2::1,0)=0 
\notag \\
(s+iA(3,2))\,\widetilde{U}(3,2 &:&:1,0)-iB(3,2)\,\widetilde{U}(4,3::1,0)=0 
\notag \\
&& \hdots
\end{eqnarray}%
and it is easy to see that the solution to these equations is 
\begin{eqnarray}
\widetilde{U}(1,0 &:&:1,0;s)=(s+iA(1,0))^{-1}  \notag \\
\,\widetilde{U}(n+1,n &:&:1,0;s)=0\qquad (n>1)
,
\end{eqnarray}%
from the same considerations as for the first six $S_{6}(\omega )$ terms.
This means that for the last two $S_{2}(\omega )$ terms the $\widetilde{U}%
_{m\lambda ;\,\overline{m-1}\lambda ::k\alpha ;\,\overline{n-1}\nu
}(-i\omega +\epsilon )$ must be from the $U_{1,0::1,0}(t)$ only, and this
places a restriction on $m$ and $\lambda $. Since the only density matrix
element in the $(1,0)$ set of the form $\rho _{m\lambda ;\,\overline{m-1}%
\lambda }$ is $\rho _{10;\,00}$ then in fact the only terms are for $m=1$
and $\lambda =0$ (namely $\widetilde{U}_{10;\,00::01,00}(-i\omega +\epsilon
) $ and $\widetilde{U}_{10;\,00::10,00}(-i\omega +\epsilon )$). Each of the
two $S_{2}(\omega )$ terms produces one contribution only.

We finally have for the spectrum%
\begin{eqnarray}
S(\omega ) &=&S_{2}(\omega )+S_{6}(\omega ) 
\end{eqnarray}
\begin{widetext}
\begin{eqnarray}
S_{2}(\omega ) &=&2\,\text{Re}\,\frac{|\mu |^{2}}{\hbar ^{2}}\sqrt{1}\sqrt{1}
\{\widetilde{\rho }_{10;\,10}(\epsilon ^{\prime })\widetilde{U}%
_{10;\,00::01;\,00}(-i\omega +\epsilon )\,+ \widetilde{\rho }%
_{01;\,10}(\epsilon ^{\prime }) \widetilde{U}_{10;\,00::10;\,00}(-i\omega
+\epsilon )\}  \notag \\
S_{6}(\omega ) &=&2\,\text{Re}\,\frac{|\mu |^{2}}{\hbar ^{2}}\,  \notag \\
&&\times \lbrack \sqrt{1}\,\widetilde{\rho }_{02;\,11}(\epsilon ^{\prime })\{%
\sqrt{1}\widetilde{U}_{10,00::02;\,01}(-i\omega +\epsilon )+\sqrt{1}%
\widetilde{U}_{11,01::02;\,01}(-i\omega +\epsilon )  \notag \\
&&+\sqrt{2}\widetilde{U}_{20,10::02;\,01}(-i\omega +\epsilon )\}  \notag \\
&&+\sqrt{2}\,\widetilde{\rho }_{02;\,20}(\epsilon ^{\prime })\{\sqrt{1}%
\widetilde{U}_{10,00::02;\,10}(-i\omega +\epsilon )+\sqrt{1}\widetilde{U}%
_{11,01::02;\,10}(-i\omega +\epsilon )  \notag \\
&&+\sqrt{2}\widetilde{U}_{20,10::02;\,10}(-i\omega +\epsilon )\}  \notag \\
&&+\sqrt{1}\,\widetilde{\rho }_{11;\,11}(\epsilon ^{\prime })\{\sqrt{1}%
\widetilde{U}_{10,00::11;\,01}(-i\omega +\epsilon )+\sqrt{1}\widetilde{U}%
_{11,01::11;\,01}(-i\omega +\epsilon )  \notag \\
&&+\sqrt{2}\widetilde{U}_{20,10::11,01}(-i\omega +\epsilon )\}  \notag \\
&&+\sqrt{2}\,\widetilde{\rho }_{11;\,20}(\epsilon ^{\prime })\{\sqrt{1}%
\widetilde{U}_{10,00::11,10}(-i\omega +\epsilon )+\sqrt{1}\widetilde{U}%
_{11,01::11,10}(-i\omega +\epsilon )  \notag \\
&&+\sqrt{2}\widetilde{U}_{20,10::11;\,10}(-i\omega +\epsilon )\}  \notag \\
&&+\sqrt{1}\,\widetilde{\rho }_{20;\,11}(\epsilon ^{\prime })\{\sqrt{1}%
\widetilde{U}_{10,00::20;\,01}(-i\omega +\epsilon )+\sqrt{1}\widetilde{U}%
_{11,01::20;\,01}(-i\omega +\epsilon )  \notag \\
&&+\sqrt{2}\widetilde{U}_{20,10::20;\,01}(-i\omega +\epsilon )\}  \notag \\
&&+\sqrt{2}\,\widetilde{\rho }_{20;\,20}(\epsilon ^{\prime })\{\sqrt{1}%
\widetilde{U}_{10,00::20,10}(-i\omega +\epsilon )+\sqrt{1}\widetilde{U}%
_{11,01::20,10}(-i\omega +\epsilon )  \notag \\
&&+\sqrt{2}\widetilde{U}_{20,10::20,10}(-i\omega +\epsilon )\}] 
.
\end{eqnarray}
\end{widetext}

\section{Spectrum for Case B}

\label{Appendix 3}

For the cascade case B we have 
\begin{equation}
\widehat{V}_{-}=\int d\Delta \,\rho _{C}(\Delta )\,\mu ^{\ast }(\Delta )%
\widehat{b}(\Delta )
,
\end{equation}%
where $\mu ^{\ast }(\Delta )$ is also a slowly varying function of $\Delta $.

The Heisenberg equations of motion for the Heisenberg operators $\widehat{b}%
(\Delta ,t)$ and $\widehat{a}(t)$ are%
\begin{eqnarray}
\frac{d}{dt}\widehat{b}(\Delta ,t) &=&-i\Delta \widehat{b}(\Delta
,t)-iW^{\ast }(\Delta )\widehat{a}(t) \\
\frac{d}{dt}\widehat{a}(t) &=&-i\omega _{c}\widehat{a}(t)-i\frac{1}{2}g\,%
\widehat{\sigma }_{-}(t)
\nonumber\\&& %
-i\int d\Delta \,\rho _{C}(\Delta )\,W(\Delta )%
\widehat{b}(\Delta ,t)
,
\end{eqnarray}%
and we easily obtain the formal solution of the first equation as%
\begin{eqnarray}
\widehat{b}(\Delta ,t)&=&\exp (-i\Delta t)\,\widehat{b}(\Delta ,0)
\nonumber\\&& %
-iW^{\ast
}(\Delta )\int_{0}^{t}d\tau \,\exp (-i\Delta \tau )\,\widehat{a}(t-\tau ).
\nonumber\\&& %
\end{eqnarray}%
If there were no coupling terms, the free evolution for $\widehat{a}(t)$
follows from the second equation to be $\widehat{a}(t)=\widehat{a}(0)\exp
(-i\omega _{c}t)$.

Hence we see that $\widehat{V}_{-}$ is the sum of a free field term $%
\widehat{V}_{-}^{F}$ and a cavity term $\widehat{V}_{-}^{C}$, where%
\begin{widetext}
\begin{eqnarray}
\widehat{V}_{-} &=&\widehat{V}_{-}^{F}+\widehat{V}_{-}^{C} \\
\widehat{V}_{-}^{F} &=&\int d\Delta \,\rho _{C}(\Delta )\,\mu ^{\ast
}(\Delta )\,\exp (-i\Delta t)\,\widehat{b}(\Delta ) \\
\widehat{V}_{-}^{C} &=&\int_{0}^{t}d\tau \,\left( -i\int d\Delta \,\rho
_{C}(\Delta )\,\mu ^{\ast }(\Delta )\,W^{\ast }(\Delta )\,\exp (-i\Delta
\tau )\right) \widehat{a}(t-\tau )
.
\end{eqnarray}
\end{widetext}

For the cavity contribution $\widehat{V}_{-}^{C}$ the quantity
$F(\tau )=\left( -i\int d\Delta \,\rho _{C}(\Delta )\,\mu ^{\ast }(\Delta
)\,W^{\ast }(\Delta )\,\exp (-i\Delta \tau )\right) $ involves slowly
varying factors $\rho _{C}(\Delta ),\,\mu ^{\ast }(\Delta )$ and$\,W^{\ast
}(\Delta )$, so it is not difficult to see that if the frequency width of
the product of these factors is $\Delta _{C}$ then the overall quantity $%
F(\tau )$ decreases to zero over a time scale $\tau _{C} \sim 1/\Delta
_{C}$. We can then invoke the Markoff approximation on the basis that $\tau
_{C}$ is small compared to evolution times for the system and approximate $%
\widehat{a}(t-\tau )$ over the small time-scale $\tau _{C}$ over which the
quantity $F(\tau )$ is non-zero by its free evolution expression $\widehat{a}%
(t-\tau )\doteqdot \widehat{a}(t)\exp (+i\omega _{c}\tau )$. Extending the $%
\tau $ integral to $\infty $ and introducing the usual integrating factor $%
\exp (-\epsilon \tau )$ gives the following Markovian result for $\widehat{V}%
_{-}^{C}$%
\begin{equation}
\widehat{V}_{-}^{C}=M^{\ast }\widehat{a}
,
\end{equation}%
where 
\begin{equation}
M^{\ast }=\int d\Delta \,\frac{\rho _{C}(\Delta )\,\mu ^{\ast }(\Delta
)\,W^{\ast }(\Delta )}{\omega _{c}-\Delta +i\epsilon }\,
\end{equation}%
is an effective dipole coupling constant. Thus the cavity contribution $%
\widehat{V}_{-}^{C}$ to the spectral quantity $\widehat{V}_{-}$ is the same
as for Case A, apart from a constant of proportionality.

\section{Spectrum for Case C}

\label{Appendix 4}

For the cascade Case C we have 
\begin{eqnarray}
\widehat{V}_{-} &=&R_{2}^{\ast }\,\widehat{\sigma }_{2}^{-}+R_{1}^{\ast }\,%
\widehat{\sigma }_{1}^{-} \\
&=&R_{2}^{\ast }\sum_{n}\,\left\vert n;1\right\rangle \left\langle
n;2\right\vert +R_{1}^{\ast }\sum_{n}\,\left\vert n;0\right\rangle
\left\langle n;1\right\vert \\
\widehat{V}_{+} &=&R_{2}\,\widehat{\sigma }_{2}^{+}+R_{1}\,\widehat{\sigma }%
_{1}^{+} \\
&=&R_{2}\sum_{m}\,\left\vert m;2\right\rangle \left\langle m;1\right\vert
+R_{1}\sum_{m}\,\left\vert m;1\right\rangle \left\langle m;0\right\vert
,
\notag \\ && %
\end{eqnarray}%
where henceforth we will choose $R_{1}$and $R_{2}$ to be real.
Hence the two-time correlation function is 
\begin{eqnarray}
&&Tr_{SR}\,\left( \widehat{\rho }_{I}\,\widehat{V}_{+}(t_{2})\,\,\widehat{V}%
_{-}(t_{1})\right)  \notag \\
&=&R_{2}^{2}\sum\limits_{m}\sum\limits_{n}\,\left\langle \widehat{S}%
_{m2;\,m1}(t_{2})\widehat{S}_{n1;\,n2}(t_{1})\right\rangle
\nonumber\\&& %
+R_{1}^{2}\sum\limits_{m}\sum\limits_{n}\,\left\langle \widehat{S}%
_{m1;\,m0}(t_{2})\widehat{S}_{n0;\,n1}(t_{1})\right\rangle  \notag \\
&&+R_{2}R_{1}\sum\limits_{m}\sum\limits_{n}\,\left\langle \widehat{S}%
_{m2;\,m1}(t_{2})\widehat{S}_{n0;\,n1}(t_{1})\right\rangle
\nonumber\\&& %
+R_{1}R_{2}\sum\limits_{m}\sum\limits_{n}\,\left\langle \widehat{S}%
_{m1;\,m0}(t_{2})\widehat{S}_{n1;\,n2}(t_{1})\right\rangle  
.
\notag \\ &&
\end{eqnarray}
For the first term using the quantum regression theorem result (\ref%
{Eq.QRThm1}) we find that for $t_{2}=t_{1}+\tau \geqslant t_{1}$%
\begin{eqnarray}
&&R_{2}^{2}\sum\limits_{m}\sum\limits_{n}\,\left\langle \widehat{S}%
_{m2;\,m1}(t_{2})\widehat{S}_{n1;\,n2}(t_{1})\right\rangle  \notag \\
&&=R_{2}^{2}\sum\limits_{m}\sum\limits_{n}\,\sum\limits_{k\alpha
}U_{m1;\,m2::n1;\,k\alpha }(\tau )\,\rho _{n2;\,k\alpha }(t_{1})
,
\nonumber\\
\end{eqnarray}%
and using (\ref{Eq.QRThm2}) for $t_{1}=t_{2}+\tau \geqslant t_{2}$
\begin{eqnarray}
&&R_{2}^{2}\sum\limits_{m}\sum\limits_{n}\,\left\langle \widehat{S}%
_{m2;\,m1}(t_{2})\widehat{S}_{n1;\,n2}(t_{1})\right\rangle  \notag \\
&&=R_{2}^{2}\sum\limits_{m}\sum\limits_{n}\sum\limits_{k\alpha
}U_{n2;\,n1::k\alpha ;\,m1}(\tau )\,\rho _{k\alpha ;\,m2}(t_{2})  \notag \\
&&=R_{2}^{2}\sum\limits_{m}\sum\limits_{n}\sum\limits_{k\alpha
}U_{m2;\,m1::k\alpha ;\,n1}(\tau )\,\rho _{k\alpha ;\,n2}(t_{2})
.
\nonumber\\
\end{eqnarray}
For the second term using the quantum regression theorem we find that for $%
t_{2}=t_{1}+\tau \geqslant t_{1}$%
\begin{eqnarray}
&&R_{1}^{2}\sum\limits_{m}\sum\limits_{n}\,\left\langle \widehat{S}%
_{m1;\,m0}(t_{2})\widehat{S}_{n0;\,n1}(t_{1})\right\rangle  \notag \\
&&=R_{1}^{2}\sum\limits_{m}\sum\limits_{n}\,\sum\limits_{k\alpha
}U_{m0;\,m1::n0;\,k\alpha }(\tau )\,\rho _{n1;\,k\alpha }(t_{1})
,
\nonumber\\
\end{eqnarray}%
and for $t_{1}=t_{2}+\tau \geqslant t_{2}$ 
\begin{eqnarray}
&&R_{1}^{2}\sum\limits_{m}\sum\limits_{n}\,\left\langle \widehat{S}%
_{m1;\,m0}(t_{2})\widehat{S}_{n0;\,n1}(t_{1})\right\rangle  \notag \\
&&=R_{1}^{2}\sum\limits_{m}\sum\limits_{n}\sum\limits_{k\alpha
}U_{n1;\,n0::k\alpha ;\,m0}(\tau )\,\rho _{k\alpha ;\,m1}(t_{2})  \notag \\
&&=R_{1}^{2}\sum\limits_{m}\sum\limits_{n}\sum\limits_{k\alpha
}U_{m1;\,m0::k\alpha ;\,n0}(\tau )\,\rho _{k\alpha ;\,n1}(t_{2})
.
\nonumber\\
\end{eqnarray}
For the third term using the quantum regression theorem we find that for $%
t_{2}=t_{1}+\tau \geqslant t_{1}$%
\begin{eqnarray}
&&R_{2}R_{1}\sum\limits_{m}\sum\limits_{n}\,\left\langle \widehat{S}%
_{m2;\,m1}(t_{2})\widehat{S}_{n0;\,n1}(t_{1})\right\rangle  \notag \\
&&=R_{2}R_{1}\sum\limits_{m}\sum\limits_{n}\,\sum\limits_{k\alpha
}U_{m1;\,m2::n0;\,k\alpha }(\tau )\,\rho _{n1;\,k\alpha }(t_{1})
,
\nonumber\\
\end{eqnarray}%
and for $t_{1}=t_{2}+\tau \geqslant t_{2}$
\begin{eqnarray}
&&R_{2}R_{1}\sum\limits_{m}\sum\limits_{n}\,\left\langle \widehat{S}%
_{m2;\,m1}(t_{2})\widehat{S}_{n0;\,n1}(t_{1})\right\rangle  \notag \\
&&=R_{2}R_{1}\sum\limits_{m}\sum\limits_{n}\sum\limits_{k\alpha
}U_{n1;\,n0::k\alpha ;\,m1}(\tau )\,\rho _{k\alpha ;\,m2}(t_{2})  \notag \\
&&=R_{2}R_{1}\sum\limits_{m}\sum\limits_{n}\sum\limits_{k\alpha
}U_{m1;\,m0::k\alpha ;\,n1}(\tau )\,\rho _{k\alpha ;\,n2}(t_{2})
.
\nonumber\\
\end{eqnarray}
For the fourth term using the quantum regression theorem we find that for $%
t_{2}=t_{1}+\tau \geqslant t_{1}$%
\begin{eqnarray}
&&R_{1}R_{2}\sum\limits_{m}\sum\limits_{n}\,\left\langle \widehat{S}%
_{m1;\,m0}(t_{2})\widehat{S}_{n1;\,n2}(t_{1})\right\rangle  \notag \\
&&=R_{1}R_{2}\sum\limits_{m}\sum\limits_{n}\,\sum\limits_{k\alpha
}U_{m0;\,m1::n1;\,k\alpha }(\tau )\,\rho _{n2;\,k\alpha }(t_{1})
,
\nonumber\\
\end{eqnarray}%
and for $t_{1}=t_{2}+\tau \geqslant t_{2}$
\begin{eqnarray}
&&R_{1}R_{2}\sum\limits_{m}\sum\limits_{n}\,\left\langle \widehat{S}%
_{m1;\,m0}(t_{2})\widehat{S}_{n1;\,n2}(t_{1})\right\rangle  \notag \\
&&=R_{1}R_{2}\sum\limits_{m}\sum\limits_{n}\sum\limits_{k\alpha
}U_{n2;\,n1::k\alpha ;\,m0}(\tau )\,\rho _{k\alpha ;\,m1}(t_{2})  \notag \\
&&=R_{1}R_{2}\sum\limits_{m}\sum\limits_{n}\sum\limits_{k\alpha
}U_{m2;\,m1::k\alpha ;\,n0}(\tau )\,\rho _{k\alpha ;\,n1}(t_{2})
.
\nonumber\\
\end{eqnarray}

For the spectrum we have%
\begin{widetext}
\begin{eqnarray*}
S(\omega ) &=&\frac{1}{\hbar ^{2}}\left[ \iint_{0}^{\infty }dt_{1}dt_{2}\exp
i\omega (t_{1}-t_{2})Tr_{S}\,\left( \widehat{\rho }_{i}\,\widehat{V}%
_{+}(t_{2})\,\,\widehat{V}_{-}(t_{1})\right) \right] _{t_{1}\geqslant
\,t_{2}} \\
&&+\frac{1}{\hbar ^{2}}\left[ \iint_{0}^{\infty }dt_{1}dt_{2}\exp i\omega
(t_{1}-t_{2})Tr_{S}\,\left( \widehat{\rho }_{i}\,\widehat{V}_{+}(t_{2})\,\,%
\widehat{V}_{-}(t_{1})\right) \right] _{_{t_{2}\geqslant \,t_{1}}}
,
\end{eqnarray*}%
and substituting gives%
\begin{eqnarray}
S(\omega ) &=&S(\omega )_{t_{1}\geqslant \,t_{2}}+S(\omega
)_{_{t_{2}\geqslant \,t_{1}}}  \notag \\
S(\omega )_{t_{1}\geqslant \,t_{2}} &=&\frac{R_{2}^{2}}{\hbar ^{2}}%
\sum\limits_{m}\sum\limits_{n}\,\sum\limits_{k\alpha }\int_{0}^{\infty
}dt_{2}\int_{t_{2}}^{\infty }dt_{1}\exp i\omega
(t_{1}-t_{2})\,U_{m2;\,m1::k\alpha ;\,n1}(\tau )\,\rho _{k\alpha
;\,n2}(t_{2})  \notag \\
&&+\frac{R_{1}^{2}}{\hbar ^{2}}\sum\limits_{m}\sum\limits_{n}\,\sum%
\limits_{k\alpha }\int_{0}^{\infty }dt_{2}\int_{t_{2}}^{\infty }dt_{1}\exp
i\omega (t_{1}-t_{2})\,U_{m1;\,m0::k\alpha ;\,n0}(\tau )\,\rho _{k\alpha
;\,n1}(t_{2})  \notag \\
&&+\frac{R_{2}R_{1}}{\hbar ^{2}}\sum\limits_{m}\sum\limits_{n}\,\sum%
\limits_{k\alpha }\int_{0}^{\infty }dt_{2}\int_{t_{2}}^{\infty }dt_{1}\exp
i\omega (t_{1}-t_{2})\,U_{m1;\,m0::k\alpha ;\,n1}(\tau )\,\rho _{k\alpha
;\,n2}(t_{2})  \notag \\
&&+\frac{R_{1}R_{2}}{\hbar ^{2}}\sum\limits_{m}\sum\limits_{n}\,\sum%
\limits_{k\alpha }\int_{0}^{\infty }dt_{2}\int_{t_{2}}^{\infty }dt_{1}\exp
i\omega (t_{1}-t_{2})\,U_{m2;\,m1::k\alpha ;\,n0}(\tau )\,\rho _{k\alpha
;\,n1}(t_{2})  \notag \\
&&  \notag \\
S(\omega )_{_{t_{2}\geqslant \,t_{1}}} &=&\frac{R_{2}^{2}}{\hbar ^{2}}%
\sum\limits_{m}\sum\limits_{n}\,\sum\limits_{k\alpha }\int_{0}^{\infty
}dt_{1}\int_{t_{1}}^{\infty }dt_{2}\exp i\omega
(t_{1}-t_{2})\,U_{m1;\,m2::n1;\,k\alpha }(\tau )\,\rho _{n2;\,k\alpha
}(t_{1})  \notag \\
&&+\frac{R_{1}^{2}}{\hbar ^{2}}\sum\limits_{m}\sum\limits_{n}\sum\limits_{k%
\alpha }\int_{0}^{\infty }dt_{1}\int_{t_{1}}^{\infty }dt_{2}\exp i\omega
(t_{1}-t_{2})\,U_{m0;\,m1::n0;\,k\alpha }(\tau )\,\rho _{n1;\,k\alpha
}(t_{1})  \notag \\
&&+\frac{R_{2}R_{1}}{\hbar ^{2}}\sum\limits_{m}\sum\limits_{n}\sum\limits_{k%
\alpha }\int_{0}^{\infty }dt_{1}\int_{t_{1}}^{\infty }dt_{2}\exp i\omega
(t_{1}-t_{2})\,U_{m1;\,m2::n0;\,k\alpha }(\tau )\,\rho _{n1;\,k\alpha
}(t_{1})  \notag \\
&&+\frac{R_{1}R_{2}}{\hbar ^{2}}\sum\limits_{m}\sum\limits_{n}\sum\limits_{k%
\alpha }\int_{0}^{\infty }dt_{1}\int_{t_{1}}^{\infty }dt_{2}\exp i\omega
(t_{1}-t_{2})\,U_{m0;\,m1::n1;\,k\alpha }(\tau )\,\rho _{n2;\,k\alpha
}(t_{1})  
.
\notag \\ &&
\end{eqnarray}%
Thus
\begin{eqnarray}
S(\omega )_{t_{1}\geqslant \,t_{2}} &=&\frac{R_{2}^{2}}{\hbar ^{2}}%
\sum\limits_{m}\sum\limits_{n}\,\sum\limits_{k\alpha }\int_{0}^{\infty
}dt_{2}\int_{0}^{\infty }d\tau \,\exp (i\omega \tau )\,U_{m2;\,m1::k\alpha
;\,n1}(\tau )\,\rho _{k\alpha ;\,n2}(t_{2})  \notag \\
&&+\frac{R_{1}^{2}}{\hbar ^{2}}\sum\limits_{m}\sum\limits_{n}\,\sum%
\limits_{k\alpha }\int_{0}^{\infty }dt_{2}\int_{0}^{\infty }d\tau \,\exp
(i\omega \tau )\,U_{m1;\,m0::k\alpha ;\,n0}(\tau )\,\rho _{k\alpha
;\,n1}(t_{2})  \notag \\
&&+\frac{R_{2}R_{1}}{\hbar ^{2}}\sum\limits_{m}\sum\limits_{n}\,\sum%
\limits_{k\alpha }\int_{0}^{\infty }dt_{2}\int_{0}^{\infty }d\tau \,\exp
(i\omega \tau )\,U_{m1;\,m0::k\alpha ;\,n1}(\tau )\,\rho _{k\alpha
;\,n2}(t_{2})  \notag \\
&&+\frac{R_{1}R_{2}}{\hbar ^{2}}\sum\limits_{m}\sum\limits_{n}\,\sum%
\limits_{k\alpha }\int_{0}^{\infty }dt_{2}\int_{0}^{\infty }d\tau \,\exp
(i\omega \tau )\,U_{m2;\,m1::k\alpha ;\,n0}(\tau )\,\rho _{k\alpha
;\,n1}(t_{2})  \notag \\
&&
\end{eqnarray}%
\begin{eqnarray}
S(\omega )_{_{t_{2}\geqslant \,t_{1}}} &=&\frac{R_{2}^{2}}{\hbar ^{2}}%
\sum\limits_{m}\sum\limits_{n}\,\sum\limits_{k\alpha }\int_{0}^{\infty
}dt_{1}\int_{0}^{\infty }d\tau \,\exp (-i\omega \tau
)\,U_{m1;\,m2::n1;\,k\alpha }(\tau )\,\rho _{n2;\,k\alpha }(t_{1})  \notag \\
&&+\frac{R_{1}^{2}}{\hbar ^{2}}\sum\limits_{m}\sum\limits_{n}\sum\limits_{k%
\alpha }\int_{0}^{\infty }dt_{1}\int_{0}^{\infty }d\tau \,\exp (-i\omega
\tau )\,U_{m0;\,m1::n0;\,k\alpha }(\tau )\,\rho _{n1;\,k\alpha }(t_{1}) 
\notag \\
&&+\frac{R_{2}R_{1}}{\hbar ^{2}}\sum\limits_{m}\sum\limits_{n}\sum\limits_{k%
\alpha }\int_{0}^{\infty }dt_{1}\int_{0}^{\infty }d\tau \,\exp (-i\omega
\tau )\,U_{m1;\,m2::n0;\,k\alpha }(\tau )\,\rho _{n1;\,k\alpha }(t_{1}) 
\notag \\
&&+\frac{R_{1}R_{2}}{\hbar ^{2}}\sum\limits_{m}\sum\limits_{n}\sum\limits_{k%
\alpha }\int_{0}^{\infty }dt_{1}\int_{0}^{\infty }d\tau \,\exp (-i\omega
\tau )\,U_{m0;\,m1::n1;\,k\alpha }(\tau )\,\rho _{n2;\,k\alpha }(t_{1}) 
.
\notag \\&&
\end{eqnarray}%
\end{widetext}
To obtain the last expressions, the integration variables have been changed
in the first region $(t_{1}\geqslant t_{2})$ to $\tau ,t_{2}$ via the
transformation $t_{1}=\tau +t_{2},t_{2}=t_{2}$ (so the Jacobian equals $+1$)
and in the second region $(t_{2}\geqslant t_{1})$ to $t_{1},\tau $ via the
transformation $t_{1}=t_{1},t_{2}=\tau +t_{1}$ (so the Jacobian equals $+1$).

Since by taking the complex conjugate of the density matrix equations we
have 
\begin{eqnarray}
\rho _{k\alpha ;\,n\nu }(t) &=&\rho _{n\nu ;\,k\alpha }(t)^{\ast } \\
U_{m\lambda ;\,\overline{m-1}\lambda ::k\alpha ;\,\overline{n-1}\nu }(\tau )
&=&U_{\overline{m-1}\lambda ;\,m\lambda ::\overline{n-1}\nu ;\,k\alpha
}(\tau )^{\ast }
,
\end{eqnarray}%
it is not difficult to see that the terms in the last equation occur in
complex conjugate pairs, thus proving that our result for the spectrum is
real.

\begin{widetext}
Hence we find that%
\begin{eqnarray}
S(\omega ) &=&2\,\text{Re}\,\frac{R_{2}^{2}}{\hbar ^{2}}\sum\limits_{m}\sum%
\limits_{n}\sum\limits_{k\alpha }\int_{0}^{\infty }dt_{2}\int_{0}^{\infty
}d\tau \,\exp (i\omega \tau )\,U_{m2;\,m1::k\alpha ;\,n1}(\tau )\,\rho
_{k\alpha ;\,n2}(t_{2})  \notag \\
&&+2\,\text{Re}\,\frac{R_{1}^{2}}{\hbar ^{2}}\sum\limits_{m}\sum\limits_{n}%
\sum\limits_{k\alpha }\int_{0}^{\infty }dt_{2}\int_{0}^{\infty }d\tau \,\exp
(i\omega \tau )\,U_{m1;\,m0::k\alpha ;\,n0}(\tau )\,\rho _{k\alpha
;\,n1}(t_{2})  \notag \\
&&+2\,\text{Re}\,\frac{R_{2}R_{1}}{\hbar ^{2}}\sum\limits_{m}\sum\limits_{n}%
\sum\limits_{k\alpha }\int_{0}^{\infty }dt_{2}\int_{0}^{\infty }d\tau \,\exp
(i\omega \tau )\,U_{m1;\,m0::k\alpha ;\,n1}(\tau )\,\rho _{k\alpha
;\,n2}(t_{2})  \notag \\
&&+2\,\text{Re}\,\frac{R_{1}R_{2}}{\hbar ^{2}}\sum\limits_{m}\sum\limits_{n}%
\sum\limits_{k\alpha }\int_{0}^{\infty }dt_{2}\int_{0}^{\infty }d\tau \,\exp
(i\omega \tau )\,U_{m2;\,m1::k\alpha ;\,n0}(\tau )\,\rho _{k\alpha
;\,n1}(t_{2})  
.
\notag \\&&
\end{eqnarray}

For each contribution the double integrals factorise, each giving a Laplace
transform---albeit for $s$ on the imaginary axis (which may need to be
written as a limiting process). We find that%
\begin{eqnarray}
S(\omega ) &=&2\,\text{Re}\,\frac{R_{2}^{2}}{\hbar ^{2}}\sum\limits_{m}\sum%
\limits_{n}\,\sum\limits_{k\alpha }\widetilde{U}_{m2;\,m1::k\alpha
;\,n1}(-i\omega +\epsilon )\,\widetilde{\rho }_{k\alpha ;\,n2}(\epsilon
^{\prime })  \notag \\
&&+2\,\text{Re}\,\frac{R_{1}^{2}}{\hbar ^{2}}\sum\limits_{m}\sum\limits_{n}%
\sum\limits_{k\alpha }\widetilde{U}_{m1;\,m0::k\alpha ;\,n0}(-i\omega
+\epsilon )\,\widetilde{\rho }_{k\alpha ;\,n1}(\epsilon ^{\prime })  \notag
\\
&&+2\,\text{Re}\,\frac{R_{2}R_{1}}{\hbar ^{2}}\sum\limits_{m}\sum\limits_{n}%
\sum\limits_{k\alpha }\widetilde{U}_{m1;\,m0::k\alpha ;\,n1}(-i\omega
+\epsilon )\,\widetilde{\rho }_{k\alpha ;\,n2}(\epsilon ^{\prime })  \notag
\\
&&+2\,\text{Re}\,\frac{R_{1}R_{2}}{\hbar ^{2}}\sum\limits_{m}\sum\limits_{n}%
\sum\limits_{k\alpha }\widetilde{U}_{m2;\,m1::k\alpha ;\,n0}(-i\omega
+\epsilon )\,\widetilde{\rho }_{k\alpha ;\,n1}(\epsilon ^{\prime })  
,
\notag\\&&
\end{eqnarray}%
\end{widetext}
where the limits $\epsilon ,\epsilon ^{\prime }\rightarrow 0$ are understood.

The initial conditions will lead to restrictions on the terms to be summed
over. Of course the quantity $\alpha $ already only sums over the three
atomic states. As shown previously for the present initial conditions, the
only $\rho _{k\alpha ;\,n2}(t)$ and $\rho _{k\alpha ;\,n1}(t)$ that could be
non-zero are those in the $(2,2)$, $(1,1)$ and $(0,0)$ coupled sets---for $%
(2,2)$ these are $\rho _{02;\,02}$, $\rho _{02;\,11}$, $\rho _{02;\,20}$, $%
\rho _{11;\,02}$, $\rho _{11;\,11}$, $\rho _{11;\,20}$, $\rho _{20;\,02}$, $%
\rho _{20;\,11}$ and $\rho _{20;\,20}$; for the $(1,1)$ set we have $\rho
_{01;\,01},\rho _{01;\,10},\rho _{10;\,01}$\ and $\rho _{10;\,10}$\ and for
the $(0,0)$\ set we have $\rho _{00;\,00}(t)$---corresponding to the system
states $\left\vert 0;2\right\rangle ,\left\vert 1;1\right\rangle $ and $%
\left\vert 0;1\right\rangle $ being the only ones populated during the decay
process. The only non-zero elements of the form $\rho _{k\alpha ;\,n2}(t)$
are $\rho _{02;\,02}$, $\rho _{11;\,02}$ and $\rho _{20;\,02}$ from the $%
(2,2)$ set. \ The only non-zero elements of the form $\rho _{k\alpha
;\,n1}(t)$ are $\rho _{02;\,11}$, $\rho _{11;\,11}$ and $\rho _{20;\,11}$
from the $(2,2)$ set and $\rho _{01;\,01}$ and $\rho _{10;\,01}$ from the $%
(1,1)$ set. \ No elements from the $(0,0)$ set are involved. Only the
Laplace transforms of these elements will appear in the expression for the
spectrum, and this restricts the related values of $k,\alpha $ and $n$. The
first and third terms in the last expression for the spectrum will thus give
three different $k,\alpha $ and $n$ contributions, whilst the second and
fourth terms will give five. Thus we may write%
\begin{widetext}
\begin{eqnarray}
S(\omega ) &=&S_{22}(\omega )+S_{11}(\omega )+S_{21}(\omega )+S_{12}(\omega )
\end{eqnarray}
\begin{eqnarray}
S_{22}(\omega ) =2\,\text{Re}\,\frac{R_{2}^{2}}{\hbar ^{2}}  
\sum\limits_{m}&&[\widetilde{U}_{m2;\,m1::20;\,01}(-i\omega +\epsilon
)\,\widetilde{\rho }_{20;\,02}(\epsilon ^{\prime })+\widetilde{U}%
_{m2;\,m1::11;\,01}(-i\omega +\epsilon )\,\widetilde{\rho }%
_{11;\,02}(\epsilon ^{\prime })  \notag \\
&&+\widetilde{U}_{m2;\,m1::02;\,01}(-i\omega +\epsilon )\,\widetilde{\rho }%
_{02;\,02}(\epsilon ^{\prime })] 
\end{eqnarray}
\begin{eqnarray}
S_{11}(\omega ) = 2\,\text{Re}\,\frac{R_{1}^{2}}{\hbar ^{2}}  
\sum\limits_{m}&&[\widetilde{U}_{m1;\,m0::02;\,10}(-i\omega +\epsilon
)\,\widetilde{\rho }_{02;\,11}(\epsilon ^{\prime })+\widetilde{U}%
_{m1;\,m0::11;\,10}(-i\omega +\epsilon )\,\widetilde{\rho }%
_{11;\,11}(\epsilon ^{\prime })  \notag \\
&&+\widetilde{U}_{m1;\,m0::20;\,10}(-i\omega +\epsilon )\,\widetilde{\rho }%
_{20;\,11}(\epsilon ^{\prime })+\widetilde{U}_{m1;\,m0::01;\,00}(-i\omega
+\epsilon )\,\widetilde{\rho }_{01;\,01}(\epsilon ^{\prime })  \notag \\
&&+\widetilde{U}_{m1;\,m0::10;\,00}(-i\omega +\epsilon )\,\widetilde{\rho }%
_{10;\,01}(\epsilon ^{\prime })] 
\end{eqnarray}
\begin{eqnarray}
S_{21}(\omega ) =2\,\text{Re}\,\frac{R_{2}R_{1}}{\hbar ^{2}}  
\sum\limits_{m}&&[\widetilde{U}_{m1;\,m0::02;\,01}(-i\omega +\epsilon
)\,\widetilde{\rho }_{02;\,02}(\epsilon ^{\prime })+\widetilde{U}%
_{m1;\,m0::11;\,01}(-i\omega +\epsilon )\,\widetilde{\rho }%
_{11;\,02}(\epsilon ^{\prime })  \notag \\
&&+\widetilde{U}_{m1;\,m0::20;\,01}(-i\omega +\epsilon )\,\widetilde{\rho }%
_{20;\,02}(\epsilon ^{\prime })] 
\end{eqnarray}
\begin{eqnarray}
S_{12}(\omega )  = 2\,\text{Re}\,\frac{R_{1}R_{2}}{\hbar ^{2}}  
\sum\limits_{m}&&[\widetilde{U}_{m2;\,m1::02;\,10}(-i\omega +\epsilon
)\,\widetilde{\rho }_{02;\,11}(\epsilon ^{\prime })+\widetilde{U}%
_{m2;\,m1::11;\,10}(-i\omega +\epsilon )\,\widetilde{\rho }%
_{11;\,11}(\epsilon ^{\prime })  \notag \\
&&+\widetilde{U}_{m2;\,m1::20;\,10}(-i\omega +\epsilon )\,\widetilde{\rho }%
_{20;\,11}(\epsilon ^{\prime })  \notag \\
&&+\widetilde{U}_{m2;\,m1::01;\,00}(-i\omega +\epsilon )\,\widetilde{\rho }%
_{01;\,01}(\epsilon ^{\prime })+\widetilde{U}_{m2;\,m1::10;\,00}(-i\omega
+\epsilon )\,\widetilde{\rho }_{10;\,01}(\epsilon ^{\prime })]  
.
\notag \\ &&
\end{eqnarray}%
This result can be rewritten to group separately the terms where the $%
\widetilde{\rho }$ are from the $(2,2)$ from those from the $(1,1)$ set:
\begin{eqnarray}
S(\omega ) &=&S_{2}^{\prime }(\omega )+S_{6}^{\prime }(\omega ) 
\end{eqnarray}
\begin{eqnarray}
S_{2}^{\prime }(\omega ) =2\,\text{Re}\,\frac{R_{1}^{2}}{\hbar ^{2}} 
\sum\limits_{m}&&[\widetilde{U}_{m1;\,m0::01;\,00}(-i\omega +\epsilon
)\,\widetilde{\rho }_{01;\,01}(\epsilon ^{\prime })+\widetilde{U}%
_{m1;\,m0::10;\,00}(-i\omega +\epsilon )\,\widetilde{\rho }%
_{10;\,01}(\epsilon ^{\prime })]  \notag \\
+2\,\text{Re}\,\frac{R_{1}R_{2}}{\hbar ^{2}}  
\sum\limits_{m}&&[\widetilde{U}_{m2;\,m1::01;\,00}(-i\omega +\epsilon
)\,\widetilde{\rho }_{01;\,01}(\epsilon ^{\prime })+\widetilde{U}%
_{m2;\,m1::10;\,00}(-i\omega +\epsilon )\,\widetilde{\rho }%
_{10;\,01}(\epsilon ^{\prime })]  
\end{eqnarray}
\begin{eqnarray}
S_{6}^{\prime }(\omega ) =2\,\text{Re}\,\frac{R_{2}^{2}}{\hbar ^{2}}%
\sum\limits_{m}&&[\widetilde{U}_{m2;\,m1::20;\,01}(-i\omega +\epsilon )\,%
\widetilde{\rho }_{20;\,02}(\epsilon ^{\prime })  \notag \\
&&+\widetilde{U}_{m2;\,m1::11;\,01}(-i\omega +\epsilon )\,\widetilde{\rho }%
_{11;\,02}(\epsilon ^{\prime })+\widetilde{U}_{m2;\,m1::02;\,01}(-i\omega
+\epsilon )\,\widetilde{\rho }_{02;\,02}(\epsilon ^{\prime })]  \notag \\
+2\,\text{Re}\,\frac{R_{1}^{2}}{\hbar ^{2}}  
\sum\limits_{m}&&[\widetilde{U}_{m1;\,m0::02;\,10}(-i\omega +\epsilon
)\,\widetilde{\rho }_{02;\,11}(\epsilon ^{\prime })+\widetilde{U}%
_{m1;\,m0::11;\,10}(-i\omega +\epsilon )\,\widetilde{\rho }%
_{11;\,11}(\epsilon ^{\prime })  \notag \\
&&+\widetilde{U}_{m1;\,m0::20;\,10}(-i\omega +\epsilon )\,\widetilde{\rho }%
_{20;\,11}(\epsilon ^{\prime })]  \notag \\
+2\,\text{Re}\,\frac{R_{2}R_{1}}{\hbar ^{2}}  
\sum\limits_{m}&&[\widetilde{U}_{m1;\,m0::02;\,01}(-i\omega +\epsilon
)\,\widetilde{\rho }_{02;\,02}(\epsilon ^{\prime })+\widetilde{U}%
_{m1;\,m0::11;\,01}(-i\omega +\epsilon )\,\widetilde{\rho }%
_{11;\,02}(\epsilon ^{\prime })  \notag \\
&&+\widetilde{U}_{m1;\,m0::20;\,01}(-i\omega +\epsilon )\,\widetilde{\rho }%
_{20;\,02}(\epsilon ^{\prime })]  \notag \\
+2\,\text{Re}\,\frac{R_{1}R_{2}}{\hbar ^{2}}%
\sum\limits_{m}&&[%
\widetilde{U}_{m2;\,m1::02;\,10}(-i\omega +\epsilon )\,\widetilde{\rho }%
_{02;\,11}(\epsilon ^{\prime })  \notag \\
&&+\widetilde{U}_{m2;\,m1::11;\,10}(-i\omega +\epsilon )\,\widetilde{\rho }%
_{11;\,11}(\epsilon ^{\prime })+\widetilde{U}_{m2;\,m1::20;\,10}(-i\omega
+\epsilon )\,\widetilde{\rho }_{20;\,11}(\epsilon ^{\prime })]  
.
\notag \\ &&
\end{eqnarray}
\end{widetext}
The $S_{2}^{\prime }(\omega )$ terms involve $\widetilde{U}$ of the form $%
\widetilde{U}_{n,m::1,0}$\ since $\rho _{01;\,00}$ and $\rho _{10;\,00}$ are
both in the $(1,0)$\ set. From the discussion for the Case A spectrum the
only non-zero $\widetilde{U}_{n,m::1,0}$ are of the form $\widetilde{U}%
_{1,0::1,0}$. Since there are no elements in the $(1,0)$\ set of the form $%
\rho _{m2;\,m1}$ then the second line terms $\widetilde{U}%
_{m2;\,m1::01;\,00}(-i\omega +\epsilon )$ and $\widetilde{U}%
_{m2;\,m1::10;\,00}(-i\omega +\epsilon )$ are both zero. As the only element
in the $(1,0)$\ set of the form $\rho _{m1;\,m0}$ \ is $\rho _{01;\,00}$
then the first line terms only involve one value of $m$, which is $0$. This
gives%
\begin{eqnarray}
S_{2}^{\prime }(\omega ) &=&2\,\text{Re}\,\frac{R_{1}^{2}}{\hbar ^{2}}[%
\widetilde{U}_{01;\,00::01;\,00}(-i\omega +\epsilon )\,\widetilde{\rho }%
_{01;\,01}(\epsilon ^{\prime })
\nonumber\\&& %
+\widetilde{U}_{01;\,00::10;\,00}(-i\omega
+\epsilon )\,\widetilde{\rho }_{10;\,01}(\epsilon ^{\prime })]  
.
\notag \\&&
\end{eqnarray}
The $S_{6}^{\prime }(\omega )$ terms involve $\widetilde{U}$ of the form $%
\widetilde{U}_{n,m::2,1}$\ since $\rho _{20;\,01}$, $\rho _{11;\,01}$, $\rho
_{02;\,01}$, $\rho _{02;\,10}$, $\rho _{11;\,10}$, $\rho _{20;\,10}$, $\rho
_{02;\,01}$, $\rho _{11;\,01}$, $\rho _{20;\,01}$, $\rho _{02;\,10}$, $\rho
_{11;\,10}$ and $\rho _{20;\,10}$ all in the $(2,1)$\ set (three repeated
terms). From the discussion for the Case A spectrum the only non-zero $%
\widetilde{U}_{n,m::2,1}$ are of the form $\widetilde{U}_{2,1::2,1}$ and $%
\widetilde{U}_{1,0::2,1}$. Now in the $(1,0)$\ set there are no elements of
the form $\rho _{m2;\,m1},$ and the only element of the form $\rho _{m1;\,m0}$
is $\rho _{01;\,00}$. However in the $(2,1)$\ set there is one element of
the form $\rho _{m2;\,m1}$---which is $\rho _{02;\,01}$, and there is also
one element of the form $\rho _{m1;\,m0}$---which is $\rho _{11;\,10}$. Thus
the first and fourth lines for $S_{6}^{\prime }(\omega )$ will only involve
single $\widetilde{U}$, namely $\widetilde{U}_{02;\,01::20;\,01}(-i\omega
+\epsilon )$, $\widetilde{U}_{02;\,01::11;\,01}(-i\omega +\epsilon )$ and $%
\widetilde{U}_{02;\,01::02;\,01}(-i\omega +\epsilon )$\ in the first line,
and $\widetilde{U}_{02;\,01::02;\,10}(-i\omega +\epsilon )$, $\widetilde{U}%
_{02;\,01::11;\,10}(-i\omega +\epsilon )$ and $\widetilde{U}%
_{02;\,01::20;\,10}(-i\omega +\epsilon )$ in the fourth line. However, in
the second and third lines two $\widetilde{U}$ are involved, namely $%
\widetilde{U}_{01;\,00::02;\,10}(-i\omega +\epsilon )$, $\widetilde{U}%
_{01;\,00::11;\,10}(-i\omega +\epsilon )$ and $\widetilde{U}%
_{01;\,00::20;\,10}(-i\omega +\epsilon )$ or $\widetilde{U}%
_{11;\,10::02;\,10}(-i\omega +\epsilon )$, $\widetilde{U}_{11;\,10::11;%
\,10}(-i\omega +\epsilon )$ and $\widetilde{U}_{11;\,10::20;\,10}(-i\omega
+\epsilon )$ for the second line, and $\widetilde{U}_{01;\,00::02;\,01}(-i%
\omega +\epsilon )$, $\widetilde{U}_{01;\,00::11;\,01}(-i\omega +\epsilon
)$ and $\widetilde{U}_{01;\,00::20;\,01}(-i\omega +\epsilon )$ or $%
\widetilde{U}_{11;\,10::02;\,01}(-i\omega +\epsilon )$, $\widetilde{U}%
_{11;\,10::11;\,01}(-i\omega +\epsilon )$ and $\widetilde{U}%
_{11;\,10::20;\,01}(-i\omega +\epsilon )$ for the third line. This gives 
\begin{widetext}
\begin{eqnarray}
S_{6}^{\prime }(\omega ) = 2\,\text{Re}\,\frac{R_{2}^{2}}{\hbar ^{2}} 
&&\lbrack \widetilde{U}_{02;\,01::20;\,01}(-i\omega +\epsilon )\,%
\widetilde{\rho }_{20;\,02}(\epsilon ^{\prime })+\widetilde{U}%
_{02;\,01::11;\,01}(-i\omega +\epsilon )\,\widetilde{\rho }%
_{11;\,02}(\epsilon ^{\prime })  
+\widetilde{U}_{02;\,01::02;\,01}(-i\omega +\epsilon )\,\widetilde{\rho }%
_{02;\,02}(\epsilon ^{\prime })]  \notag \\
+2\,\text{Re}\,\frac{R_{1}^{2}}{\hbar ^{2}}&&[\{\widetilde{U}%
_{01;\,00::02;\,10}(-i\omega +\epsilon )\,+\widetilde{U}_{11;\,10::02;%
\,10}(-i\omega +\epsilon )\}\,\widetilde{\rho }_{02;\,11}(\epsilon ^{\prime
})  \notag \\
&&+\{\widetilde{U}_{01;\,00::11;\,10}(-i\omega +\epsilon )+\widetilde{U}%
_{11;\,10::11;\,10}(-i\omega +\epsilon )\}\,\widetilde{\rho }%
_{11;\,11}(\epsilon ^{\prime })  \notag \\
&&+\{\widetilde{U}_{01;\,00::20;\,10}(-i\omega +\epsilon )+\widetilde{U}%
_{11;\,10::20;\,10}(-i\omega +\epsilon )\}\,\widetilde{\rho }%
_{20;\,11}(\epsilon ^{\prime })]  \notag \\
+2\,\text{Re}\,\frac{R_{2}R_{1}}{\hbar ^{2}}  
&&\lbrack \{\widetilde{U}_{01;\,00::02;\,01}(-i\omega +\epsilon )+%
\widetilde{U}_{11;\,10::02;\,01}(-i\omega +\epsilon )\}\,\widetilde{\rho }%
_{02;\,02}(\epsilon ^{\prime })  \notag \\
&&+\{\widetilde{U}_{01;\,00::11;\,01}(-i\omega +\epsilon )+\widetilde{U}%
_{11;\,10::11;\,01}(-i\omega +\epsilon )\}\,\widetilde{\rho }%
_{11;\,02}(\epsilon ^{\prime })  \notag \\
&&+\{\widetilde{U}_{01;\,00::20;\,01}(-i\omega +\epsilon )+\widetilde{U}%
_{11;\,10::20;\,01}(-i\omega +\epsilon )\}\,\widetilde{\rho }%
_{20;\,02}(\epsilon ^{\prime })]  \notag \\
+2\,\text{Re}\,\frac{R_{1}R_{2}}{\hbar ^{2}}  
&&\lbrack \widetilde{U}_{02;\,01::02;\,10}(-i\omega +\epsilon )\,%
\widetilde{\rho }_{02;\,11}(\epsilon ^{\prime })+\widetilde{U}%
_{02;\,01::11;\,10}(-i\omega +\epsilon )\,\widetilde{\rho }%
_{11;\,11}(\epsilon ^{\prime })  \notag \\
&&+\widetilde{U}_{02;\,01::20;\,10}(-i\omega +\epsilon )\,\widetilde{\rho }%
_{20;\,11}(\epsilon ^{\prime })]
.
\end{eqnarray}
Combining the results for $S_{2}^{\prime }(\omega )$ and $S_{6}^{\prime
}(\omega )$ we obtain for the spectrum 
\begin{eqnarray}
S(\omega ) =2\,\text{Re}\,\frac{R_{2}^{2}}{\hbar ^{2}} && \lbrack 
\widetilde{U}_{02;\,01::20;\,01}(-i\omega +\epsilon )\,\widetilde{\rho }%
_{20;\,02}(\epsilon ^{\prime })+\widetilde{U}_{02;\,01::11;\,01}(-i\omega
+\epsilon )\,\widetilde{\rho }_{11;\,02}(\epsilon ^{\prime })  
+\widetilde{U}_{02;\,01::02;\,01}(-i\omega +\epsilon )\,\widetilde{\rho }%
_{02;\,02}(\epsilon ^{\prime })]  \notag \\
+2\,\text{Re}\,\frac{R_{1}^{2}}{\hbar ^{2}} && \lbrack \widetilde{U}%
_{01;\,00::01;\,00}(-i\omega +\epsilon )\,\widetilde{\rho }%
_{01;\,01}(\epsilon ^{\prime })+\widetilde{U}_{01;\,00::10;\,00}(-i\omega
+\epsilon )\,\widetilde{\rho }_{10;\,01}(\epsilon ^{\prime })  \notag \\
&&+\{\widetilde{U}_{01;\,00::02;\,10}(-i\omega +\epsilon )\,+\widetilde{U}%
_{11;\,10::02;\,10}(-i\omega +\epsilon )\}\,\widetilde{\rho }%
_{02;\,11}(\epsilon ^{\prime })  \notag \\
&&+\{\widetilde{U}_{01;\,00::11;\,10}(-i\omega +\epsilon )+\widetilde{U}%
_{11;\,10::11;\,10}(-i\omega +\epsilon )\}\,\widetilde{\rho }%
_{11;\,11}(\epsilon ^{\prime })  \notag \\
&&+\{\widetilde{U}_{01;\,00::20;\,10}(-i\omega +\epsilon )+\widetilde{U}%
_{11;\,10::20;\,10}(-i\omega +\epsilon )\}\,\widetilde{\rho }%
_{20;\,11}(\epsilon ^{\prime })]  \notag \\
+2\,\text{Re}\,\frac{R_{2}R_{1}}{\hbar ^{2}} && \lbrack \{\widetilde{U}%
_{01;\,00::02;\,01}(-i\omega +\epsilon )+\widetilde{U}_{11;\,10::02;\,01}(-i%
\omega +\epsilon )\}\,\widetilde{\rho }_{02;\,02}(\epsilon ^{\prime }) 
\notag \\
&&+\{\widetilde{U}_{01;\,00::11;\,01}(-i\omega +\epsilon )+\widetilde{U}%
_{11;\,10::11;\,01}(-i\omega +\epsilon )\}\,\widetilde{\rho }%
_{11;\,02}(\epsilon ^{\prime })  \notag \\
&&+\{\widetilde{U}_{01;\,00::20;\,01}(-i\omega +\epsilon )+\widetilde{U}%
_{11;\,10::20;\,01}(-i\omega +\epsilon )\}\,\widetilde{\rho }%
_{20;\,02}(\epsilon ^{\prime })]  \notag \\
+2\,\text{Re}\,\frac{R_{1}R_{2}}{\hbar ^{2}} && \lbrack \widetilde{U}%
_{02;\,01::02;\,10}(-i\omega +\epsilon )\,\widetilde{\rho }%
_{02;\,11}(\epsilon ^{\prime })+\widetilde{U}_{02;\,01::11;\,10}(-i\omega
+\epsilon )\,\widetilde{\rho }_{11;\,11}(\epsilon ^{\prime })  \notag \\
&&+\widetilde{U}_{02;\,01::20;\,10}(-i\omega +\epsilon )\,\widetilde{\rho }%
_{20;\,11}(\epsilon ^{\prime })]
.
\end{eqnarray}
\end{widetext}


\begin{thebibliography}{99}
\section*{References}

\bibitem{Purcell46a} E. M. Purcell, Phys. Rev. \textbf{69} 681 (1946).

\bibitem{Berman94a} P.R. Berman, editor. \textit{Cavity Quantum
Electrodynamics, }(Academic Press, New York, 1994).

\bibitem{Lambropoulos00a} P. Lambropoulos, G.M. Nikolopoulos, T.R. Nielsen
and S. Bay, Rep. Prog. Phys. \textbf{63} 455 (2000).

\bibitem{Walther06a} H. Walther, B.T.H. Varcoe, B-G. Englert and T. Becker,
Rep. Prog. Phys. \textbf{69} 1325 (2006).

\bibitem{Weisskopf30a} V. Weisskopf and E. Wigner, Z. Phys. \textbf{63} 54
(1930).

\bibitem{JaynesCummings63a} E.T. Jaynes and F.W. Cummings, Proc. Inst.
Elect. Eng. \textbf{51} 89 (1963).

\bibitem{Cohen-Tannoudji77a} C. Cohen-Tannoudji, in \textit{Frontiers in
Laser Spectroscopy}, eds. R. Balian, S. Haroche and S. Liberman
(North-Holland, Amsterdam, 1977), Vol.\ 1, p1.

\bibitem{Sanchez83a} J.J. Sanchez-Mondragon, N.B. Narozhny and J.H. Eberly,
Phys. Rev. Lett. \textbf{51} 550 (1983).

\bibitem{Eberly77a} J.H. Eberly and K. Wodkiewicz, J. Opt. Soc. Amer. B 
\textbf{2} 1252 (1977).

\bibitem{Agarwal86a} G.S. Agarwal and R.R. Puri, Phys. Rev. A \textbf{33},
1757 (1986).

\bibitem{Lewenstein88a} M. Lewenstein, J. Zakrzewski and T. Mossberg, Phys.
Rev. A \textbf{38} 808 (1988).

\bibitem{Lax63a67a} M. Lax, Phys. Rev. \textbf{129}, 2342 (1963), 
\textit{ibid.}\ Phys. Rev. \textbf{157} 213 (1967).

\bibitem{Carmichael89a} H.J. Carmichael, R.J. Brecha, M.G. Raizen, H.J.
Kimble and P.R. Rice, Phys. Rev. A \textbf{40} 5516 (1991).

\bibitem{Childs94a} J.J. Childs, K. An, R.R. Dasari and M.S. Feld, in 
\textit{Cavity Quantum Electrodynamics, }edited by P.R. Berman (Academic
Press, New York, 1994), p.325.

\bibitem{Carmichael94a} H.J. Carmichael, L. Tian, W. Ren and P. Alsing, in 
\textit{Cavity Quantum Electrodynamics, }edited by P.R. Berman (Academic
Press, New York, 1994), p.381.

\bibitem{Thompson92a} R.J. Thompson, G. Rempe and H.J. Kimble, Phys. Rev.
Lett. \textbf{68} 1132 (1992).

\bibitem{Boca04a} A. Boca, R. Miller, K.M. Birnbaum, A.D. Boozer, J.
McKeever and H.J. Kimble. Phys. Rev. Lett. \textbf{93} 233603 (2004).

\bibitem{Brune96a} M. Brune, F. Schmidt-Kaler, A. Maali, J. Dreyer, E.
Hagley, J.M. Raimond and S. Haroche. Phys. Rev. Lett. \textbf{76} 1800
(1996).

\bibitem{Rempe87a} G. Rempe, H. Walther and N. Klein, Phys. Rev. Lett. 
\textbf{58} 353 (1987).

\bibitem{Eberly80a} J.H. Eberly, N.B. Narozhny and J.J. Sanchez-Mondragon.
Phys. Rev. Lett. \textbf{44} 1323 (1980).

\bibitem{Barnett86a} S.M. Barnett and P.L. Knight, Phys. Rev. A \textbf{33}
2444 (1986).

\bibitem{Mollow69a72a} B.R. Mollow, Phys. Rev \textbf{188} 1969 (1969), 
\textit{ibid.}\ Phys. Rev. A \textbf{5} 2217 (1972).

\bibitem{Agarwal91a} G.S. Agarwal, R.K. Bullough and N. Nayak, Opt. Comm. 
\textbf{85} 202 (1991).

\bibitem{John94a} S. John and T. Quang, Phys. Rev. A \textbf{50} 1764 (1994).

\bibitem{Ashraf94a} M.M. Ashraf, Phys. Rev. A \textbf{50} 741 (1994).

\bibitem{Bay98a} S. Bay and P. Lambropoulos, Opt. Comm. \textbf{146} 130
(1998).

\bibitem{Paspalakis99a} E. Paspalakis, D.G. Angelakis and P.L. Knight, .Opt.
Comm. \textbf{172} 229 (1999).

\bibitem{Zhou05a} Q-C. Zhou, S-N. Zhu and N-B Ming, J. Phys. B: At. Mol.
Opt. Phys. \textbf{38} 4309 (2005).

\bibitem{Garraway06a} B.M. Garraway and B.J Dalton, J. Phys. B: At. Mol.
Opt. Phys. \textbf{39} S767 (2006).

\bibitem{Dalton01a} B.J. Dalton, S.M. Barnett and B.M. Garraway, Phys. Rev.
A \textbf{64} 053813 (2001).

\bibitem{Swain02a} S. Swain and Z. Ficek, editors. Special Issue: \textit{%
Quantum Interference}. J. Mod. Opt. \textbf{49} 1/2 (2002).

\bibitem{Glauber65a} R.J. Glauber, in \textit{Quantum Optics and
Electronics}, edited by C. DeWitt, A. Blandin and C. Cohen-Tannoudji
(Gordon and Breach, London, 1965), p65.

\bibitem{Cresser83a} J.D. Cresser, Phys. Rep. \textbf{94} 47 (1983).

\bibitem{Dalton99a} B.J. Dalton, S.M. Barnett and P.L. Knight, J. Mod. Opt. 
\textbf{46} 1315 (1999).

\bibitem{Brown01a} S. Brown and B.J. Dalton, J. Mod. Opt. \textbf{48} 597
(2001).

\bibitem{Dalton96a} B.J. Dalton, E.S. Guerra and P.L. Knight, Phys. Rev. A 
\textbf{54} 2292 (1996).

\bibitem{Dalton97a} B.J. Dalton and M. Babiker, Phys. Rev. A \textbf{56} 905
(1997).

\bibitem{Louisell73a} W.H. Louisell, \textit{Quantum Statistical 
Properties of Radiation }(Wiley, New York, 1973), p57.

\bibitem{Linington06a} I.E. Linington and B.M. Garraway, J. Phys. B: At.
Mol. Opt. Phys. \textbf{39} 3383 (2006).

\bibitem{Walls94a} D.F. Walls and G.J. Milburn, \textit{Quantum Optics}
(Springer-Verlag, Berlin, 1994), p118.

\bibitem{Dalton79a} B.J. Dalton, \textit{Lecture Notes in Quantum Optics }
(University of Queensland, Brisbane, 1979). Unpublished.

\end{thebibliography}
\end{document}